%% file: Draft.tex
\documentclass[a4paper,11pt]{article}

\input{./auxi/packages.tex}

\input{./auxi/commands.tex}
\input{./figures/images.tex}

\hypersetup{hypertexnames=false}

\begin{document} 

\thispagestyle{empty}

\vspace*{-.6in}
\begin{flushright}
    
\end{flushright}

\vspace{1cm}
{\large
\begin{center}
    {\Large \bf The pinning-field line defect \\
    at finite temperature
    }\\
\end{center}}

\vspace{0.5cm}

\begin{center}
    {Julien Barrat$^{a,}\footnote{julien.barrat@princeton.edu}$ and Simone Giombi$^{a,}\footnote{sgiombi@princeton.edu}$}\\[0.5cm] 
    { \small
    $^{a}$ \textit{Department of Physics, Princeton University, Princeton, NJ 08544, USA}
    }
\vspace{1cm} 

   \bf Abstract
\end{center}

\begin{abstract}
\noindent
We study the pinning-field line defect in the critical $\mathrm{O}(N)$ model at finite temperature, describing a static symmetry-breaking impurity at a quantum critical point.
Using the operator product expansion (OPE), analyticity, and thermal periodicity, we construct a manifestly periodic expansion relating heavy-operator asymptotics to low-lying thermal OPE data.
We determine thermal correlators and one-point coefficients in the $\veps$-expansion and, combining these results with the two-dimensional Ising endpoint, obtain the Pad\'e estimate $\vev{\hphi} \sim 1.701/\beta^{\smash{\Dh_{\smash{\hphi}}}}$ for the lowest defect operator in the three-dimensional Ising model.
We conjecture that truncating the periodic OPE to a few low-lying operators accurately approximates defect correlators.
To test this, we numerically solve the large-$N$ thermal saddle and extract defect one-point coefficients across $2<d<4$.
In three dimensions, we compute the tilt and longitudinal singlet two-point functions, finding very good agreement with the periodic OPE truncated to the identity and lowest singlet.
Finally, we extend the analysis to finite spatial volume using antipodal defects on the sphere.
\end{abstract}

\newpage

\setcounter{footnote}{0}
\setcounter{tocdepth}{3}

\noindent

\tableofcontents

\setcounter{page}{1}

\input{sections/Introduction}
\input{sections/LineDefectCorrelatorsInThermalGeometries}
\input{sections/EpsExpansionAndInterpolationTo3d}
\input{sections/LargeN}
\input{sections/Discussion}

\acknowledgments

We are particularly grateful to Daniele Artico, Deniz Bozkurt, Zohar Komargodski, Enrico Marchetto, Alessio Miscioscia, Elli Pomoni, Fedor Popov, Nika Sokolova for valuable discussions related to this work.
Artificial intelligence tools were used to assist with calculations; all reported results have been independently verified by the authors.
JB is funded by the Deutsche Forschungsgemeinschaft (DFG, German Research Foundation) – Projektnummer 570367913.

\appendix

\input{appendices/FreeTheory}
\input{appendices/FeynmanDiagrams}
\input{appendices/WittenDiagrams}
\input{appendices/Numerics}

\bibliography{auxi/biblio}
\bibliographystyle{auxi/JHEP}

\end{document}

%% file: auxi/packages.tex
\usepackage{./auxi/jheppub}
\usepackage[utf8]{inputenc}
\usepackage[ngerman, english]{babel}
\usepackage[T1]{fontenc}
\usepackage[titletoc]{appendix}
\usepackage{titletoc}
\usepackage{graphicx}
\usepackage{subcaption} %
\usepackage[usenames,dvipsnames]{xcolor}
\usepackage{amsmath,amssymb,amsthm}
\usepackage{bm}
\usepackage{float}
\usepackage[format=plain,labelfont={bf,it},textfont=it]{caption}
\usepackage{multirow}
\usepackage{slashed} %
\usepackage{tabularx}
\usepackage{bbold}
\usepackage[shortlabels]{enumitem}
\usepackage{tikz}
\usetikzlibrary{positioning,arrows,calc}
\usetikzlibrary{arrows.meta}
\usetikzlibrary{decorations.pathmorphing}
\usetikzlibrary{decorations.markings}
\usetikzlibrary{shapes.geometric}
\usetikzlibrary{patterns}
\tikzset{
  snake it/.style={
    decorate, 
    decoration=snake,
    segment length=3
  }
}
\usepackage{xparse}
\usepackage{pdfpages} %
\usepackage{titlesec} %
\usepackage{fancyhdr} %
\usepackage{emptypage} %
\usepackage{dsfont} %
\usepackage{tcolorbox}

%% file: auxi/commands.tex
\definecolor{DarkBlueGrey}{RGB}{76,94,107}
\definecolor{MediumBlueGrey}{RGB}{110,135,153}
\definecolor{LightBlueGrey}{RGB}{134,163,184}
\definecolor{WCOrange}{RGB}{242,146,29}
\definecolor{SCRed}{RGB}{179,48,48}
\definecolor{SCRed2}{RGB}{230,100,100}
\definecolor{VertexColor}{RGB}{242,146,29}
\definecolor{GluonColor}{RGB}{255,172,172}
\definecolor{SEColor}{RGB}{134,163,184}

\definecolor{BGBox}{RGB}{255,254,230}
\definecolor{PlaneColor}{RGB}{230,230,230}
\definecolor{BlobColor}{RGB}{190,180,230}

\newcommand{\pd}{\partial}

\newcommand{\vev}[1]{\langle\, #1 \, \rangle}

\def\veps{\varepsilon}

\newcommand{\Oh}{{\hat{\mathcal{O}}}}
\newcommand{\Dh}{{\hat{\Delta}}}

\def\Bm{{\mathcal{B}}}
\def\Cm{{\mathcal{C}}}
\def\Dm{{\mathcal{D}}}

\def\Hm{{\mathcal{H}}}

\def\Nm{{\mathcal{N}}}
\def\Om{{\mathcal{O}}}

\def\Vm{{\mathcal{V}}}

\usepackage{bbm}

\def\veps{\varepsilon}

\newcommand\hphi{{\hat{\phi}}}

\newif\ifstartcompletesineup
\newif\ifendcompletesineup
\pgfkeys{
    /pgf/decoration/.cd,
    start up/.is if=startcompletesineup,
    start up=true,
    start up/.default=true,
    start down/.style={/pgf/decoration/start up=false},
    end up/.is if=endcompletesineup,
    end up=true,
    end up/.default=true,
    end down/.style={/pgf/decoration/end up=false}
}
\pgfdeclaredecoration{complete sines}{initial}
{
    \state{initial}[
        width=+0pt,
        next state=upsine,
        persistent precomputation={
            \ifstartcompletesineup
                \pgfkeys{/pgf/decoration automaton/next state=upsine}
                \ifendcompletesineup
                    \pgfmathsetmacro\matchinglength{
                        0.5*\pgfdecoratedinputsegmentlength / (ceil(0.5* \pgfdecoratedinputsegmentlength / \pgfdecorationsegmentlength) )
                    }
                \else
                    \pgfmathsetmacro\matchinglength{
                        0.5 * \pgfdecoratedinputsegmentlength / (ceil(0.5 * \pgfdecoratedinputsegmentlength / \pgfdecorationsegmentlength ) - 0.499)
                    }
                \fi
            \else
                \pgfkeys{/pgf/decoration automaton/next state=downsine}
                \ifendcompletesineup
                    \pgfmathsetmacro\matchinglength{
                        0.5* \pgfdecoratedinputsegmentlength / (ceil(0.5 * \pgfdecoratedinputsegmentlength / \pgfdecorationsegmentlength ) - 0.4999)
                    }
                \else
                    \pgfmathsetmacro\matchinglength{
                        0.5 * \pgfdecoratedinputsegmentlength / (ceil(0.5 * \pgfdecoratedinputsegmentlength / \pgfdecorationsegmentlength ) )
                    }
                \fi
            \fi
            \setlength{\pgfdecorationsegmentlength}{\matchinglength pt}
        }] {}
    \state{downsine}[width=\pgfdecorationsegmentlength,next state=upsine]{
        \pgfpathsine{\pgfpoint{0.5\pgfdecorationsegmentlength}{0.5\pgfdecorationsegmentamplitude}}
        \pgfpathcosine{\pgfpoint{0.5\pgfdecorationsegmentlength}{-0.5\pgfdecorationsegmentamplitude}}
    }
    \state{upsine}[width=\pgfdecorationsegmentlength,next state=downsine]{
        \pgfpathsine{\pgfpoint{0.5\pgfdecorationsegmentlength}{-0.5\pgfdecorationsegmentamplitude}}
        \pgfpathcosine{\pgfpoint{0.5\pgfdecorationsegmentlength}{0.5\pgfdecorationsegmentamplitude}}
}
    \state{final}{}
}

\tikzset{
corner/.style={line width=1pt,dashed,draw=black,dash pattern=on 6pt off 4pt},
scalar/.style={line width=1pt,draw=black},
gluon/.style={line width=1pt,decorate, draw=GluonColor,
    decoration={complete sines,aspect=0,amplitude=1.25mm,segment length=1.5mm,start up,end up}},
gluontwo/.style={line width=1pt,decorate, draw=GluonColor,
    decoration={complete sines,aspect=0,amplitude=.7mm,segment length=1mm,start up,end up}},
ghost/.style={line width=1pt,loosely dotted,draw=black},
wilson/.style={line width=2pt,draw=black},
 }
\NewDocumentCommand\semiloop{O{black}mmmO{}O{above}}
{%
\draw[#1] let \p1 = ($(#3)-(#2)$) in (#3) arc (#4:({#4+180}):({0.5*veclen(\x1,\y1)})node[midway, #6] {#5};)
}

%% file: figures/images.tex
\newcommand\x{.75}

\newcommand\vertexsize{.07}

\newcommand{\InfiniteStraightLineDefect}{\begin{tikzpicture}[scale = 1.1, baseline={([yshift=-.4ex]current bounding box.center)}]
\draw[scalar] (0,0) -- (0,2.5) -- (2.5,2.5) -- (2.5,0) -- (0,0);
\draw[scalar] (2.5,2.5) -- (3.5,3) -- (3.5,.5) -- (2.5,0);
\draw[scalar] (0,2.5) -- (1,3) -- (3.5,3);
\draw[scalar,dashed] (0,0) -- (1,.5) -- (3.5,.5);
\draw[scalar,dashed] (1,.5) -- (1,3);
\draw[scalar,dashed] (0+1.2,0+1.2) -- (1+1.2,.5+1.2);
\draw[scalar,dashed] (1+.9,1.1) -- (1+.9,3.25);
\draw[scalar,dashed] (1.4,.5+1.05) -- (3.4,.5+1.05);
\draw[fill=WCOrange,WCOrange] (1.9,1.55) circle (.1);
\end{tikzpicture}}

\newcommand{\AntipodalDefects}{\begin{tikzpicture}[scale = 1, baseline={([yshift=-.4ex]current bounding box.center)}]
\draw[scalar] (0,0) circle (2);
\begin{scope}[rotate=0]
\draw[scalar,dashed]
    plot[domain=0:180,samples=60]
    ({1.55*cos(\x)},{.32*sin(\x)+1.20});
\draw[scalar]
    plot[domain=180:360,samples=60]
    ({1.55*cos(\x)},{.32*sin(\x)+1.20});
\draw[scalar,dashed]
    plot[domain=0:180,samples=60]
    ({1.87*cos(\x)},{.42*sin(\x)+.62});
\draw[scalar]
    plot[domain=180:360,samples=60]
    ({1.87*cos(\x)},{.42*sin(\x)+.62});
\draw[scalar,dashed]
    plot[domain=0:180,samples=60]
    ({2*cos(\x)},{.48*sin(\x)});
\draw[scalar]
    plot[domain=180:360,samples=60]
    ({2*cos(\x)},{.48*sin(\x)});
\draw[scalar,dashed]
    plot[domain=0:180,samples=60]
    ({1.87*cos(\x)},{.42*sin(\x)-.62});
\draw[scalar]
    plot[domain=180:360,samples=60]
    ({1.87*cos(\x)},{.42*sin(\x)-.62});
\draw[scalar,dashed]
    plot[domain=0:180,samples=60]
    ({1.55*cos(\x)},{.32*sin(\x)-1.20});
\draw[scalar]
    plot[domain=180:360,samples=60]
    ({1.55*cos(\x)},{.32*sin(\x)-1.20});
\end{scope}
\draw[fill=WCOrange,WCOrange] (0,1.75) circle (.1);
\draw[fill=WCOrange,WCOrange] (0,-1.75) circle (.1);
\end{tikzpicture}}

\newcommand{\AnalyticStructureInfiniteVolume}{\begin{tikzpicture}[scale = 1, baseline={([yshift=-.4ex]current bounding box.center)}]
\draw[scalar,-Latex] (0,0) -- (6.2,0);
\draw[scalar,-Latex] (3,-2) -- (3,2.2);
\node[] at (5.75,1.8) {$\tau$};
\draw[scalar] (5.4,1.9) -- (5.4,1.45);
\draw[scalar] (5.4,1.45) -- (6,1.45);
\draw[-, decorate, decoration={zigzag, segment length=6, amplitude=2},line width=1,SCRed] (1,-1.9) -- (1,1.9);
\draw[fill=WCOrange,WCOrange] (1,0) circle (.075);
\node[] at (.55,-.2) {\tiny $\color{WCOrange}{-2\beta}$};
\draw[-, decorate, decoration={zigzag, segment length=6, amplitude=2},line width=1,SCRed] (3,-1.9) -- (3,1.9);
\draw[fill=WCOrange,WCOrange] (3,0) circle (.075);
\node[] at (2+.75,-.2) {\tiny $\color{WCOrange}{0}$};
\draw[-, decorate, decoration={zigzag, segment length=6, amplitude=2},line width=1,SCRed] (2,-1.9) -- (2,1.9);
\draw[fill=WCOrange,WCOrange] (2,0) circle (.075);
\node[] at (1+.65,-.2) {\tiny $\color{WCOrange}{-\beta}$};
\draw[-, decorate, decoration={zigzag, segment length=6, amplitude=2},line width=1,SCRed] (4,-1.9) -- (4,1.9);
\draw[fill=WCOrange,WCOrange] (4,0) circle (.075);
\node[] at (4+.3,-.2) {\tiny $\color{WCOrange}{\beta}$};
\draw[-, decorate, decoration={zigzag, segment length=6, amplitude=2},line width=1,SCRed] (5,-1.9) -- (5,1.9);
\draw[fill=WCOrange,WCOrange] (5,0) circle (.075);
\node[] at (5+.35,-.2) {\tiny $\color{WCOrange}{2\beta}$};
\end{tikzpicture}}

\newcommand{\AnalyticStructureFiniteVolume}{\begin{tikzpicture}[scale = 1, baseline={([yshift=-.4ex]current bounding box.center)}]
\draw[scalar,-Latex] (0,0) -- (6.2,0);
\draw[scalar,-Latex] (3,-2) -- (3,2.2);
\node[] at (5.75,1.8) {$\tau$};
\draw[scalar] (5.4,1.9) -- (5.4,1.45);
\draw[scalar] (5.4,1.45) -- (6,1.45);
\node[] at (0,1.33) {\tiny $\color{WCOrange}{2 \pi i R}$};
\node[] at (0,-1.33) {\tiny $\color{WCOrange}{-2\pi i R}$};
\draw[-, decorate, decoration={zigzag, segment length=6, amplitude=2},line width=1,SCRed] (1,-1.9) -- (1,1.9);
\draw[fill=WCOrange,WCOrange] (1,0) circle (.075);
\draw[fill=WCOrange,WCOrange] (1,1.33) circle (.075);
\draw[fill=WCOrange,WCOrange] (1,-1.33) circle (.075);
\node[] at (.55,-.2) {\tiny $\color{WCOrange}{-2\beta}$};
\draw[-, decorate, decoration={zigzag, segment length=6, amplitude=2},line width=1,SCRed] (3,-1.9) -- (3,1.9);
\draw[fill=WCOrange,WCOrange] (3,0) circle (.075);
\draw[fill=WCOrange,WCOrange] (3,1.33) circle (.075);
\draw[fill=WCOrange,WCOrange] (3,-1.33) circle (.075);
\node[] at (2+.75,-.2) {\tiny $\color{WCOrange}{0}$};
\draw[-, decorate, decoration={zigzag, segment length=6, amplitude=2},line width=1,SCRed] (2,-1.9) -- (2,1.9);
\draw[fill=WCOrange,WCOrange] (2,0) circle (.075);
\draw[fill=WCOrange,WCOrange] (2,1.33) circle (.075);
\draw[fill=WCOrange,WCOrange] (2,-1.33) circle (.075);
\node[] at (1+.65,-.2) {\tiny $\color{WCOrange}{-\beta}$};
\draw[-, decorate, decoration={zigzag, segment length=6, amplitude=2},line width=1,SCRed] (4,-1.9) -- (4,1.9);
\draw[fill=WCOrange,WCOrange] (4,0) circle (.075);
\draw[fill=WCOrange,WCOrange] (4,1.33) circle (.075);
\draw[fill=WCOrange,WCOrange] (4,-1.33) circle (.075);
\node[] at (4+.3,-.2) {\tiny $\color{WCOrange}{\beta}$};
\draw[-, decorate, decoration={zigzag, segment length=6, amplitude=2},line width=1,SCRed] (5,-1.9) -- (5,1.9);
\draw[fill=WCOrange,WCOrange] (5,0) circle (.075);
\draw[fill=WCOrange,WCOrange] (5,1.33) circle (.075);
\draw[fill=WCOrange,WCOrange] (5,-1.33) circle (.075);
\node[] at (5+.35,-.2) {\tiny $\color{WCOrange}{2\beta}$};
\end{tikzpicture}}

\newcommand{\ContourFiniteVolume}{\begin{tikzpicture}[scale = 1, baseline={([yshift=-.4ex]current bounding box.center)}]
\draw[scalar,-Latex] (0,0) -- (6.2,0);
\draw[scalar,-Latex] (3,-2) -- (3,2.2);
\node[] at (5.75,1.8) {$\tau$};
\draw[scalar] (5.4,1.9) -- (5.4,1.45);
\draw[scalar] (5.4,1.45) -- (6,1.45);
\draw[-, decorate, decoration={zigzag, segment length=6, amplitude=2},line width=1,SCRed] (1,-1.9) -- (1,1.9);
\draw[fill=WCOrange,WCOrange] (1,0) circle (.075);
\draw[fill=WCOrange,WCOrange] (1,1.33) circle (.075);
\draw[fill=WCOrange,WCOrange] (1,-1.33) circle (.075);
\draw[-, decorate, decoration={zigzag, segment length=6, amplitude=2},line width=1,SCRed] (3,-1.9) -- (3,1.9);
\draw[fill=WCOrange,WCOrange] (3,0) circle (.075);
\draw[fill=WCOrange,WCOrange] (3,1.33) circle (.075);
\draw[fill=WCOrange,WCOrange] (3,-1.33) circle (.075);
\draw[-, decorate, decoration={zigzag, segment length=6, amplitude=2},line width=1,SCRed] (2,-1.9) -- (2,1.9);
\draw[fill=WCOrange,WCOrange] (2,0) circle (.075);
\draw[fill=WCOrange,WCOrange] (2,1.33) circle (.075);
\draw[fill=WCOrange,WCOrange] (2,-1.33) circle (.075);
\draw[-, decorate, decoration={zigzag, segment length=6, amplitude=2},line width=1,SCRed] (4,-1.9) -- (4,1.9);
\draw[fill=WCOrange,WCOrange] (4,0) circle (.075);
\draw[fill=WCOrange,WCOrange] (4,1.33) circle (.075);
\draw[fill=WCOrange,WCOrange] (4,-1.33) circle (.075);
\draw[-, decorate, decoration={zigzag, segment length=6, amplitude=2},line width=1,SCRed] (5,-1.9) -- (5,1.9);
\draw[fill=WCOrange,WCOrange] (5,0) circle (.075);
\draw[fill=WCOrange,WCOrange] (5,1.33) circle (.075);
\draw[fill=WCOrange,WCOrange] (5,-1.33) circle (.075);
\draw[fill=MediumBlueGrey,MediumBlueGrey,line width=1] (1-.25,1.33/2-.25) -- (1-.25,-1.33/2+.25);
\draw[fill=MediumBlueGrey,MediumBlueGrey,line width=1] (1+.25,1.33/2-.25) -- (1+.25,-1.33/2+.25);
\draw[MediumBlueGrey,line width=1] (1-.25,1.33/2-.25) arc (180:0:.25);
\draw[MediumBlueGrey,line width=1] (1+.25,-1.33/2+.25) arc (0:-180:.25);
\draw[fill=MediumBlueGrey,MediumBlueGrey,line width=1] (2-.25,1.33/2-.25) -- (2-.25,-1.33/2+.25);
\draw[fill=MediumBlueGrey,MediumBlueGrey,line width=1] (2+.25,1.33/2-.25) -- (2+.25,-1.33/2+.25);
\draw[MediumBlueGrey,line width=1] (2-.25,1.33/2-.25) arc (180:0:.25);
\draw[MediumBlueGrey,line width=1] (2+.25,-1.33/2+.25) arc (0:-180:.25);
\draw[fill=MediumBlueGrey,MediumBlueGrey,line width=1] (3-.25,1.33/2-.25) -- (3-.25,-1.33/2+.25);
\draw[fill=MediumBlueGrey,MediumBlueGrey,line width=1] (3+.25,1.33/2-.25) -- (3+.25,-1.33/2+.25);
\draw[MediumBlueGrey,line width=1] (3-.25,1.33/2-.25) arc (180:0:.25);
\draw[MediumBlueGrey,line width=1] (3+.25,-1.33/2+.25) arc (0:-180:.25);
\draw[fill=MediumBlueGrey,MediumBlueGrey,line width=1] (4-.25,1.33/2-.25) -- (4-.25,-1.33/2+.25);
\draw[fill=MediumBlueGrey,MediumBlueGrey,line width=1] (4+.25,1.33/2-.25) -- (4+.25,-1.33/2+.25);
\draw[MediumBlueGrey,line width=1] (4-.25,1.33/2-.25) arc (180:0:.25);
\draw[MediumBlueGrey,line width=1] (4+.25,-1.33/2+.25) arc (0:-180:.25);
\draw[fill=MediumBlueGrey,MediumBlueGrey,line width=1] (5-.25,1.33/2-.25) -- (5-.25,-1.33/2+.25);
\draw[fill=MediumBlueGrey,MediumBlueGrey,line width=1] (5+.25,1.33/2-.25) -- (5+.25,-1.33/2+.25);
\draw[MediumBlueGrey,line width=1] (5-.25,1.33/2-.25) arc (180:0:.25);
\draw[MediumBlueGrey,line width=1] (5+.25,-1.33/2+.25) arc (0:-180:.25);
\foreach \x in {1,...,5}{
  \draw[MediumBlueGrey, line width=.75pt, -Latex]
    (\x-.25,.2) -- (\x-.25,.4);
}
\end{tikzpicture}}

\newcommand{\DiagOneZeroTFiniteV}{\begin{tikzpicture}[scale = 1, baseline={([yshift=-.8ex]current bounding box.center)}]
\draw[wilson] (0,1) -- (1.5,1);
\draw[wilson] (0,0) -- (1.5,0);
\draw[scalar] (.25,0) arc (180:0:.5);
\draw[fill=white] (.25,0) circle (.075);
\draw[fill=white] (1.25,0) circle (.075);
\end{tikzpicture}}

\newcommand{\DiagTwoZeroTFiniteV}{\begin{tikzpicture}[scale = 1, baseline={([yshift=-.8ex]current bounding box.center)}]
\draw[wilson] (0,1) -- (1.5,1);
\draw[wilson] (0,0) -- (1.5,0);
\draw[scalar] (.25,0) arc (180:0:.5);
\draw[scalar] (.5,0) -- (.75,.5);
\draw[scalar] (1,0) -- (.75,.5);
\draw[fill=white] (.25,0) circle (.075);
\draw[fill=white] (1.25,0) circle (.075);
\draw[fill=WCOrange,WCOrange] (.75,.5) circle (.075);
\draw[fill=WCOrange,WCOrange] (.5,0) circle (.075);
\draw[fill=WCOrange,WCOrange] (1,0) circle (.075);
\end{tikzpicture}}

\newcommand{\DiagThreeZeroTFiniteV}{\begin{tikzpicture}[scale = 1, baseline={([yshift=-.8ex]current bounding box.center)}]
\draw[wilson] (0,1) -- (1.5,1);
\draw[wilson] (0,0) -- (1.5,0);
\draw[scalar] (.25,0) arc (180:0:.5);
\draw[scalar] (.5,1) -- (.75,.5);
\draw[scalar] (1,1) -- (.75,.5);
\draw[fill=white] (.25,0) circle (.075);
\draw[fill=white] (1.25,0) circle (.075);
\draw[fill=WCOrange,WCOrange] (.75,.5) circle (.075);
\draw[fill=WCOrange,WCOrange] (.5,1) circle (.075);
\draw[fill=WCOrange,WCOrange] (1,1) circle (.075);
\end{tikzpicture}}

\newcommand{\DiagFourZeroTFiniteV}{\begin{tikzpicture}[scale = 1, baseline={([yshift=-.8ex]current bounding box.center)}]
\draw[wilson] (0,1) -- (1.5,1);
\draw[wilson] (0,0) -- (1.5,0);
\draw[scalar] (.25,0) arc (180:0:.5);
\draw[scalar] (.75,0) -- (.75,1);
\draw[fill=white] (.25,0) circle (.075);
\draw[fill=white] (1.25,0) circle (.075);
\draw[fill=WCOrange,WCOrange] (.75,1) circle (.075);
\draw[fill=WCOrange,WCOrange] (.75,0) circle (.075);
\draw[fill=WCOrange,WCOrange] (.75,.5) circle (.075);
\end{tikzpicture}}

\newcommand{\DiagOneZeroTFiniteVNS}{\begin{tikzpicture}[scale = 1, baseline={([yshift=-.8ex]current bounding box.center)}]
\draw[wilson] (0,1) -- (1.5,1);
\draw[wilson] (0,0) -- (1.5,0);
\draw[scalar] (.75,0) -- (.75,1);
\draw[fill=white] (.75,1) circle (.075);
\draw[fill=white] (.75,0) circle (.075);
\end{tikzpicture}}

\newcommand{\DiagOneFiniteTFiniteV}{\begin{tikzpicture}[scale = 1, baseline={([yshift=-.8ex]current bounding box.center)}]
\draw[wilson] (.75,-.125) ellipse (.75cm and .15cm);
\draw[wilson] (.75,1-.125) ellipse (.75cm and .15cm);
\draw[scalar] (.25,0) arc (180:0:.5);
\draw[fill=white] (.25,0) circle (.075);
\draw[fill=white] (1.25,0) circle (.075);
\end{tikzpicture}}

\newcommand{\DiagTwoFiniteTFiniteV}{\begin{tikzpicture}[scale = 1, baseline={([yshift=-.8ex]current bounding box.center)}]
\draw[wilson] (.75,-.125) ellipse (.75cm and .15cm);
\draw[wilson] (.75,1-.125) ellipse (.75cm and .15cm);
\draw[scalar] (.25,0) arc (180:0:.5);
\draw[scalar] (.5,0) -- (.75,.5);
\draw[scalar] (1,0) -- (.75,.5);
\draw[fill=white] (.25,0) circle (.075);
\draw[fill=white] (1.25,0) circle (.075);
\draw[fill=WCOrange,WCOrange] (.75,.5) circle (.075);
\draw[fill=WCOrange,WCOrange] (.5,0) circle (.075);
\draw[fill=WCOrange,WCOrange] (1,0) circle (.075);
\end{tikzpicture}}

\newcommand{\DiagThreeFiniteTFiniteV}{\begin{tikzpicture}[scale = 1, baseline={([yshift=-.8ex]current bounding box.center)}]
\draw[wilson] (.75,-.125) ellipse (.75cm and .15cm);
\draw[wilson] (.75,1-.125) ellipse (.75cm and .15cm);
\draw[scalar] (.25,-.1) arc (170:10:.5);
\draw[scalar] (.5,.7) -- (.75,.3);
\draw[scalar] (1,.7) -- (.75,.3);
\draw[fill=white] (.25,0) circle (.075);
\draw[fill=white] (1.25,0) circle (.075);
\draw[fill=WCOrange,WCOrange] (.75,.3) circle (.075);
\draw[fill=WCOrange,WCOrange] (.5,.725) circle (.075);
\draw[fill=WCOrange,WCOrange] (1,.725) circle (.075);
\end{tikzpicture}}

\newcommand{\DiagFourFiniteTFiniteV}{\begin{tikzpicture}[scale = 1, baseline={([yshift=-.8ex]current bounding box.center)}]
\draw[wilson] (.75,-.125) ellipse (.75cm and .15cm);
\draw[wilson] (.75,1-.125) ellipse (.75cm and .15cm);
\draw[scalar] (.25,-.1) arc (170:10:.5);
\draw[scalar] (.75,0) -- (.75,.7);
\draw[fill=white] (.25,0) circle (.075);
\draw[fill=white] (1.25,0) circle (.075);
\draw[fill=WCOrange,WCOrange] (.75,.725) circle (.075);
\draw[fill=WCOrange,WCOrange] (.75,.025) circle (.075);
\draw[fill=WCOrange,WCOrange] (.75,.3) circle (.075);
\end{tikzpicture}}

\newcommand{\DiagFiveFiniteTFiniteV}{\begin{tikzpicture}[scale = 1, baseline={([yshift=-.8ex]current bounding box.center)}]
\draw[wilson] (.75,-.125) ellipse (.75cm and .15cm);
\draw[wilson] (.75,1-.125) ellipse (.75cm and .15cm);
\draw[scalar] (.25,-.1) arc (170:10:.5);
\draw[scalar] (.75,.45) circle (.175);
\draw[fill=white] (.25,0) circle (.075);
\draw[fill=white] (1.25,0) circle (.075);
\draw[fill=WCOrange,WCOrange] (.75,.3) circle (.075);
\end{tikzpicture}}

\newcommand{\DiagPhiOne}{\begin{tikzpicture}[scale = 1, baseline={([yshift=-.8ex]current bounding box.center)}]
\draw[wilson] (.75,-.1) ellipse (.75cm and .15cm);
\draw[scalar] (.75,0) arc (200:-20:.25);
\draw[fill=white] (.75,.055) circle (.075);
\draw[fill=WCOrange,WCOrange] (1.225,.012) circle (.075);
\end{tikzpicture}}

\newcommand{\DiagOneFiniteT}{\begin{tikzpicture}[scale = 1, baseline={([yshift=-.8ex]current bounding box.center)}]
\draw[wilson] (.75,-.1) ellipse (.75cm and .15cm);
\draw[scalar] (.25,0) arc (180:0:.5);
\draw[fill=white] (.25,0) circle (.075);
\draw[fill=white] (1.25,0) circle (.075);
\end{tikzpicture}}

\newcommand{\DiagTwoFiniteT}{\begin{tikzpicture}[scale = 1, baseline={([yshift=-.8ex]current bounding box.center)}]
\draw[wilson] (.75,-.1) ellipse (.75cm and .15cm);
\draw[scalar] (.25,0) arc (180:0:.5);
\draw[scalar] (.5,.05) -- (.75,.5);
\draw[scalar] (1,.05) -- (.75,.5);
\draw[fill=white] (.25,0) circle (.075);
\draw[fill=white] (1.25,0) circle (.075);
\draw[fill=WCOrange,WCOrange] (.75,.5) circle (.075);
\draw[fill=WCOrange,WCOrange] (.5,.05) circle (.075);
\draw[fill=WCOrange,WCOrange] (1,.05) circle (.075);
\end{tikzpicture}}

\newcommand{\DiagThreeFiniteT}{\begin{tikzpicture}[scale = 1, baseline={([yshift=-1.9ex]current bounding box.center)}]
\draw[wilson] (.75,-.1) ellipse (.75cm and .15cm);
\draw[scalar] (.25,0) arc (180:0:.5);
\draw[scalar] (.75,.7) circle (.2);
\draw[fill=white] (.25,0) circle (.075);
\draw[fill=white] (1.25,0) circle (.075);
\draw[fill=WCOrange,WCOrange] (.75,.5) circle (.075);
\end{tikzpicture}}

\newcommand{\DiagOneZeroT}{\begin{tikzpicture}[scale = 1, baseline={([yshift=-.8ex]current bounding box.center)}]
\draw[wilson] (0,0) -- (1.5,0);
\draw[scalar] (.25,0) arc (180:0:.5);
\draw[fill=white] (.25,0) circle (.075);
\draw[fill=white] (1.25,0) circle (.075);
\end{tikzpicture}}

\newcommand{\DiagTwoZeroT}{\begin{tikzpicture}[scale = 1, baseline={([yshift=-.8ex]current bounding box.center)}]
\draw[wilson] (0,0) -- (1.5,0);
\draw[scalar] (.25,0) arc (180:0:.5);
\draw[scalar] (.5,0) -- (.75,.5);
\draw[scalar] (1,0) -- (.75,.5);
\draw[fill=white] (.25,0) circle (.075);
\draw[fill=white] (1.25,0) circle (.075);
\draw[fill=WCOrange,WCOrange] (.75,.5) circle (.075);
\draw[fill=WCOrange,WCOrange] (.5,0) circle (.075);
\draw[fill=WCOrange,WCOrange] (1,0) circle (.075);
\end{tikzpicture}}

\newcommand{\VertexFourScalars}{\begin{tikzpicture}[baseline={([yshift=-.35ex]current bounding box.center)}]
\draw[scalar] (0,0) -- (1.25,1);
\draw[scalar] (0,1) -- (1.25,0);
\draw[fill=VertexColor,VertexColor] (.625,.5) circle (\vertexsize);
\node[anchor=west] at (1.3,1) {\footnotesize $L$};
\node[anchor=west] at (1.3,0) {\footnotesize $K$};
\node[anchor=east] at (-0.05,1) {\footnotesize $I$};
\node[anchor=east] at (-0.05,0) {\footnotesize $J$};
\end{tikzpicture}}

\newcommand{\VertexDefect}{\begin{tikzpicture}[baseline={([yshift=-.35ex]current bounding box.center)}]
\draw[wilson] (1,-.15) ellipse (.75cm and .15cm);
\draw[scalar] (1,0) -- (1,.75);
\draw[fill=VertexColor,VertexColor] (1,0) circle (\vertexsize);
\node[anchor=east] at (1.2,1) {\footnotesize $I$};
\end{tikzpicture}}

%% file: sections/Introduction.tex
\section{Introduction}
\label{sec:Introduction}

Conformal field theories (CFTs) provide a universal description of continuous phase transitions and play a central role in the study of strongly-interacting quantum systems~\cite{pelissetto2002critical,Sachdev:2008jho,El-Showk:2012cjh}.
Their scope extends naturally to systems containing defects~\cite{Billo:2013jda,Giombi:2021uae,Giombi:2022vnz,Giombi:2023dqs}, which describe localized impurities in condensed matter~\cite{Affleck:1991tk,Affleck:1995ge,sachdev1999quantum,Sachdev:2003yk,potok2007observation,assaad2013pinning} as well as extended probes such as Wilson and 't Hooft lines in gauge theories~\cite{Wilson:1974sk,tHooft:1977nqb}.
At a conformal defect fixed point, these systems preserve a subgroup of the ambient conformal symmetry~\cite{Cardy:1991tv,McAvity:1995zd,Billo:2016cpy,Lauria:2018klo}; in particular, a straight line supports local operators organized by one-dimensional conformal symmetry~\cite{Giombi:2017cqn}.
Recent developments in analytic and numerical bootstrap techniques have considerably advanced our understanding of their spectra and interactions~\cite{Liendo:2012hy,Gaiotto:2013nva,Gliozzi:2015qsa,Liendo:2018ukf,Bissi:2018mcq,Kaviraj:2018tfd,Liendo:2019jpu,Drukker:2022pxk,Barrat:2022psm,Bianchi:2022ppi,Lanzetta:2025xfw,Girault:2025kzt}.
A particularly simple yet non-trivial example is the pinning-field line defect in the critical $\mathrm{O}(N)$ model, generated by a magnetic field localized at a spatial point~\cite{allais2014magnetic,ParisenToldin:2016szc,Cuomo:2021kfm}.
This setup provides a model of a static symmetry-breaking impurity, with potential realizations in quantum simulators~\cite{Ebadi:2020ldi,Manovitz:2024hif}, and offers a setting in which perturbative expansions~\cite{Cuomo:2021kfm,Bianchi:2022sbz,Nishioka:2022qmj,Barrat:2023ivo,Bianchi:2023gkk,Dey:2024ilw,Belton:2025hbu,Kravchuk:2025evf}, large-$N$ methods~\cite{Cuomo:2021kfm}, and non-perturbative approaches~\cite{allais2014magnetic,ParisenToldin:2016szc,Hu:2023ghk} can be compared.

Introducing a non-zero temperature enriches this setting and connects its universal properties to experimentally accessible thermal fluctuations and response functions~\cite{Chubukov:1993aau,Damle:1997rxu}.
For impurities near a quantum critical point, defect correlators characterize local fluctuations and susceptibilities~\cite{Vojta:2000tld,Sachdev:2003yk}, while in holographic theories they provide probes of thermal states described by black-hole geometries~\cite{deBoer:2008gu}.
The inverse temperature $\beta=1/T$ introduces a scale that breaks conformal invariance of the state without altering the underlying local CFT data.
Symmetry-allowed bulk and defect operators can consequently acquire non-zero one-point functions, and their two-point functions are no longer kinematically fixed by symmetry alone~\cite{Katz:2014rla}.
Here we consider static impurities whose Euclidean worldlines wrap the thermal circle, analogously to Polyakov loops in gauge theories~\cite{Polyakov:1978vu}, as shown in Figure~\ref{fig:AntipodalDefects}.
This configuration preserves translations along the thermal circle and spatial rotations around the defect, providing a particularly symmetric setting in which to study thermal defect dynamics~\cite{Barrat:2024aoa}.

Although the thermal state breaks conformal invariance, the OPE retains the scaling dimensions and three-point coefficients of the zero-temperature theory~\cite{Petkou:1998fb,El-Showk:2011yvt,Iliesiu:2018fao}.
Thermal correlators therefore combine vacuum CFT data with new information encoded in thermal one-point functions~\cite{Katz:2014rla,Marchetto:2023fcw}.
These contributions are constrained by the Kubo--Martin--Schwinger (KMS) condition~\cite{Kubo:1957mj,Martin:1959jp}, which relates different Euclidean orderings and takes the form of a crossing equation for thermal two-point functions~\cite{El-Showk:2011yvt,Iliesiu:2018fao}.
This observation underlies the thermal bootstrap program~\cite{Iliesiu:2018fao,Petkou:2018ynm,Manenti:2019wxs,Marchetto:2023xap,Barrat:2025nvu,Guo:2026xyl}, which has yielded quantitative predictions for the $3d$ Ising model and critical $\mathrm{O}(N)$ theories~\cite{Iliesiu:2018zlz,Petkou:2018ynm,Barrat:2025wbi}, together with applications to holographic CFTs~\cite{Parnachev:2020fna,Alday:2020eua,Karlsson:2022osn,Dodelson:2022yvn,Parisini:2022wkb,Dodelson:2023vrw,Esper:2023jeq,Parisini:2023nbd,Buric:2025anb,Buric:2025fye,Niarchos:2025cdg,Barrat:2025twb,Arnaudo:2026der,Niarchos:2026vow,Buric:2026cmn,Barrat:2026jfg,Arnaudo:2026axe,Kakkar:2026czr}.
Unlike the squared OPE coefficients entering vacuum four-point functions of identical scalars, the thermal coefficients are here products of three-point coefficients and thermal one-point functions and need not be positive.
This complicates the direct application of standard numerical bootstrap techniques, while motivating complementary approaches based on analyticity, spectral representations, and dispersive methods.

Thermal impurity physics has a long history~\cite{Vojta:2000tld,Sachdev:2003yk}, and recent work has formulated thermal observables in the language of defect CFT~\cite{Barrat:2024aoa,Li:2026udh,Artico:2026rye} and investigated defect correlators in holographic theories~\cite{Giombi:2026kdz}.
Here we focus on determining local two-point functions and thermal one-point data for the pinning-field line defect.
Two-point correlators depend kinematically on the single dimensionless separation $\tau/\beta$, and their OPE converges throughout the thermal circle $0<\tau<\beta$.
Preserved transverse rotations and internal symmetries restrict the contributing one-point functions to singlets, while translation invariance eliminates expectation values of defect descendants.\footnote{At finite volume, curvature descendants however contribute.}
These simplifications make thermal line defects a promising setting for bootstrap methods.
Moreover, the thermal one-point coefficients also determine bulk profiles (in the presence of the defect) through the bulk-defect OPE~\cite{Barrat:2024aoa} and are the only non-zero temperature data entering the block expansion of higher-point functions.
The pinning-field defect is particularly suitable for this program because perturbation theory, large-$N$ methods, and non-perturbative studies provide complementary information about its zero-temperature spectrum and interactions~\cite{Cuomo:2021kfm,ParisenToldin:2016szc,Hu:2023ghk}.

\bigskip

The outline of the paper is as follows.
In Section~\ref{sec:LineDefectCorrelatorsInThermalGeometries}, we develop the framework for thermal line-defect correlators from first principles, deriving their symmetry constraints, OPE structure, and analytic properties by adapting and extending familiar results from bulk CFTs at finite temperature.
We develop dispersive thermal-bootstrap methods for line defects and organize two-point functions into a manifestly periodic expansion,
\begin{equation}
    g(\tau)
    =
    \frac{1}{\beta^{2\Dh_{\hphi}}}
    \sum_{\Oh} a_{\Oh} F_{\Oh}(\tau)
    + \tilde{g} (\tau)\,.
    \label{eq:Intro_PeriodicOPE}
\end{equation}
The periodic blocks $F_{\Oh} (\tau)$ encode the contributions of individual exchanged operators and admit explicit representations, provided in~\eqref{eq:PeriodicBlocks_InfiniteV} and~\eqref{eq:PeriodicBlock_FiniteV}, while $\tilde{g} (\tau)$ accounts for contributions not captured by the term-wise dispersive reconstruction.
Re-expanding these blocks in the ordinary OPE shows that heavy operators can be asymptotically estimated by functions of the light-operator data, reproducing and extending results known in bulk theories~\cite{Iliesiu:2018fao,Marchetto:2023xap,Barrat:2025nvu,Barrat:2026jfg}.
We also investigate the extension to finite spatial volume, considering two defects wrapping the thermal circle at antipodal points of $S_R^{d-1}$.
This configuration has a zero-temperature limit conformally related to the infinite straight line (unlike a single thermal line on the sphere) and introduces an additional dependence on the ratio $\beta/R$.

In Section~\ref{sec:epsExpansionAndInterpolationTo3d}, we apply this framework to the pinning-field line defect near four dimensions, determining perturbative correlators of tilt and longitudinal singlet operators, extracting thermal OPE data, and comparing them with direct diagrammatic calculations.
Thermal screening produces non-analytic dependence on $\veps$ that is missed by naive fixed-order perturbation theory.
Combining $d=4-\veps$ results with the two-dimensional Ising theory yields Padé estimates for the leading thermal coefficient of the magnetic impurity in the $3d$ Ising model.
This motivates an approximation obtained by retaining only the identity and the leading singlet operator in the periodic block expansion,
\begin{equation}
    g(\tau)
    \sim
    \frac{1}{\beta^{\smash{2\Dh_{\hphi}}}}
    \left(
    F_{\hat{\mathds{1}}}(\tau)
    +
    a_{\smash{\hphi^1}} F_{\smash{\hphi^1}}(\tau)
    +
    c_{\smash{\hphi^1}}
    \right)\,,
    \label{eq:Intro_TruncatedPeriodicOPE}
\end{equation}
with $c_{\smash{\hphi^1}}$ a constant undetermined by our framework and related to the impurity susceptibility.
In Section~\ref{sec:LargeN}, we test this truncation against numerical solutions of the leading large-$N$ equations.
We first extract thermal defect one-point coefficients from bulk profiles across dimensions and investigate their finite-volume dependence in the antipodal configuration.
At infinite volume, we then determine the tilt and longitudinal singlet correlators independently and compare them with the truncated periodic OPE.
Retaining the leading singlet gives very good agreement over the tested Euclidean-time intervals; in the longitudinal channel, this comparison includes an additional fitted constant, which we relate to a renormalized static susceptibility.
This motivates the truncation~\eqref{eq:Intro_TruncatedPeriodicOPE} and poses the question of whether the large-$\Dh$ behavior of thermal OPE coefficients is dominated by free double-twist operators.
These results provide quantitative benchmarks for future thermal defect bootstrap studies of the three-dimensional Ising impurity.
In Section~\ref{sec:DiscussionAndOutlook}, we discuss the implications of this work and list potential future directions we find interesting.

%% file: sections/LineDefectCorrelatorsInThermalGeometries.tex
\section{Line defect correlators at finite temperature}
\label{sec:LineDefectCorrelatorsInThermalGeometries}

In this section, we study general properties of correlators of line defect operators at finite temperature.
We begin with a review of the zero-temperature line defect CFT, and provide a summary of results for the pinning-field line defect that will be relevant in subsequent sections.
We then consider the effects of temperature, and summarize the breaking of symmetry on the correlator and on the OPE.
We discuss the spectral decomposition of two-point functions of scalar defect operators and the constraints that follow from it.
We then present a periodic version of the OPE obtained from dispersive methods and show how it forms a complete bootstrap problem.
Finally, we discuss asymptotic estimations of OPE coefficients, following from the periodic OPE.

\subsection{Line defect CFT at zero temperature}
\label{subsec:ReviewOfZeroTemperatureLineDefectCFT}

Our starting point is a CFT at zero temperature, in which we insert an infinite straight line defect operator.
We review here known facts about these defect CFTs.
We explain how a one-dimensional CFT arises from this configuration, and discuss its properties.
Specializing to the pinning-field line defect, we review results for the $\veps$-expansion and in the large $N$ limit that are useful for the rest of this work.

\subsubsection{Generalities}
\label{subsubsec:Generalities}

\paragraph{Symmetries.}
Our starting point is a $d$-dimensional CFT in flat space $\mathbb{R}^d$ with a line defect extending along the Euclidean time direction $\tau$.
Its presence breaks the conformal symmetry~\cite{Billo:2016cpy}:
\begin{equation}
    \mathrm{SO}(d+1,1)
    \longrightarrow
    \mathrm{SO}(2,1)
    \times
    \mathrm{SO} (d-1)\,.
    \label{eq:ZeroT_SymmetryBreaking_SO}
\end{equation}
On the right-hand side, $\mathrm{SO}(2,1)$ is the one-dimensional conformal symmetry preserved along the line, while $\mathrm{SO}(d-1)$ corresponds to rotations around the defect.
The defect may also break global symmetries.
We denote such a breaking by
\begin{equation}
    G \longrightarrow G'\,.
    \label{eq:ZeroT_SymmetryBreaking_G}
\end{equation}

\paragraph{Defect operators and correlators.}
The line defect gives rise to local excitations, which we call defect operators and denote by $\Oh$.
They can be probed, for instance, by bringing a bulk operator close to the line defect~\cite{Billo:2016cpy}:
\begin{equation}
    \phi(\tau,\bm{x}_\perp)
    =
    \frac{\mu_{\phi\hat{\mathds{1}}}}{r^{\smash{\Delta_\phi}}}
    +
    \sum_{\smash{\Oh}} \mu_{\phi\Oh} r^{\smash{\Dh_{\Oh} - \Delta_\phi - s}} \bm{x}_\perp^{i_1} \ldots \bm{x}_\perp^{i_s} \left[ \Oh^{i_1 \ldots i_s} (\tau) + \text{descendants} \right],
    \label{eq:ZeroT_BulkDefectOPE}
\end{equation}
where $r = |\bm{x}^\perp|$ is the distance transverse to the defect.
This expansion is known as the bulk-defect OPE.
The operators $\Oh$ carry quantum numbers $\Dh$ (the scaling dimensions, associated with $\mathrm{SO}(2,1)$) and $s$ (the transverse spin, associated with $\mathrm{SO}(d-1)$ and encoded by the indices $i_1, \ldots, i_s = 1, \ldots, d-1$).

Correlation functions of defect operators can be defined as
\begin{equation}
    \vev{X}_{1d}
    =
        \frac{\vev{\Dm X}_{\text{bulk}}}{\vev{\Dm}},
    \qquad
    \vev{\mathds{1}}_{1d}
    =
    1.
    \label{eq:ZeroT_CorrelationFunctions}
\end{equation}
This relation presents two equivalent perspectives: on the left-hand side, the correlators of defect operators are viewed as those of a one-dimensional CFT, while on the right-hand side they are defined from the point of view of the bulk theory in the presence of the defect.
In the remainder of this work, we adopt the notation of the left-hand side and drop the subscript, unless stated otherwise.

As for bulk operators, conformal symmetry constrains the kinematics of defect correlators to take the form
\begin{align}
    \vev{\Oh (0)} &= \delta_{\Oh \hat{\mathds{1}}},
    \label{eq:ZeroT_OnePointFunctions} \\
    \vev{\hphi(\tau) \hphi(0)} &= \frac{1}{\tau^{\smash{2\Dh_{\smash{\hphi}}}}},
    \label{eq:ZeroT_TwoPointFunctions} \\
    \vev{\Oh_1 (\tau_1) \Oh_2 (\tau_2) \Oh_3 (\tau_3)}
    &=
    \frac{f_{\smash{\Oh_1\Oh_2\Oh_3}}}{|\tau_{12}|^{\smash{\Dh_1+\Dh_2-\Dh_3}} |\tau_{23}|^{\smash{\Dh_2+\Dh_3-\Dh_1}} |\tau_{31}|^{\smash{\Dh_3+\Dh_1-\Dh_2}}},
    \label{eq:ZeroT_ThreePointFunctions} \\
    \vev{\hphi(\tau_1) \hphi(\tau_2) \hphi(\tau_3) \hphi(\tau_4)}
    &=
    \frac{F(\chi)}{\tau_{13}^{\smash{2\Dh_{\smash{\hphi}}}} \tau_{24}^{\smash{2\Dh_{\smash{\hphi}}}}},
    \label{eq:ZeroT_FourPointFunctions}
\end{align}
where $F(\chi)$ is a function of the four-point cross-ratio,
\begin{equation}
    \chi = \frac{\tau_{12} \tau_{34}}{\tau_{13} \tau_{24}},
    \label{eq:ZeroT_CrossRatio}
\end{equation}
and satisfies the crossing-symmetry constraint
\begin{equation}
    F(\chi) = F(1-\chi),
    \label{eq:ZeroT_Crossing}
\end{equation}
which follows from exchanging the order of the identical operators.
In this work, we focus on how two-point functions~\eqref{eq:ZeroT_TwoPointFunctions} are modified when the temperature is turned on.

\begin{figure}
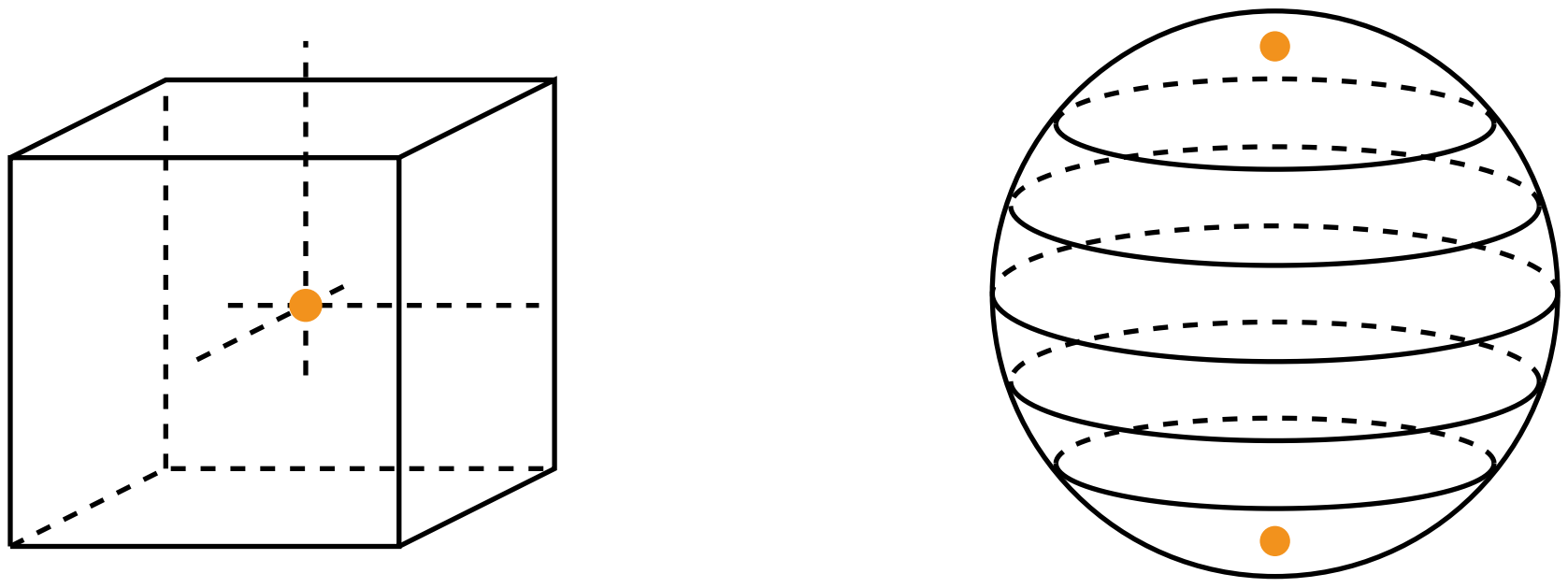

    \centering
    \InfiniteStraightLineDefect \qquad\qquad\qquad\qquad
    \AntipodalDefects
    \caption{Depiction of line defects in spatial manifolds.
    \textbf{Left}: On $\mathbb{R}^d$, a line defect extending in time corresponds to a point in the associated space.
    \textbf{Right}: At finite volume, specifically on $\mathbb{R} \times S_R^{d-1}$, the Weyl-transform of the infinite straight line defect is a pair of defects, which are points in space and extend in the time direction from $-\infty$ to $+\infty$.
    The defects are located at the North and South poles of the sphere.}
    \label{fig:AntipodalDefects}
\end{figure}

\paragraph{Finite volume.}
It is well known that a CFT on flat space $\mathbb{R}^d$ is conformally equivalent to the same theory on $\mathbb{R} \times S_R^{d-1}$.
In the presence of a line defect, the appropriate Weyl transformation to finite volume produces \textit{two} infinite temporal line defects, located at the North and South poles of the sphere.
We refer to this setup as the antipodal configuration and depict it in Figure~\ref{fig:AntipodalDefects}.

It will be important below to note that the time direction on $\mathbb{R}^d$ is not generated by the same operator as the time direction on $\mathbb{R} \times S_R^{d-1}$, due to the mixing induced by the Weyl transformation.
In particular, cylinder-time translations are generated by flat-space dilatations $D/R$, rather than by the flat-space time translation generator $P_\tau$.
Consequently the two thermal identifications define different ensembles.
At zero temperature, since the one-dimensional defect conformal symmetry is preserved by the Weyl transformation, correlators remain kinematically constrained in an analogous way.
For instance, the two-point function of scalar operators inserted on either one of the defects takes the form
\begin{equation}
   \vev{\hphi_N (\tau) \hphi_N (0)}
   =
   \vev{\hphi_S (\tau) \hphi_S (0)}
   =
   \left[ 2 R \sinh \left( \frac{\tau}{2R} \right) \right]^{\smash{-2\Dh_\hphi}}\,.
   \label{eq:ZeroT_FiniteV_TwoPointFunctions}
\end{equation}
where we set $\tau > 0$ to avoid including an absolute value in the right-hand side.
Here, $\hphi_N (\tau)$ and $\hphi_S (\tau)$ refer to the fields respectively placed on the North and South defects.
This two-point function can be visualized as this diagram:
\begin{equation}
    \DiagOneZeroTFiniteV\,,
\end{equation}
with the second defect being entirely disconnected.

One may also consider the mixed correlator between one North and one South operators.
We will not consider this setup at finite temperature, however let us mention for completeness that it is given by the diagram
\begin{equation}
    \DiagOneZeroTFiniteVNS\,,
\end{equation}
which leads to the correlator
\begin{equation}
   \vev{\hphi_N (\tau) \hphi_S (0)}
   =
   \left[ 2 R \cosh \left( \frac{\tau}{2R} \right) \right]^{\smash{-2\Dh_\hphi}}\,.
   \label{eq:ZeroT_FiniteV_TwoPointFunctions_NS}
\end{equation}

One may wonder what a configuration with a \textit{single} defect located, say, at the North pole of $S_R^{d-1}$ corresponds to at infinite volume.
Under the inverse Weyl transformation to flat space, this does not map to a complete straight line, but rather to a semi-infinite line extending radially from the origin.
Indeed, the cylinder coordinate $\tau\in\mathbb{R}$ is mapped to the radial coordinate $r=e^{\tau/R}$, while fixing a point on the sphere fixes the angular direction.
The North- and South-pole defects of the antipodal configuration therefore map to the two opposite rays which together form the complete straight defect in $\mathbb{R}^d$.
Although we do not consider the single-defect configuration here, it should be possible to capture its kinematics using similar tools as the ones presented in this paper.\footnote{We are thankful to Yifan Wang for discussions on that aspect.}
In the $R\to\infty$ cylinder limit, the neighborhood of that defect becomes an ordinary infinite straight line, similarly to the antipodal configuration.

\subsubsection{The pinning-field line defect}
\label{subsubsec:ThePinningFieldLineDefect}

We now review the specific case of the pinning-field line defect at zero temperature, and summarize the observables known at zero temperature that will play a role once the temperature is turned on.

\paragraph{Action.}
The pinning-field line defect is introduced by adding a symmetry-breaking term localized on the line to the $\mathrm{O}(N)$ action:
\begin{equation}
    S
    =
    \int_{-\infty}^\infty \mathrm{d}\tau \int \mathrm{d}^{d-1} \bm{x}_\perp
    \left[ \frac{1}{2} (\pd_\mu \phi^I)^2 + \frac{\lambda}{4!} (\phi^I \phi^I)^2 \right]
    +
    h \int_{-\infty}^\infty \mathrm{d}\tau\, \phi^1 (\tau,0)\,.
    \label{eq:ON_Action}
\end{equation}
At $h=0$, the fields $\phi^I(\tau,\bm{x}_\perp)$ transform as scalars in the fundamental representation of $\mathrm{O}(N)$, with $I=1,\ldots,N$.
The defect coupling $h$ singles out the component $\phi^1$ and explicitly breaks the bulk $\mathrm{O}(N)$ symmetry to $\mathrm{O}(N-1)$ acting on the transverse components $\phi^a$, with $a=2,\ldots,N$.

At finite spatial volume, the antipodal configuration is described by inserting two defects:
\begin{equation}
    S
    =
    S_\text{bulk}
    +
    h \int_{-\infty}^\infty \mathrm{d}\tau \left( \phi^1 (\tau,N) + \phi^1 (\tau,S) \right)\,.
    \label{eq:ON_Action_FiniteV}
\end{equation}
Here, the second argument in $\phi^1 (\tau,X)$ indicates that the defects are inserted on the North and South poles of the sphere.

\paragraph{Spectrum.}
As mentioned above, the $\mathrm{O}(N)$ symmetry breaks to $\mathrm{O}(N-1)$ when $h \neq 0$.\footnote{In this work, we set $h > 0$ without loss of generality.}
The corresponding broken internal rotations give rise to tilt operators, which we denote by $t^a (\tau)$.
These special operators have protected scaling dimension,
\begin{equation}
   \Dh_t
   =
   1\,,
   \label{eq:Tilt_ScalingDim}
\end{equation}
and we normalize them such that their two-point functions are given by\footnote{The normalization of the tilt operators is physical once they are defined through the Ward identity associated with the broken internal symmetry (See~\eqref{eq:TiltWardIdentity}).
For the pinning-field line defect, their two-point coefficient measures the response to a local rotation of the pinning direction in $\mathrm{O}(N)$ space.
For supersymmetric Wilson lines, the analogous protected normalization is related to the Bremsstrahlung function~\cite{Forini:2010ek,Correa:2012at}.}
\begin{equation}
   \vev{t^a(\tau) t^b (0)}
   =
   \frac{\delta^{ab}}{\tau^2}\,.
   \label{eq:Tilt_ZeroT_TwoPointFunction}
\end{equation}
For $N=1$, there is no broken internal rotation and therefore no tilt operator.

We denote the next-to-lowest-lying operator in the spectrum by $\hat{\phi}^1$.
For general $d$ and $N$, its scaling dimension is not known exactly.
For $N=1$, it is the lowest non-trivial defect operator in the absence of the tilt.
At $d=3$ and $N=1$, several methods yield the estimates
\begin{align}
   \text{Monte Carlo}: & \quad \Dh_{\smash{\hphi^1}} = 1.40(3)~\text{\cite{allais2014magnetic}}, 1.52(6)~\text{\cite{ParisenToldin:2016szc}}, 1.60(5)~\text{\cite{ParisenToldin:2016szc}} \,, \label{eq:ZeroT_ReferenceValue_MC} \\
   \text{Padé}: & \quad \Dh_{\smash{\hphi^1}} \sim 1.55~\text{\cite{Cuomo:2021kfm}}\,, \label{eq:ZeroT_ReferenceValue_Pade} \\
   \text{Fuzzy sphere}: & \quad \Dh_{\smash{\hphi^1}} = 1.63(6)~\text{\cite{Hu:2023ghk}} \,, \label{eq:ZeroT_ReferenceValue_FuzzySphere} \\
   \text{Endpoint bootstrap}: & \quad \Dh_{\smash{\hphi^1}} \sim 1.6081~\text{\cite{Lanzetta:2025xfw}}\,. \label{eq:ZeroT_ReferenceValue_EndpointBootstrap}
\end{align}
We retain the $\mathrm{O}(N)$ index in $\hphi^1$, even for $N=1$, to avoid confusing this operator with the tilt operators present for $N>1$.
As reported in~\cite{Cuomo:2021kfm} and discussed in more detail below, the Padé estimates suggest only a weak dependence of its scaling dimension on $N$.
This is further supported by recent fuzzy-sphere work for $N>1$ and $d=3$~\cite{Yang:2026qtk}, which concluded
\begin{align}
    N=2: \quad \Dh_{\smash{\hphi^1}} &= 1.609(11)\,, \label{eq:ZeroT_Dh_hphi1_N2} \\
    N=3: \quad \Dh_{\smash{\hphi^1}} &= 1.635(58)\,, \label{eq:ZeroT_Dh_hphi1_N3} \\
    N=4: \quad \Dh_{\smash{\hphi^1}} &= 1.666(38)\,. \label{eq:ZeroT_Dh_hphi1_N4}
\end{align}

Another important operator is the displacement $D^i$, associated with the breaking of translations transverse to the line.
It is defined through the Ward identity~\cite{Billo:2016cpy}
\begin{equation}
   \partial_\mu T^{\mu i}(\tau,\bm{x}_\perp)
   =
   \delta^{(d-1)}(\bm{x}_\perp)\, \sqrt{C_D} D^i (\tau)\,,
   \label{eq:ZeroT_WardIdentity_Displacement}
\end{equation}
which fixes its normalization once that of the stress tensor is chosen.
Here the index $i=1,\ldots,d-1$ refers to the spacetime directions transverse to the defect.
The coefficient $C_D$ is physical defect CFT data characterizing the response of the defect to infinitesimal transverse deformations.
With this convention, the two-point function takes the form
\begin{equation}
   \vev{D^i(\tau) D^j(0)}
   =
   \frac{\delta^{ij}}{\tau^4}\,.
   \label{eq:ZeroT_WardIdentity_TwoPointFunction}
\end{equation}
For the pinning-field defect, the displacement also admits a simple microscopic interpretation.
A transverse deformation of the defect position varies the interaction $h\int \mathrm{d}\tau\,\phi_1$ through an insertion of $h\,\partial_i\phi_1$, so that at the defect fixed point
$D^i \sim h_\star \partial_i\phi_1|_{\text{defect}}$, with the precise normalization fixed by the Ward identity.
The displacement is a singlet under $\mathrm{O}(N-1)$ and transforms as a vector under transverse rotations.
We do not consider correlators of the displacement operator in this work, however they can be treated in the same way as scalars.

The rest of the spectrum is largely unknown for general $d$ and $N$.
For $d=3$ and $N=1$, the fuzzy-sphere and endpoint-bootstrap methods identify another scalar operator with dimension
\begin{align}
   \text{Fuzzy sphere}: & \quad \Dh_{s_+} = 3.12(10)~\text{\cite{Hu:2023ghk}}\,,
   \label{eq:ZeroT_ReferenceValue_sp_FuzzySphere} \\
   \text{Endpoint bootstrap}: & \quad \Dh_{s_+} \sim 3.1081~\text{\cite{Lanzetta:2025xfw}}\,,
   \label{eq:ZeroT_ReferenceValue_sp_EndpointBootstrap}
\end{align}
which can be interpreted as the lowest double-twist operator built from $\hphi^1$, namely
\begin{equation}
   \Dh_{s_+} = 2 \Dh_{\smash{\hphi^1}} + \gamma_{s_+}\,,
   \label{eq:ZeroT_spAsDoubleTwist}
\end{equation}
with a relatively small anomalous dimension $\gamma_{s_+}$.
For $N>1$, there is a companion operator, $s_-$, which may similarly be viewed as a double-twist operator built from the tilts $t^a (\tau)$:
\begin{equation}
   \Dh_{s_-} = 2 + \gamma_{s_-}\,.
   \label{eq:ZeroT_smAsDoubleTwist}
\end{equation}
No direct estimate of $\gamma_{s_-}$ is currently available at $d=3$; below, we instead collect indications from the $d=4-\veps$ expansion and from the large-$N$ limit.
Since $s_-$ is interpreted as a $n=0$ member of the double-twist family $[tt]_n$, there is no reason in general to expect $\gamma_{s_-}$ to be parametrically small, as known from familiar bulk examples such as the three-dimensional Ising model.

\paragraph{$\veps$-expansion.}
The action~\eqref{eq:ON_Action} admits a conformal fixed point for non-integer spacetime dimension in the range $2<d<4$.\footnote{Strictly speaking, the continuation to non-integer $\veps$ is non-unitary: evanescent operators, which vanish at the integer-dimensional limit, can create negative-norm states and violate reflection positivity, including at the Wilson–Fisher fixed point~\cite{Hogervorst:2015akt}. Positivity arguments therefore require care when applied at non-integer dimension.}
For the bulk theory, this is the Wilson-Fisher fixed point, which at one loop is given by~\cite{Wilson:1971dc}
\begin{equation}
   \frac{\lambda_\star}{(4\pi)^2} = \frac{3\veps}{N+8} + \mathrm{O}(\veps^2)\,.
   \label{eq:ZeroT_FixedPoint_Bulk}
\end{equation}
To preserve $\mathrm{SO}(2,1)$, and hence obtain a one-dimensional defect CFT, the defect coupling must likewise be tuned to its fixed-point value~\cite{Cuomo:2021kfm}:
\begin{equation}
   h_\star^2 = N + 8 + \mathrm{O}(\veps)\,.
   \label{eq:ZeroT_FixedPoint_Defect}
\end{equation}
Notice that the defect conformal fixed point remains at a nontrivial value even in the free-theory limit $\veps=0$.
A systematic analysis of line-defect beta functions and the stability of their fixed points in general scalar and scalar–fermion theories was presented in~\cite{Pannell:2023pwz}.
To perform perturbative computations, one must first renormalize the operators and couplings before evaluating them at their fixed-point values.
Feynman rules associated to the action~\eqref{eq:ON_Action} with fixed point~\eqref{eq:ZeroT_FixedPoint_Bulk}-\eqref{eq:ZeroT_FixedPoint_Defect} are gathered in Appendix~\ref{subsec:FeynmanRules}.

Let us now discuss the low-lying spectrum from this perturbative perspective.
There are two operators of classical dimension one: the tilt $t^a$ and $\hphi^1$.
An explicit realization of the tilt is given by
\begin{equation}
   t^a (\tau) = \frac{Z_t^{-1}}{\sqrt{n_t}} \phi^a (\tau,\mathbf{0})\,,
   \label{eq:Tilt_PerturbativeDefinition}
\end{equation}
where $Z_t$ is the renormalization constant of the operator, while $n_t$ is the normalization constant ensuring~\eqref{eq:Tilt_ZeroT_TwoPointFunction}.
Both quantities were determined in~\cite{Gimenez-Grau:2022czc} and read
\begin{align}
   Z_t &= 1 - \frac{\lambda}{(4\pi)^2} \frac{h^2}{12 \veps} + \mathrm{O} (\lambda^2)\,, \label{eq:Tilt_RenormalizationConstant} \\
   n_t &= \frac{\Gamma(d/2-1)}{4 \pi^{d/2}} \left[ 1 - \left( 1 + \gamma_E + \log(\pi) \right) \frac{\veps}{2} + \mathrm{O}(\veps^2) \right]\,.
   \label{eq:Tilt_NormalizationConstant}
\end{align}
One can check perturbatively that the scaling dimension of this operator remains protected, $\Dh_t = 1$.

We define the longitudinal singlet $\hphi^1$ as
\begin{equation}
   \hphi^1 (\tau)
   =
   \frac{Z_{\smash{\hphi}^1}^{-1}}{\sqrt{n_{\smash{\hphi^1}}}} \phi^1 (\tau,\mathbf{0})\,,
   \label{eq:phi1_PerturbativeDefinition}
\end{equation}
with (re)normalization constants~\cite{Gimenez-Grau:2022czc}
\begin{align}
   Z_{\smash{\hphi^1}} &= 1 - \frac{\lambda}{(4\pi)^2} \frac{h^2}{4 \veps} + \mathrm{O} (\lambda^2)\,, \label{eq:phi1_RenormalizationConstant} \\
   n_{\smash{\hphi^1}} &= \frac{\Gamma(d/2-1)}{4 \pi^{d/2}} \left[ 1 - 3 \left( 1 + \gamma_E + \log(\pi) \right) \frac{\veps}{2} + \mathrm{O}(\veps^2) \right]\,.
   \label{eq:phi1_NormalizationConstant}
\end{align}
Its scaling dimension was computed up to two loops in~\cite{Cuomo:2021kfm} and reads
\begin{equation}
   \Dh_{\smash{\hphi^1}} = 1 + \veps - \frac{3N^2 + 49N + 194}{2(N+8)^2} \veps^2 + \mathrm{O}(\veps^3)\,.
   \label{eq:phi1_ScalingDim}
\end{equation}
Curiously, as noted in~\cite{Cuomo:2021kfm}, the coefficient at $\mathrm{O}(\veps)$ is independent of $N$, while the coefficient at $\mathrm{O}(\veps^2)$ depends only weakly on $N$.
This is in phase with the non-perturbative estimations~\eqref{eq:ZeroT_Dh_hphi1_N2}-\eqref{eq:ZeroT_Dh_hphi1_N4}.

For future reference, let us mention that the finite-volume antipodal configuration should reproduce these results after taking into account the appropriate Weyl transformation.
One can check diagrammatically that this is indeed the case.
The relevant non-vanishing diagrams for the two-point function of North (or South) tilt operators are
\begin{equation}
    \vev{t_N^a (\tau) t_N^b (0)}
    =
    \delta^{ab} \frac{Z_t^{-2}}{n_t}
    \left(\
    \DiagOneZeroTFiniteV\
    +
    \DiagTwoZeroTFiniteV\
    +
    \DiagThreeZeroTFiniteV\
    +
    \DiagFourZeroTFiniteV\
    + \ldots
    \right)\,,
    \label{eq:Diagrams_FiniteV_ZeroT}
\end{equation}
which yield
\begin{equation}
    \vev{t_N^a (\tau) t_N^b (0)}
    =
    \frac{\delta^{ab}}{\left[ 2 R \sinh \left( \frac{\tau}{2R} \right) \right]^2}
    +
    \mathrm{O}(\veps^2)\,,
    \label{eq:Tilt_TwoPoint_FiniteV_ZeroT}
\end{equation}
providing a check (up to $\mathrm{O}(\veps)$) that the scaling dimension of the tilt is protected.

The displacement operator is defined perturbatively as
\begin{equation}
   D^i (\tau)
   =
   \frac{Z_D^{-1}}{\sqrt{n_D}} \left. \pd^i \phi^1 (\tau,\bm{x}_\perp) \right|_{\bm{x}_\perp = 0}\,.
   \label{eq:D_PerturbativeDefinition}
\end{equation}
The corresponding perturbative constants are
\begin{align}
   Z_D &= 1 - \frac{\lambda}{(4\pi)^2} \frac{h^2}{12 \veps} + \mathrm{O} (\lambda^2)\,, \label{eq:D_RenormalizationConstant} \\
   n_D &=  \frac{\Gamma(d/2-1)}{2 \pi^{d/2}} \left[ 1 - \left( \frac{7}{3} + \gamma_E + \log(\pi) \right) \frac{\veps}{2} + \mathrm{O}(\veps^2) \right]\,.
   \label{eq:D_NormalizationConstant}
\end{align}
The fact that its scaling dimension is protected to the value $\Dh_D=2$ was also checked explicitly at one loop in~\cite{Gimenez-Grau:2022czc}.
Its normalization is related to the displacement two-point coefficient $C_D$ through
\begin{equation}
   C_D = \frac{N+8}{2\pi^2} + \mathrm{O}(\veps)\,.
\end{equation}

At classical dimension two, we also find two scalar operators, which flow across dimensions to the $s_\pm$ operators discussed above.
They can be realized jointly as~\cite{Cuomo:2021kfm}
\begin{equation}
   s_\pm
   =
   \frac{Z_{\smash{s_\pm}}^{-1}}{\sqrt{\smash{n_{s_\pm}}}} \left( A_\pm \phi^I \phi^I - (\phi^1)^2 \right)\,,
   \label{eq:spm_PerturbativeDefinition}
\end{equation}
where we define, for convenience,
\begin{equation}
   A_\pm
   =
   \frac{N + 16 \mp \eta}{4 (N+8)}\,, \qquad
   \eta
   =
   \sqrt{N (N+40) + 320}\,.
   \label{eq:Apm_Definition}
\end{equation}
The normalization constant is given by~\cite{Gimenez-Grau:2022czc}
\begin{align}
   n_{s_\pm}
   &=
   \frac{N A_\pm^2 - 2 A_\pm + 1}{8 \pi^4} + \mathrm{O}(\veps)\,.
   \label{eq:spm_NormalizationConstant}
\end{align}
The corresponding renormalization matrix can be found in Eqs.~(3.41)--(3.43) of~\cite{Gimenez-Grau:2022czc}.
Following the conventions introduced in~\eqref{eq:ZeroT_smAsDoubleTwist} and~\eqref{eq:ZeroT_spAsDoubleTwist}, we identify $s_-$ non-perturbatively with the $n=0$ double-twist operator built from $t^a$, while $s_+$ is identified with the lowest double-twist operator built from $\hphi^1$.
Their anomalous dimensions can then be written as
\begin{align}
   \gamma_{s_-} &= \frac{3N + 20 - \eta}{2 (N+8)} \veps + \mathrm{O}(\veps^2) \,, \label{eq:sm_AnomalousDim} \\
   \gamma_{s_+} &= - \frac{N + 12 - \eta}{2(N+8)} \veps + \mathrm{O}(\veps^2) \,.
   \label{eq:sp_AnomalousDim}
\end{align}
For $N=1$, this gives
\begin{equation}
   \gamma_{s_+} = \frac{\veps}{3} + \mathrm{O}(\veps^2) \,,
   \label{eq:sp_AnomalousDim_Ising}
\end{equation}
recalling that $s_-$ does not exist.
In the large-$N$ limit, one instead finds
\begin{align}
   \gamma_{s_-} &= \left(1 - \frac{8}{N} + \mathrm{O} (N^{-2}) \right) \veps + \mathrm{O}(\veps^2) \,, \label{eq:sm_AnomalousDim_LargeN} \\
   \gamma_{s_+} &= \left( \frac{4}{N} + \mathrm{O} (N^{-2}) \right) \veps + \mathrm{O}(\veps^2) \,.
   \label{eq:sp_AnomalousDim_LargeN}
\end{align}
Beyond the low-lying operators reviewed above, the perturbative defect spectrum has been studied more extensively in~\cite{Belton:2025hbu}, where the complete spectrum through classical dimension three is determined at $O(\veps)$, as well as several sectors of operators with classical dimension four.
Their analysis resolves operator mixing and provides additional OPE data.

\paragraph{$2d$ theory.}
The two-dimensional case $\veps=2$ is particularly transparent for $N=1$,
where the pinning field flows to a factorizing Ising interface with
identical fixed-spin boundary conditions on its two sides, either
$+$ or $-$~\cite{Cardy:1989ir,Cuomo:2021kfm}.
The two half-plane theories therefore decouple.
In the antipodal cylinder geometry, they become two independent strips of width $\pi R$.
The defect spectrum is then given by the tensor product of two Ising vacuum Virasoro modules.

Denoting the corresponding boundary stress tensors by $T_L$ and $T_R$, the lowest nontrivial operators can be identified, up to normalization, as
\begin{align}
    \hphi^{1}
    &\overset{d\to 2}{\longrightarrow}
    T_L+T_R\,, \label{eq:hphi1In2d} \\
    D
    &\overset{d\to2}{\longrightarrow}
    T_L-T_R\,,
\end{align}
with $\Dh_{\smash{\hphi^1}} = \Dh_D = 2$.
These combinations are respectively even and odd under reflection across the interface.
Although they are Virasoro descendants, they are primaries of the global defect conformal group.
As mentioned before, the tilt operators $t^a$ and the singlet $s_-$ are absent for $N=1$.

\begin{figure}
    \centering
    \includegraphics[width=0.8\linewidth]{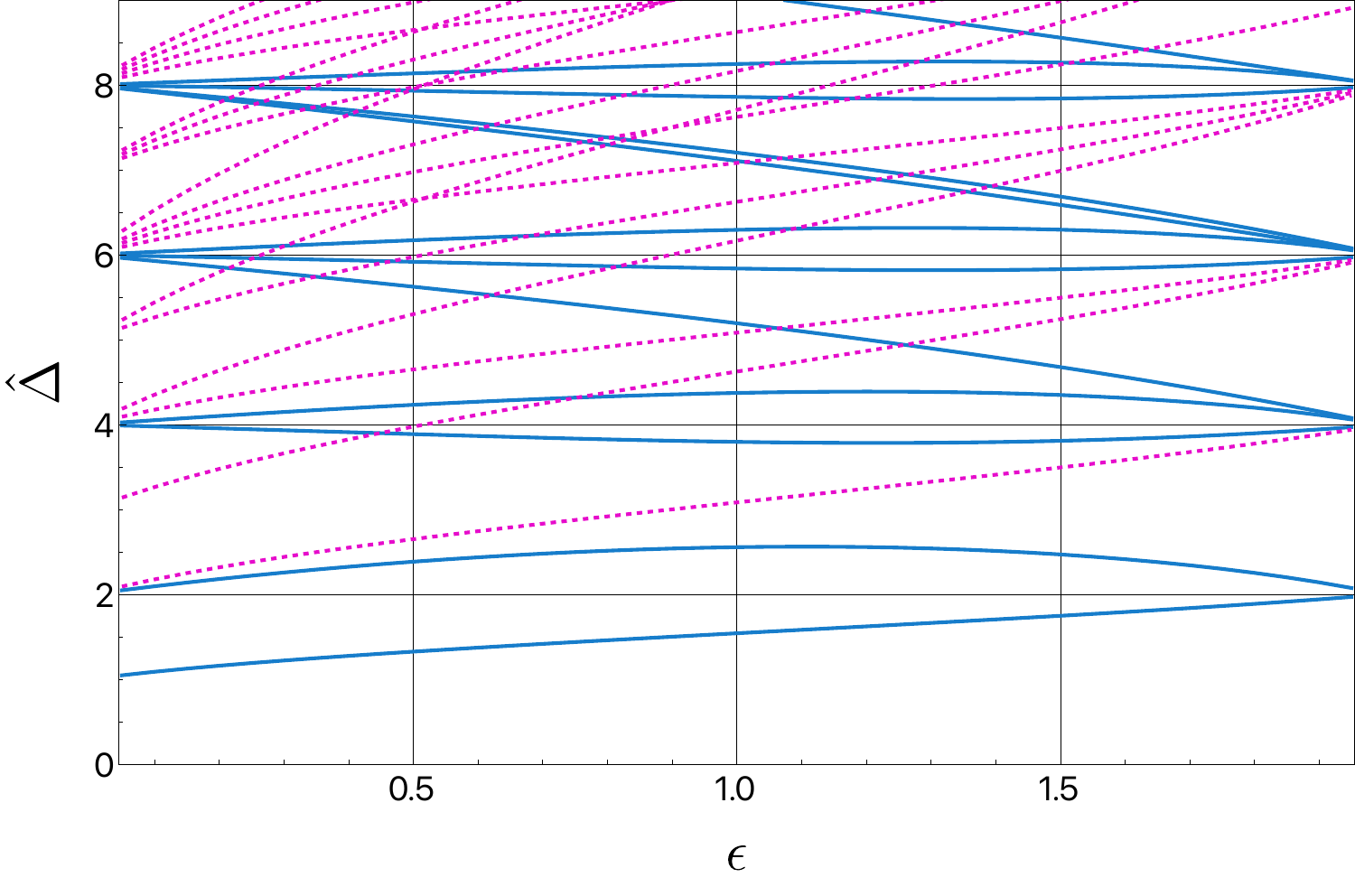}
    \caption{Low-lying spectrum of $\mathrm{O}(N-1)$ singlet representations at large $N$ across dimensions, with $d=4-\veps$.
    The blue solid lines are the single-particle operators determined by Equation~\eqref{eq:LargeN_Spectrum}~\cite{Cuomo:2021kfm}.
    The pink dashed curves represent classical multi-twist operators built from the single-particle states.
    The lowest operator, $\hphi^1$, flows from $\Dh_{\smash{\hphi^1}} (\veps=0) = 1$ to $\Dh_{\smash{\hphi^1}} (\veps=2) = 2$.}
    \label{fig:LargeNSpectrum}
\end{figure}

The next parity-even global-primary sector has dimension four and is spanned by
\begin{equation}
    T_L T_R
    \quad \text{and} \quad
    \Lambda_L+\Lambda_R\,,
    \qquad
    \text{with }
    \Lambda_X \equiv {:T_X T_X:} - \frac{3}{10}\,\partial_\tau^2 T_X\,.
\end{equation}
Identifying $s_+$ with an operator that continuously flows into this sector gives
$\Dh_{s_+}\to 4$, which also matches the large $N$ analysis presented in Figure~\ref{fig:LargeNSpectrum}.

\paragraph{Large $N$.}
We now consider the large $N$ limit of the model, following~\cite{Cuomo:2021kfm}.
In this limit, the theory simplifies considerably and allows the defect spectrum to be determined directly at fixed spacetime dimension.
After performing a Hubbard--Stratonovich transformation, the (critical) action becomes
\begin{equation}
   S
   =
   \int \mathrm{d}^d x
   \left[
       \frac{1}{2} (\pd \phi^I)^2
       +
       \frac{s}{2} \phi^I \phi^I
       +
       \sqrt{N} J^I \phi^I
   \right]\,,
   \label{eq:ZeroT_LargeN_Action}
\end{equation}
where
\begin{equation}
   J^I = \delta^{1I}\,h\,\delta^{(d-1)}(\bm{x}_\perp)\,.
   \label{eq:J_Definition}
\end{equation}
Note that the normalization of $h$ differs from that used in~\eqref{eq:ON_Action} by a factor of $\sqrt{N}$.

The leading large-$N$ solution is determined self-consistently by
\begin{align}
   \left[-\pd^2+\vev{s(x)} \right] \vev{\phi^1 (x)} &= - \sqrt{N} J^1\,,
   \label{eq:LargeN_Fphi} \\
   \left[-\pd_x^2+\vev{s(x)} \right] G_\perp (x,y)
   &=\delta^{(d)}(x-y)\,,
   \label{eq:LargeN_Gphi} \\
   G_\perp (x,x)+\frac{\vev{\phi^1 (x)}^2}{N} &=0\,.
   \label{eq:LargeN_Gap}
\end{align}
The first equation relates the bulk one-point function $\vev{s(x)}$ to $\vev{\phi^1(x)}$; the second relates the transverse bulk propagator,
\begin{equation}
    \vev{\phi^a(x) \phi^b(y)}
    =
    \delta^{ab} G_\perp (x,y)\,,
\end{equation}
to $\vev{s(x)}$; the third imposes the gap equation, relating the coincident propagator $G_\perp (x,x)$, understood to be suitably regularized, to $\vev{\phi^1(x)}$.
Conformal invariance fixes
\begin{equation}
   \vev{s(x)} = \frac{\mu_{s \hat{\mathds{1}}}}{r^2}\,,
   \qquad
   \vev{\phi^1 (x)} = \frac{\mu_{\phi \hat{\mathds{1}}}}{r^{(d-2)/2}}\,,
   \label{eq:ZeroT_LargeN_ConformalSymmetry}
\end{equation}
and Equations~\eqref{eq:LargeN_Fphi}-\eqref{eq:LargeN_Gap} determine
\begin{align}
   \mu_{s \hat{\mathds{1}}}
   &=
   \frac{(4-d)(d-2)}{4}
   \overset{d=3}{=}
   \frac{1}{4}\,,
   \label{eq:ZeroT_LargeN_BulkOnePoint_s}
   \\
   \mu_{\phi \hat{\mathds{1}}}
   &\overset{d=3}{\sim}
   -0.210744 \sqrt{N}\,.
\end{align}
In the second line, $\mu_{\phi \hat{\mathds{1}}}$ can be determined numerically to high precision for arbitrary $d$, but no functional form is known.
Note that $\mu_{s \hat{\mathds{1}}}$ is denoted by $s_0$ in~\cite{Cuomo:2021kfm}.
Also, the sign of $\mu_{\phi \hat{\mathds{1}}}$ depends on the sign convention chosen for the coupling $h$.

The $\mathrm{O}(N-1)$-singlet defect spectrum\footnote{We focus on this sector since only $\mathrm{O}(N-1)$-singlets can be exchanged in the two-point function of operators at finite temperature.} arises from fluctuations $s(x) = \vev{s(x)} + \delta s (x)$.
At leading order, the associated inverse propagator is
\begin{equation}
    \frac{G_{\delta s \delta s}^{-1} (x,y)}{N}
    =
    - \frac{1}{2} G_\perp (x,y)^2
    - \frac{\vev{\phi^1 (x)}}{\sqrt{N}} G_\perp (x,y)
    \frac{\vev{\phi^1 (y)}}{\sqrt{N}}\,.
    \label{eq:ZeroT_LargeN_ds_Inverse}
\end{equation}
It is convenient to Weyl transform the problem to $\mathrm{AdS}_2\times S^{d-2}$, where the saddle becomes homogeneous and $G_\perp$ admits a spectral decomposition.
Denoting the spherical-harmonic multiplicity as
\begin{equation}
    D_\ell
    =
    \frac{(2\ell+d-3)\Gamma(\ell+d-3)}{\Gamma(d-2)\Gamma(\ell+1)}\,,
    \qquad
    \lambda_\ell=\ell(\ell+d-3)\,,
    \qquad
    \operatorname{Vol}(S^{d-2})
    =
    \frac{2\pi^{(d-1)/2}}
    {\Gamma\!\left(\frac{d-1}{2}\right)}
    \label{eq:ZeroT_LargeN_Definitions}
\end{equation}
the dimensions of scalar single-trace $\mathrm{O}(N-1)$ singlets are determined by the zeros of the function $B_0$, defined through
\begin{equation}
    0
    =
    B_0\left(i\left(\Dh-\frac{1}{2}\right)\right)
    =
    \frac{\tilde{J}^{\,2}}
    {\operatorname{Vol}(S^{d-2})^2\,\Dh(\Dh-1)}
    -
    \frac{1}{2}
    \sum_{\ell=0}^\infty
    \frac{D_\ell}{\operatorname{Vol}(S^{d-2})}\,
    \tilde{B}_{\lambda_\ell}
    \left(i\left(\Dh-\frac{1}{2}\right)\right)\,.
\label{eq:LargeN_Spectrum}
\end{equation}
Here $\widetilde B_{m^2}(\nu)$ is the spectral density of the
$\mathrm{AdS}_2$ scalar bubble,
\begin{equation}
    \widetilde B_{m^2}(\nu)
    =
    \int_{\mathrm{AdS}_2}\mathrm{d}^2 X\,
    \left[G_{m^2}^{\mathrm{AdS}_2}(X,Y)\right]^2
    \Omega_\nu(X,Y)\,,
    \label{eq:ZeroT_LargeN_BubbleIntegral}
\end{equation}
while the tilded quantities denote the $\mathrm{AdS}_2$ counterparts of the flat-space quantities, defined alongside the AdS harmonic function $\Omega_\nu(X,Y)$ in~\cite{Cuomo:2021kfm} and/or in Appendix~\ref{app:LargeN_WittenDiagrams}.
The zeros of~\eqref{eq:LargeN_Spectrum} therefore determine the single-particle singlet defect spectrum for arbitrary $2<d<4$.
We plot this spectrum in Figure~\ref{fig:LargeNSpectrum}, together with the lowest classical multi-twist operators, formed by merging single-particle operators in a classical way.

Note as an aside that, for a high-enough given dimension, there are \textit{two} states that seem to be attached to the double-twist of the tilt operator.
We denote them as $[tt]_{n,\pm}$, with scaling dimensions
\begin{equation}
    \Dh_{\smash{[tt]_{n,\pm}}}
    =
    2 + 2n + \gamma_{\smash{[tt]_{n,\pm}}}\,.
    \label{eq:LargeN_DoubleTwistTilt}
\end{equation}
One can determine analytically from~\eqref{eq:LargeN_Spectrum} their leading large-$n$ asymptotic corrections.
These are symmetric and can be expressed as
\begin{equation}
    \gamma_{\smash{[tt]_{n,\pm}}}
    =
    \pm \frac{2^{d-3/2} \Gamma((d-1)/2) \sin(\pi(d-2)/2)}{\pi^{3/2} (2+2n)^{(d-2)/2}}
    + \ldots\,.
    \label{eq:LargeN_AnomalousDims}
\end{equation}
Higher-order corrections break the $\pm$ symmetry.
This approximation is particularly accurate for $d \geq 3$ for the $-$ states.
It also suggests identifying $\hphi^1 (\tau)$ as $[tt]_{0,-}$ for $d \ll 4$, while $s_-$ is associated to $[tt]_{0,+}$.
Equation~\eqref{eq:LargeN_DoubleTwistTilt} differs from the $N=1$ splitting~\eqref{eq:sp_AnomalousDim_Ising}, in which the tilt does not exist and $s_+$ is realized as the double-twist operator of $\hphi^1$.

The $d = 2$ case has a useful free-field interpretation.
Along the critical branch, the non-linear sigma-model coupling vanishes, while the pinning field fixes the order-parameter orientation on the line.
The remaining $N-1$ transverse fluctuations become free massless scalars with Dirichlet boundary conditions on the two decoupled half-planes.
Their boundary normal derivatives have dimension one and furnish the tilt sector, while singlet bilinears generate dimension-two operators.
This is consistent with the large-$N$ spectrum shown in Figure~\ref{fig:LargeNSpectrum}.

\subsection{Line defect CFT in thermal geometries}
\label{subsec:ThermalDefectCFT}

We now consider correlators of line-defect operators at finite temperature.
Specifically, we wrap the line defect around the thermal circle and study two-point functions of scalar defect operators,
\begin{equation}
   g(\tau)
   =
   \vev{\hphi(\tau) \hphi(0)}\,,
   \label{eq:g_Definition}
\end{equation}
where $\hphi(\tau)$ denotes a generic scalar defect operator.
We first describe the resulting symmetry breaking, both for a flat spatial manifold and for the sphere, and discuss its consequences for the OPE and the associated thermal blocks.
We then discuss properties of defect correlators familiar from bulk thermal CFTs.
Finally, we exploit their analytic structure to derive a manifestly periodic formulation of the OPE and use it to obtain asymptotic expressions for the double-twist OPE coefficients.

\subsubsection{Symmetries and OPE}
\label{subsubsec:SymmetriesAndOPE}

We begin by studying the symmetries of the system in the two thermal geometries and their consequences for the thermal block expansions.

\begin{figure}
    \centering
    \begin{minipage}{.5\textwidth}
        \centering
        \includegraphics[width=0.95\linewidth]{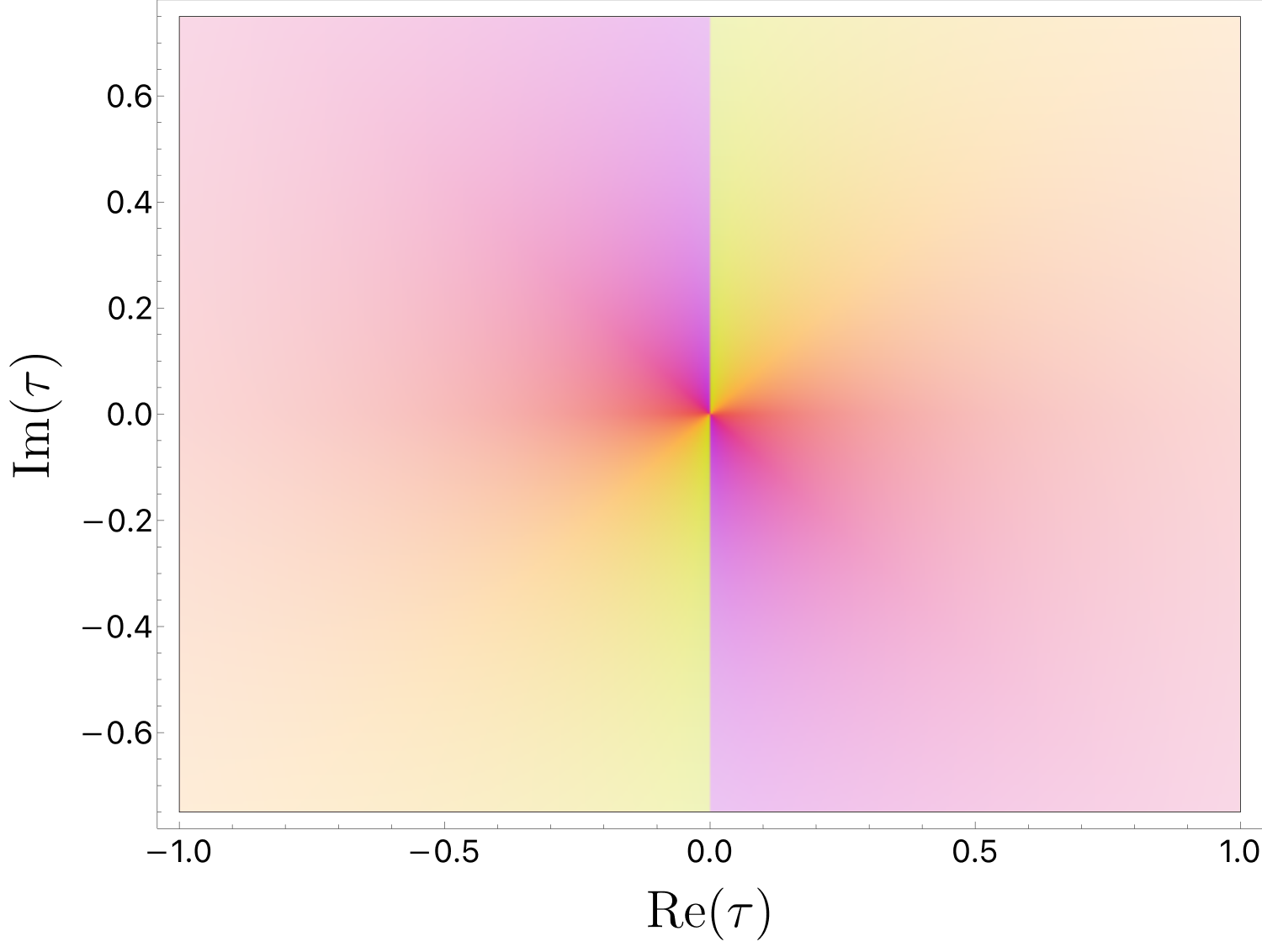}
    \end{minipage}%
    \begin{minipage}{0.5\textwidth}
        \centering
        \includegraphics[width=0.95\linewidth]{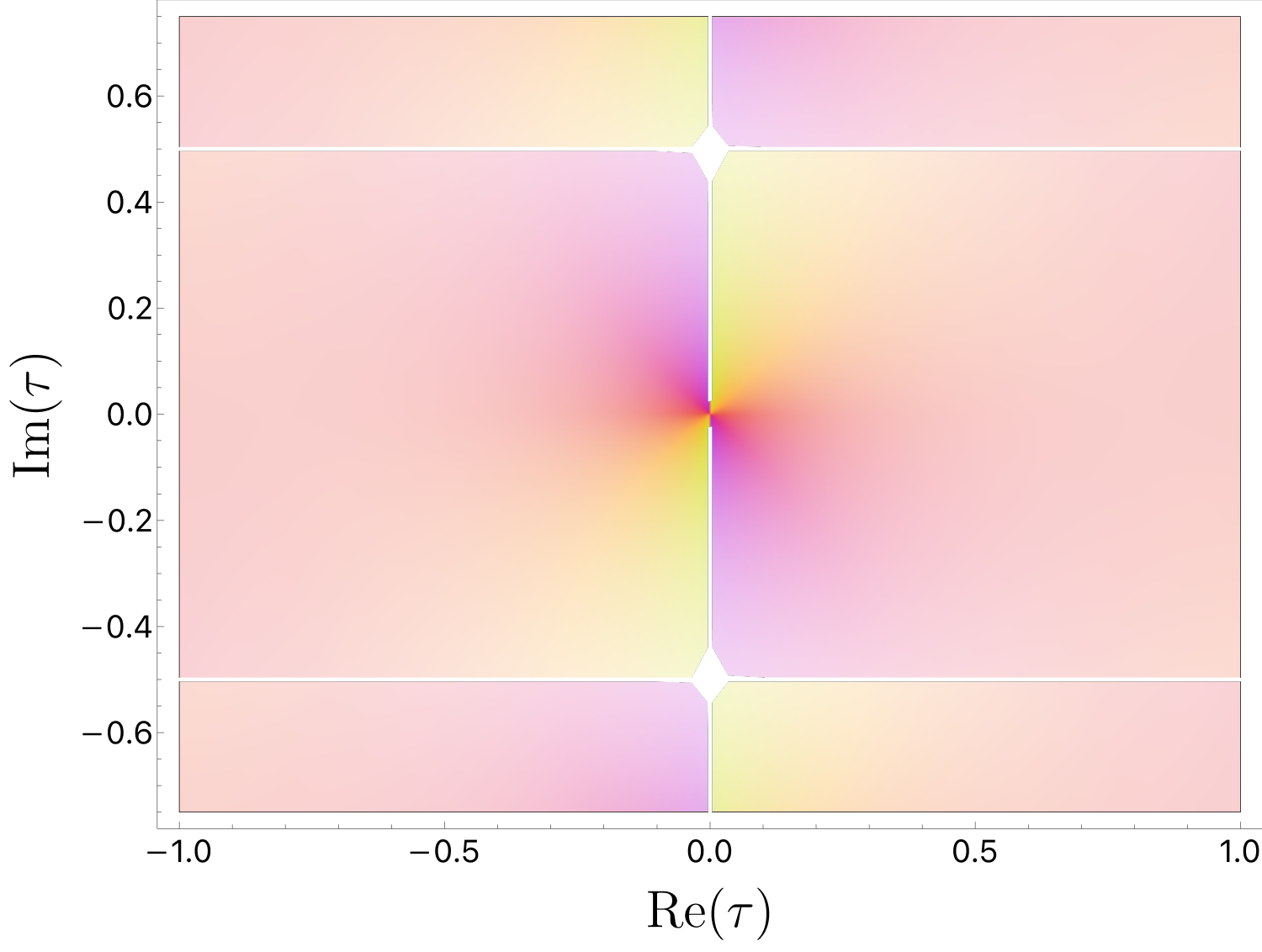}
    \end{minipage}
    \caption{Analytic structure of thermal blocks at infinite (\textbf{left}) and finite (\textbf{right}) volume in the complex-$\tau$ plane, according to their respective definitions~\eqref{eq:ThermalBlocks_InfiniteV} and~\eqref{eq:ThermalBlocks_FiniteV}.
    For complex separation, the powers in these expressions are understood as $(\tau^2/\beta^2)^p$ and $[4R^2\sinh^2(\tau/(2R))/\beta^2]^p$, respectively.
    Here we set $\beta = 2 \pi R = 1$.
    Ordinary thermal blocks are not periodic, and thus the cut located at $\operatorname{Re}(\tau) = 0$ is not reproduced at $\operatorname{Re}(\tau) = m \beta$ with $m \in \mathbb{Z}$.
    The horizontal cuts at finite volume are artifacts of the choice of continuation for the hypergeometric function present in~\eqref{eq:ThermalBlocks_FiniteV}.
    We include them here to emphasize that this expression should be used in the interval $-\pi R < \operatorname{Im}(\tau) < \pi R$.}
    \label{fig:ThermalBlocks_AnalyticStructure}
\end{figure}

\paragraph{Infinite volume.}
If the spatial manifold is $\mathbb{R}^{d-1}$, introducing the temperature $T=1/\beta$ breaks the defect conformal symmetry~\cite{Barrat:2024aoa}:
\begin{equation}
    \mathrm{SO}(2,1) \times \mathrm{SO}(d-1)
    \longrightarrow
    \mathrm{U}(1)_\tau \times \mathrm{SO}(d-1)\,.
    \label{eq:SymmetryBreaking_SO}
\end{equation}
Dilatations and one-dimensional special conformal transformations are broken, while temporal translations and spatial rotations are preserved.
The thermal identification acts along the straight defect coordinate,
\begin{equation}
    \tau \sim \tau + \beta\,,
    \label{eq:ThermalIdentification}
\end{equation}
with associated translation generator $P_\tau$.
The global internal symmetries are unaffected by the finite temperature.
Using the notation of~\eqref{eq:ZeroT_SymmetryBreaking_G}, we simply have
\begin{equation}
    G' \longrightarrow G'\,.
    \label{eq:SymmetryBreaking_G}
\end{equation}
Considering the OPE between two identical operators:
\begin{equation}
    \hphi(\tau) \hphi(0)
    =
    \sum_{\Oh} f_{\hphi\hphi\Oh}\, |\tau|^{\Dh - 2\Dh_{\hphi}} \Oh(\tau) + \text{descendants}\,,
\end{equation}
we can write a thermal block expansion for the correlator $g(\tau)$.
Here $f_{\smash{\hphi\hphi\Oh}}$ is the zero-temperature three-point coefficient defined through~\eqref{eq:ZeroT_ThreePointFunctions}.
The thermal one-point functions are fixed by translational invariance to take the form
\begin{equation}
   \vev{\Oh (0)}
   =
   \frac{b_{\smash{\Oh}}}{\beta^{\smash{\Dh_{\Oh}}}}\,,
   \label{eq:b_Definition_InfiniteV}
\end{equation}
Only scalar operators ($s = 0$) can acquire a thermal one-point function.
Descendants also have vanishing expectation values as a consequence of translation invariance.
The correlator can thus be expressed as
\begin{equation}
   g(\tau)
   =
   \frac{1}{\beta^{\smash{2\Dh_{\hphi}}}}
   \sum_{\Oh} a_\Oh\, f_\Oh (\tau)\,,
   \label{eq:OPE_InfiniteV}
\end{equation}
where we defined the thermal blocks,
\begin{equation}
   f_\Oh (\tau)
   =
   \left( \frac{\tau}{\beta} \right)^{2p}\,,
   \label{eq:ThermalBlocks_InfiniteV}
\end{equation}
and with
\begin{equation}
   p = \frac{\Dh_\Oh}{2} - \Dh_{\smash{\hphi}}\,,
   \label{eq:p_Definition}
\end{equation}
as well as the combination of thermal OPE coefficients
\begin{equation}
   a_{\smash{\Oh}}
   =
   f_{\smash{\hphi\hphi\Oh}} b_{\smash{\Oh}}\,.
   \label{eq:a_Definition}
\end{equation}
The absence of descendants explains why the thermal blocks take the particularly simple form~\eqref{eq:ThermalBlocks_InfiniteV}.
They can also be obtained by considering the limit $x=0$ of the bulk blocks, see for instance~(2.12) in~\cite{Iliesiu:2018fao}.
Their analytic structure in the complex-$\tau$ plane is displayed in Figure~\ref{fig:ThermalBlocks_AnalyticStructure}.

In Section~\ref{sec:LargeN}, we also study one-point functions of bulk operators at large $N$ in order to extract OPE data.
We summarize here the consequences of the symmetry breaking~\eqref{eq:SymmetryBreaking_SO} for these correlators.
As shown in~\cite{Barrat:2024aoa}, they depend on a single variable and can be expanded using the bulk-defect OPE~\eqref{eq:ZeroT_BulkDefectOPE}:
\begin{equation}
    \vev{\Om (x)}
    =
    \frac{F_\Om (z)}{\beta^{\smash{\Delta_\Om}}}\,,
    \qquad
    z = \frac{r}{\beta}\,,
    \label{eq:ThermalBulkOnePointFunctions}
\end{equation}
with
\begin{equation}
    F_\Om (z)
    =
    \sum_{\smash{\Oh}} \mu_{\smash{\Om\Oh}} b_{\smash{\Oh}} z^{\Dh_{\Oh}-\Delta_\Om}\,,
    \label{eq:ThermalBulkOnePointFunctions_OPE}
\end{equation}
where $\mu_{\smash{\Om\Oh}}$ denotes the coefficient appearing in the bulk-defect two-point function~\cite{Billo:2016cpy}
\begin{equation}
    \vev{\Om (\tau,\bm{x}_\perp) \Oh (0)}_{T=0}
    =
    \mu_{\Om\Oh}
    \frac{r^{\smash{\Dh_\Oh - \Delta_\Om}}}{(\tau^2 + r^2)^{\smash{\Dh_\Oh}}}\,.
    \label{eq:ZeroT_BulkDefect_OPECoefficients}
\end{equation}

\paragraph{Finite volume.}
At finite volume, the symmetry breaking is again described by~\eqref{eq:SymmetryBreaking_SO}.
However, translations along cylinder time are generated by flat-space dilatations $D/R$.
Descendants are therefore allowed to contribute, since they are not annihilated by $P_\tau$.
As a result, the thermal block expansion takes a more complicated form.
Restricting ourselves to two operators living both on either the North or South pole, the correlator can be written as
\begin{equation}
    g(\tau)
    =
    \frac{1}{\beta^{\smash{2\Dh_{\hphi}}}}
    \sum_{\Oh} a_\Oh (q)\, f_\Oh (\tau)\,,
    \label{eq:OPE_FiniteV}
\end{equation}
where we have introduced the dimensionless variable $0 \leq q \leq 1$,
\begin{equation}
    q = e^{-\beta/R}\,.
    \label{eq:q_Definition}
\end{equation}
The limit $q\to 1$ corresponds to finite temperature and infinite volume ($R\to\infty$), while $q\to 0$ corresponds to zero temperature at finite volume ($\beta\to\infty$).
This dependence originates from the one-point function coefficient, since the zero-temperature three-point coefficient is temperature-independent:
\begin{equation}
    \vev{\Oh (0)}
    =
    \frac{b_{\smash{\Oh}} (q)}{\beta^{\smash{\Dh_{\Oh}}}}\,.
    \label{eq:b_Definition_FiniteV}
\end{equation}
These blocks were obtained by resumming the defect descendants on the cylinder.
They coincide with zero-temperature heavy-light-light-heavy blocks in the $d=1$ limit~\cite{Hogervorst:2013kva}.
The thermal blocks also depend on $q$ (through the ratio $R/\beta$) and, due to the contribution of descendants, take the form
\begin{equation}
    f_\Oh (\tau)
    =
    \left[ \frac{2 R}{\beta} \sinh \left( \frac{\tau}{2R} \right) \right]^{2p}
    {}_2 F_1 \left( \frac{\Dh_{\smash{\Oh}}}{2}, \frac{\Dh_{\smash{\Oh}}}{2}; \Dh_{\smash{\Oh}} + \frac{1}{2} ;  -\sinh^2 \left( \frac{\tau}{2R} \right) \right)\,.
    \label{eq:ThermalBlocks_FiniteV}
\end{equation}
The analytically continued expression is well-defined for real $0<\tau<\beta$ at any $R>0$, and reduces to~\eqref{eq:ThermalBlocks_InfiniteV} as $R\to\infty$.
Note that~\eqref{eq:ThermalBlocks_FiniteV} is valid in the OPE regime, and that the hypergeometric function introduces horizontal branch cuts in the complex plane of $\tau$ (see Figure~\ref{fig:ThermalBlocks_AnalyticStructure}), which are not expected to be present in the full correlator.

At finite volume, the one-point function of bulk operators~\eqref{eq:ThermalBulkOnePointFunctions} becomes
\begin{equation}
    \vev{\Om (x)}
    =
    \frac{F_\Om (\theta)}{\beta^{\Delta_\Om}}\,,
    \label{eq:ThermalBulkOnePointFunctions_FiniteVolume}
\end{equation}
with the bulk-defect OPE around the North pole taking the form
\begin{equation}
    F_\Om (\theta)
    =
    \sum_{\Oh}
    \mu_{\Om\Oh} b_{\Oh}(q)
    \left(\frac{R\sin (\theta)}{\beta}\right)^{
        \Dh_{\Oh}-\Delta_\Om}
    {}_2F_1\!\left(
        \frac{\Dh_{\Oh}}{2},
        \frac{\Dh_{\Oh}}{2};
        \Dh_{\Oh}+\frac12;
        \sin^2 (\theta)
    \right)\,.
    \label{eq:ThermalBulkOnePointFunctions_OPE_FiniteVolume}
\end{equation}
The hypergeometric function resums the defect descendants induced by the Weyl transformation to the cylinder.

\subsubsection{Additional constraints}
\label{subsubsec:AdditionalConstraints}

In this section, we use the density matrix formulation of correlation functions to determine constraints on two-point functions of defect operators, in close analogy with bulk thermal theories.
We work in the geometry $S_\beta^1 \times S_R^{d-1}$; the corresponding statements for the spatial manifold $\mathbb{R}^{d-1}$ follow by taking the limit $R \to \infty$.

\paragraph{Defect density matrix.}
Euclidean time translations are generated by the Hamiltonian $H_D$, so compactifying $\tau$ with period $\beta$ amounts to identifying configurations after evolution by $e^{-\beta H_D}$.
Correlation functions of defect operators can therefore be written as thermal traces over $\mathcal H_D$, the full Hilbert space of the theory in the presence of the prescribed defect configuration:
\begin{equation}
    \vev{\Oh_1(\tau_1) \cdots \Oh_n(\tau_n)}
    =
    \text{tr}_{\Hm_D}
    \left[
    \rho_D\,
    \Oh_1(\tau_1)\cdots
    \Oh_n(\tau_n)
    \right]\,,
    \qquad
    \rho_D
    =
    \frac{e^{-\beta H_D}}{Z_D}\,,
    \label{eq:DefectDensityMatrix}
\end{equation}
with $Z_D=\text{tr}_{\Hm_D} \left[ e^{-\beta H_D} \right]$.
In the following, we use this representation to derive properties of defect correlators.

\paragraph{Spectral representation and periodicity.}
We now turn~\eqref{eq:DefectDensityMatrix} into a spectral representation for the two-point functions $g(\tau)$.
To do so, we insert a complete set of energy eigenstates $|n\rangle$ of the Hamiltonian, such that $H_D |n\rangle = E_n |n\rangle$.
For two identical Hermitian defect operators $\Oh$ and $0<\tau<\beta$, this gives
\begin{align}
    g(\tau)
    &=
    \vev{\Oh (\tau) \Oh (0)}_\beta
    \notag \\
    &=
    \frac{1}{Z_D}
    \sum_{m,n}
    e^{-\beta E_m}
    e^{-\tau (E_n - E_m)}
    \bigl|
    \vev{ m|\Oh |n}
    \bigr|^2\,.
    \label{eq:SpectralDecomposition}
\end{align}
Introducing the positive spectral measure,
\begin{equation}
    \rho(\omega)
    =
    \frac{1}{Z_D}
    \sum_{m,n}
    e^{-\beta E_m}
    \bigl|
    \vev{m|\Oh|n}
    \bigr|^2
    \delta\!\left(\omega - E_n + E_m\right)\,,
    \label{eq:rhoPositive_Definition}
\end{equation}
the correlator takes the Laplace form
\begin{equation}
    g(\tau)
    =
    \int_{-\infty}^{\infty}
    \mathrm{d}\omega\,
    \rho(\omega) e^{-\omega\tau}\,.
    \label{eq:LaplaceForm}
\end{equation}
Exchanging $m$ and $n$ yields the detailed-balance relation
$\rho(-\omega)=e^{-\beta\omega}\rho(\omega)$, from which the KMS symmetry (the finite-temperature analog of~\eqref{eq:ZeroT_Crossing}) follows immediately~\cite{Kubo:1957mj,Martin:1959jp}:
\begin{equation}
    g(\tau)
    =
    g(\beta - \tau)\,.
    \label{eq:KMS}
\end{equation}
For a fixed Hamiltonian and a complete operator algebra, the full KMS condition for all observables uniquely fixes the Gibbs density matrix to be~\eqref{eq:DefectDensityMatrix}, provided that the thermal trace exists.
The KMS condition allows to equate two thermal block expansions~\eqref{eq:OPE_InfiniteV}: one around $\tau \sim 0$, which we call $s$-channel, and one around $\tau \sim \beta$, to which we refer as $t$-channel.

\paragraph{Positivity, holomorphy, and analytic structure.}
The spectral representation makes both positivity and analyticity manifest.
For $0<\tau<\beta$, every term in the spectral sum is non-negative, and therefore the correlator is positive:
\begin{equation}
   g(\tau) \geq 0\,.
   \label{eq:Positivity}
\end{equation}
More generally, after complexifying the Euclidean separation $\tau$ in~\eqref{eq:SpectralDecomposition}, $g(\tau)$ is manifestly holomorphic in the open strip $0 < \operatorname{Re}(\tau) < \beta$, provided the thermal trace converges.
The two boundaries of this strip correspond, upon analytic continuation, to the two Lorentzian Wightman orderings.

\begin{figure}[t]
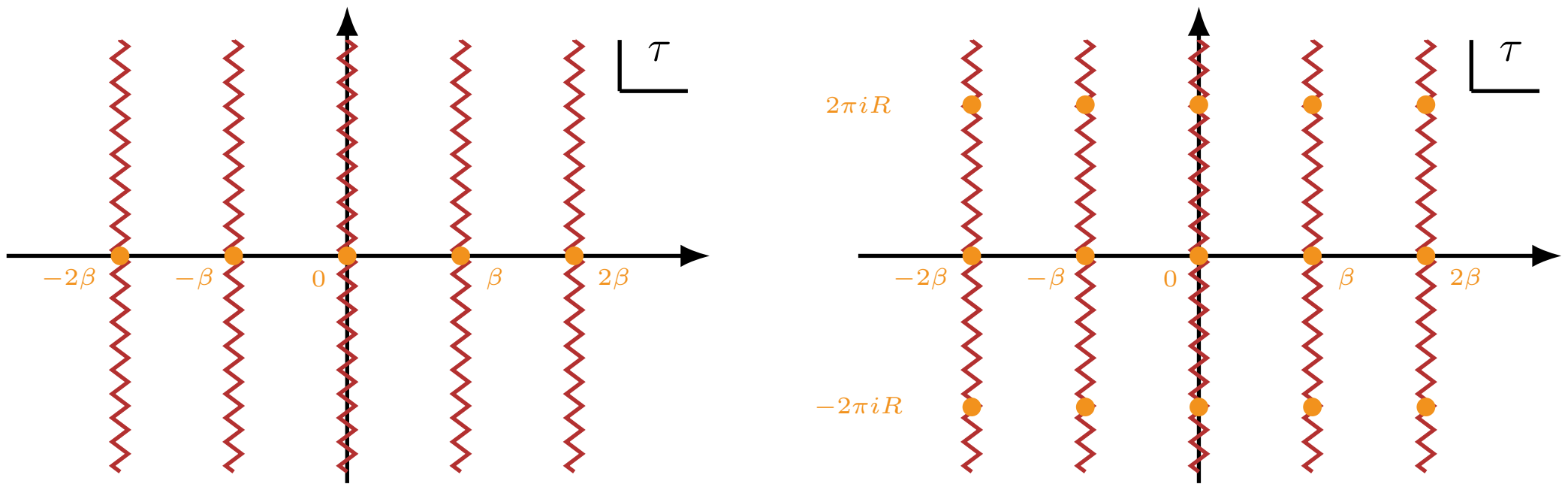

    \centering
    \AnalyticStructureInfiniteVolume
    \qquad
    \AnalyticStructureFiniteVolume
    \caption{
    Analytic structure of thermal two-point functions of line defect operators in the complex $\tau$-plane.
    \textbf{Left}: At infinite volume, the Euclidean correlator consists of cuts along the imaginary directions located at collision points $\operatorname{Re}(\tau) = m\beta$ on the real axis and extending from $-\infty$ to $+\infty$.
    The orange dots indicate singularities.
    \textbf{Right:} The same structure persists for correlators in the antipodal configuration, with the difference that there are additional singularities along the imaginary axis located at $m\beta + 2\pi i n R$.
    Note that the correlator is not periodic in the imaginary direction.
    }
    \label{fig:AnalyticStructure}
\end{figure}

The holomorphy of $g(\tau)$ in the open thermal strip implies that singularities can only arise on its boundaries, and periodicity then replicates them at $\operatorname{Re}(\tau)=m\beta$, with $m\in\mathbb{Z}$.
These singular loci originate from operator collisions at the points $\tau=m\beta$ and extend, after analytic continuation in the imaginary direction, into vertical branch cuts.
Thus, at infinite volume, the analytic structure consists of vertical cuts attached to the collision points $\tau=m\beta$.
At finite volume, the same structure persists, but additional singularities appear along the imaginary direction due to the compact spatial geometry, located at $\tau=m\beta+2\pi i n R$, with $m,n\in\mathbb{Z}$.
The analytic structure associated to both spatial manifolds is represented in Figure~\ref{fig:AnalyticStructure}.

\paragraph{Boundedness.}
The spectral representation also implies a useful boundedness property within the thermal strip.
For $0< \operatorname{Re}(\tau) <\beta$, we have
\begin{equation}
    g(\tau)
    =
    \frac{1}{Z_D}
    \sum_{m,n}
    e^{-\beta E_m}
    e^{-\operatorname{Re}(\tau) (E_n - E_m)}
    e^{-i \operatorname{Im}(\tau) (E_n - E_m)}
    \bigl|
    \vev{m|\Oh|n}
    \bigr|^2\,.
    \label{eq:Boundedness_Step}
\end{equation}
Since the dependence on $\operatorname{Im}(\tau)$ is a pure phase, one can use the triangle inequality to write
\begin{equation}
    |g(\sigma+i t)|
    \leq
    \frac{1}{Z_D}
    \sum_{m,n}
    e^{-\beta E_m}
    e^{- \sigma (E_n - E_m)}
    \bigl|
    \vev{m|\Oh|n}
    \bigr|^2
    =
    g(\sigma)\,.
    \label{eq:Boundedness}
\end{equation}
Hence the correlator is uniformly bounded along every vertical line inside the thermal strip.
In particular, on any closed substrip $\epsilon \leq \operatorname{Re}(\tau) \leq \beta-\epsilon$, it is bounded by a finite constant.
Possible divergences can occur only as one approaches the boundaries
$\operatorname{Re}(\tau)=0,\beta$, where the short-distance singularities of the local operators reside.

\paragraph{Clustering.}
Clustering concerns the large real-time separation of the connected correlator,
\begin{equation}
   g_{\rm conn} (\tau)
   =
   \vev{\Oh(\tau) \Oh(0)}
   -
   \vev{\Oh}^2\,.
   \label{eq:ConnectedCorrelator_Definition}
\end{equation}
In terms of the connected spectral density, one may write
\begin{equation}
   g_{\rm conn}(\sigma + i t)
   =
   \int_{-\infty}^{\infty}
   \mathrm{d}\omega\,
   \rho_{\rm conn} (\omega)e^{-\omega (\sigma + i t)}\,.
   \label{eq:ConnectedCorrelator_UsingSpectralDensity}
\end{equation}
If the connected spectral measure has no zero-frequency contribution and is sufficiently regular, the Riemann--Lebesgue lemma implies
\begin{equation}
   g_{\rm conn}(\sigma + i t)\xrightarrow{t\to \pm \infty} 0\,,
   \label{eq:Clustering}
\end{equation}
so that the full correlator clusters to $\vev{\Oh}^2$.
This behavior is natural in the infinite-volume theory, where the spectrum is continuous.
At finite spatial volume, however, the spectrum is discrete and the correlator is a sum of oscillatory exponentials; generic recurrences therefore prevent pointwise clustering at arbitrarily late times.
Note that clustering requires additional assumptions on the spectrum and, unlike positivity, holomorphy, and boundedness, does not follow from the existence of the thermal density matrix alone.

\paragraph{Impurity susceptibility.}
A useful integrated observable is the impurity susceptibility associated with a scalar defect operator $\Oh$~\cite{Kubo:1957mj,Vojta:2000tld}.
Introducing a constant source $h_\Oh$ through the deformation $S \to S_h = S-h_\Oh \int_0^\beta \mathrm{d}\tau\,\Oh(\tau)$, we define
\begin{equation}
    \chi_\Oh
    =
    \left. \frac{\partial \vev{\Oh}_h}{\partial h_\Oh} \right|_{h_\Oh = 0}
    =
    \int_0^\beta \mathrm{d}\tau\,
    \left(
    g (\tau)
    -
    \vev{\Oh}^2
    \right)\,.
    \label{eq:DefectSusceptibility}
\end{equation}
Thus, the susceptibility is the integral of the (connected) thermal two-point function along the defect, which measures the linear response of the defect one-point function to a constant source coupled to $\Oh$.

Note that, in the case of the tilt operator $t^a$, the susceptibility can be shown to vanish\footnote{At finite volume, the susceptibility of the tilt is defined by including the contribution from the North-North, South-South, and North-South correlators.
Thus only the full combination vanishes, and an isolated North-North two-point function of tilt operators does not have vanishing susceptibility on its own. \label{fn:Susceptibility_FiniteV}} by using its Ward identity~\cite{Drukker:2022pxk,Girault:2025kzt,Belton:2025ief},
\begin{equation}
    \pd_\mu J^a_\mu (\tau,\bm{x}_\perp)
    =
    C_t\, \delta^{d-1} (\bm{x}_\perp)\, t^a (\tau)\,,
    \label{eq:TiltWardIdentity}
\end{equation}
which can be recast as
\begin{equation}
    \lim\limits_{\bm{x}_\perp \to 0}
    | \bm{x}_\perp |^{d-3} \bm{x}_\perp^i\, J^a_i (\tau,\bm{x}_\perp)
    =
    \frac{C_t\, t^a (\tau)}{\operatorname{Vol} (S^{d-2})}\,.
\end{equation}
We normalize the current $J_\mu^a$ (and its associated generator $\delta^a$) with $C_t$ to account for the unit normalization of the tilt~\eqref{eq:Tilt_ZeroT_TwoPointFunction}.
Assuming that the integrated current flux vanishes at spatial infinity, the integrated Ward identity gives
\begin{equation}
    \int_0^\beta \mathrm{d}\tau\,
    \vev{t^a(\tau) J_i^b(0,\bm{x}_\perp)}
    =
    - \frac{1}{C_t} \vev{\delta^a J_i^b(0,\bm{x}_\perp)}
    =
    0\,.
\end{equation}
Here $\delta^a$ denotes the internal symmetry variation generated by the current $J_\mu^a$.
Consequently, we obtain
\begin{equation}
    \delta^{ab} \chi_t
    =
    \frac{1}{C_t}
    \lim_{\bm{x}_\perp\to 0}
    \operatorname{Vol}(S^{d-2})
    |\bm{x}_\perp|^{d-3} x_\perp^i
    \int_0^\beta \mathrm{d}\tau\,
    \vev{t^a(\tau) J_i^b(0,\bm{x}_\perp)}
    =
    0\,,
    \label{eq:TiltSusceptibilityIsZero}
\end{equation}
where we used the fact that the one-point function of the tilt operator vanishes.
Note that we define the regulated impurity susceptibility through this order of limit and integral.

\subsection{Periodic OPE and asymptotics}
\label{subsec:PeriodicOPEAndAsymptotics}

We now discuss how to construct a representation of $g(\tau)$ that is simultaneously periodic and consistent with the OPE.
The methods developed here largely build on previous work at zero temperature, including both bulk only~\cite{Caron-Huot:2017vep,Carmi:2019cub,Bissi:2019kkx,Caron-Huot:2020adz} and defect setups~\cite{Lemos:2017vnx,Barrat:2021yvp,Barrat:2022psm,Bianchi:2022ppi,Bonomi:2024lky,Bianchi:2026ukq}.
This was adapted to bulk theories at finite temperature in~\cite{Barrat:2025nvu,Barrat:2025twb,Barrat:2026jfg}, while the finite-volume considerations are new.
We study the properties of the resulting periodic blocks and show how they can be used to derive asymptotic expressions for the OPE coefficients of heavy operators.

\subsubsection{Periodic blocks}
\label{subsubsec:PeriodicBlocks}

In this section, we determine periodic blocks using dispersive methods.
We start with the infinite-volume case before adapting the derivation to the antipodal configuration.

\paragraph{Dispersive method.}
We work here at infinite spatial volume and consider $0<\operatorname{Re}(\tau)<\beta$.
Letting $\Cm$ be a small clockwise contour enclosing $\tau$, Cauchy's formula can be written using a kernel adapted to thermal periodicity:
\begin{equation}
   g(\tau)
   =
   \sum_{m=-\infty}^\infty
   \oint_{\Cm}
   \frac{\mathrm{d}\tau'}{2\pi i}
   \frac{g(\tau')}{\tau+m\beta-\tau'}\,.
   \label{eq:Cauchy}
\end{equation}
Only the term with $m=0$ has a pole inside $\Cm$.
The sum over images, understood with a symmetric summation prescription, can be performed explicitly:
\begin{equation}
   \sum_{m=-\infty}^\infty
   \frac{1}{\tau+m\beta-\tau'}
   =
   \frac{\pi}{\beta}
   \cot\left(\frac{\pi(\tau-\tau')}{\beta}\right)\,.
   \label{eq:ResumCauchyKernel}
\end{equation}
We often keep the image sum explicit, as it makes the contributions from the individual thermal images more transparent.

We now deform the Cauchy contour onto the cuts shown in Figure~\ref{fig:AnalyticStructure} and use periodicity to express all cut contributions in terms of the representative cut at $\operatorname{Re}(\tau')=\beta$.\footnote{One could choose any cut located at $\operatorname{Re}(\tau')=m\beta$ for this procedure.
We choose $m=1$ as it corresponds to the $t$-channel of the OPE, which we will use as input in order to obtain periodic blocks.}
This gives the dispersion relation
\begin{equation}
   g(\tau)
   =
   \sum_{m=-\infty}^\infty
   \int_{\beta-i\infty}^{\beta+i\infty}
   \frac{\mathrm{d}\tau'}{2\pi i}
   \frac{\operatorname{Disc}\!\left[g(\tau')\right]}{\tau+m\beta-\tau'}
   +
   g_{\text{arcs}}(\tau)\,,
   \label{eq:DR}
\end{equation}
where the cut is oriented upwards and the discontinuity is defined as the right boundary value minus the left boundary value,
\begin{equation}
   \underset{\operatorname{Re}(\tau')=\beta}{\operatorname{Disc}}
   \left[g(\tau')\right]
   =
   \lim_{\veps \to 0^+}
   \left[
   g \left( \beta + i\,\operatorname{Im}(\tau') + \veps \right)
   -
   g \left( \beta + i\,\operatorname{Im}(\tau') - \veps \right)
   \right]\,.
   \label{eq:Disc_Definition}
\end{equation}
Whenever necessary, the cut integrals are understood with an endpoint regulator together with an appropriate subtraction prescription.
The term $g_{\text{arcs}}(\tau)$ collects the contributions from the large closing contours.
In fact, assuming the properties of $g(\tau)$ listed in Section~\ref{subsubsec:AdditionalConstraints} (including clustering), this term can at most be a constant associated to the disconnected part of the correlator:
\begin{equation}
    g_{\text{arcs}}(\tau)
    =
    \vev{\hphi}^2\,.
    \label{eq:ConstantArcs}
\end{equation}
This vanishes in the case of the tilt operator.

Next, we insert the $t$-channel OPE, expanded around $\tau'=\beta$.
Interchanging the OPE sum with the discontinuity, the integral, and the image sum gives
\begin{equation}
   g(\tau)
   =
   \frac{1}{\beta^{2\Dh_{\smash{\hphi}}}}
   \sum_{\Oh} a_\Oh
   \sum_{m=-\infty}^\infty
   \int_{\beta-i\infty}^{\beta+i\infty}
   \frac{\mathrm{d}\tau'}{2\pi i}
   \frac{\operatorname{Disc}\!\left[f_\Oh(\beta-\tau')\right]}{\tau+m\beta-\tau'}
   +
   \tilde{g}(\tau)\,.
   \label{eq:DRWithOPE}
\end{equation}
Here $\tilde{g}(\tau)$ collects the constant~\eqref{eq:ConstantArcs} but also any remainder associated with exchanging the OPE sum with the integral and sum of images.\footnote{ The commutation can fail if the OPE does not converge sufficiently uniformly along the integration contour.}
We discuss the handling of such cases in Section~\ref{subsubsec:PropertiesOfPeriodicBlocks}.

The discontinuity of individual blocks is given by
\begin{equation}
   \underset{\operatorname{Re}(\tau')=\beta}{\operatorname{Disc}}
   \!\left[f_\Oh(\beta-\tau')\right]
   =
   2i\,\sin(\pi p)\,
   \left[ - \frac{(\beta-\tau')^2}{\beta^2} \right]^p
   \operatorname{sgn}\!\left(\operatorname{Im}(\tau')\right)\,.
   \label{eq:DiscBlocks_InfiniteV}
\end{equation}
Inserting this expression into the dispersion relation and evaluating the regulated integral and image sum yields\footnote{The periodic blocks $F_\Oh(\tau)$ should not be confused with the bulk one-point functions $F_\Om (z)$.}
\begin{equation}
   g(\tau)
   =
   \frac{1}{\beta^{2\Dh_{\smash{\hphi}}}}
   \sum_{\Oh} a_\Oh\,F_\Oh(\tau)
   +
   \tilde{g}(\tau)\,,
   \label{eq:PeriodicOPE_InfiniteV}
\end{equation}
where we have introduced the periodic blocks $F_\Oh(\tau)$, given by
\begin{align}
   F_\Oh(\tau)
   &=
   \sum_{m=-\infty}^\infty f_{\Oh} (\tau + m\beta)
   \notag \\
   &=
   \zeta_H \left( -2p, \frac{\tau}{\beta} \right)
   +
   \zeta_H \left( -2p, 1-\frac{\tau}{\beta} \right)\,,
   \label{eq:PeriodicBlocks_InfiniteV}
\end{align}
with $\zeta_H (s,x)$ the Hurwitz-$\zeta$ function.
The first representation is manifestly periodic and valid wherever the image sum converges, while the second provides its analytic continuation on the fundamental thermal interval $0< \text{Re}(\tau)<\beta$.
It should be understood that going from the first line to the second one in~\eqref{eq:PeriodicBlocks_InfiniteV} corresponds to a prescription that selects one thermal strip.
Note that the periodic blocks~\eqref{eq:PeriodicBlocks_InfiniteV} agree with their bulk counterpart when one sets the spatial separation to zero~\cite{Barrat:2025nvu}.
Additionally, free theories of Gaussian fields are automatically reproduced by including only the periodic contribution of the identity in~\eqref{eq:PeriodicOPE_InfiniteV}, with $\tilde{g} (\tau) = 0$.\footnote{Note that, for composite fields, more periodic blocks may be required.}

\begin{figure}
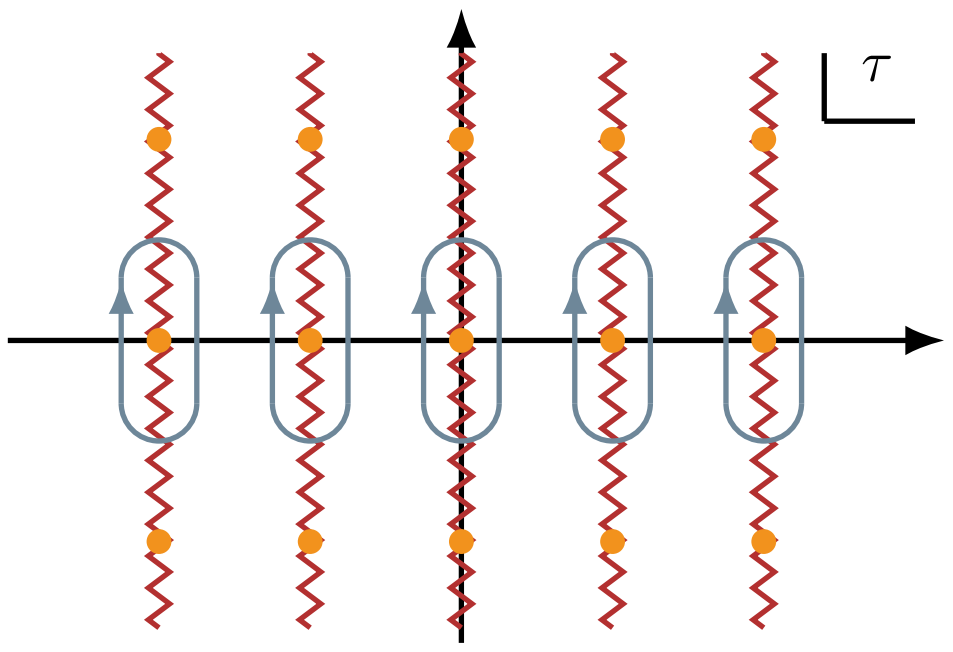

    \centering
        \ContourFiniteVolume
    \caption{Clockwise contours (solid dark blue lines) used for determining the periodic blocks in the antipodal configuration on $S_\beta^1 \times S_R^{d-1}$.
    The contribution of the rest of the cuts is taken into account in the remainder function $\tilde{g}(\tau)$.
    As explained in the main text, the motivation for such a splitting is to preserve the property~\eqref{eq:DoubleTwistPeriodicBlocksAreZero} that periodic blocks associated to classical double-twist operators vanish.}
    \label{fig:FiniteV_Contours}
\end{figure}

\paragraph{Finite volume.}
The same method can be applied at finite volume, up to certain subtleties that we explain now.
The discontinuity of the $t$-channel blocks is given by
\begin{align}
    \underset{\operatorname{Re}(\tau)=\beta}{\operatorname{Disc}}
   \!\left[f_\Oh(\beta-\tau)\right]
   &=
   2i\,\sin(\pi p)\, \operatorname{sgn} \left[ \sin \left( \frac{\operatorname{Im}(\tau)}{R} \right) \right]
   \left[ - \frac{4R^2}{\beta^2} \sinh^2 \left( \frac{\beta - \tau}{2R} \right) \right]^p \notag \\
   &\phantom{=\ } \times
   {}_2F_1 \left( \frac{\Dh_{\Oh}}{2}, \frac{\Dh_{\Oh}}{2}; \Dh_{\Oh} + \frac{1}{2}; - \sinh^2 \left( \frac{\beta - \tau}{2R} \right) \right)\,.
   \label{eq:Disc_FiniteV}
\end{align}
This expression cannot be naively inserted in~\eqref{eq:DRWithOPE} because of the horizontal cuts in the hypergeometric function shown in Figure~\ref{fig:ThermalBlocks_AnalyticStructure}.
The appropriate procedure is to consider in the dispersion solely the contribution of the cut for $-\pi R < \operatorname{Im}(\tau) < \pi R$ and pack the contribution of the rest of the cuts in the remainder $\tilde{g}(\tau)$, as illustrated in Figure~\ref{fig:FiniteV_Contours}.
The main motivation behind this choice is to preserve the property of periodic blocks vanishing for classical double-twist, as we explain in more detail in Section~\ref{subsubsec:PropertiesOfPeriodicBlocks}.
We obtain periodic blocks, which can be expressed in the form of the integral
\begin{align}
    F_\Oh (\tau)
    &=
    - 2 \sin(\pi p) \left( \frac{R}{\beta} \right)^{2p+1}
    \int_0^\pi \mathrm{d}v
    \left(
    \frac{\sinh \left( 2\pi v R/\beta \right)}{\cosh \left( 2\pi v R/\beta \right) - \cos \left( 2\pi \tau/\beta \right)} -
    1
    \right) \notag \\
    &\phantom{=\ } \times\,
    \left[ 2 \sin(v/2) \right]^{2p}\,
    _{2}F_1 \left( \frac{\Dh_{\Oh}}{2}, \frac{\Dh_{\Oh}}{2}; \Dh_{\Oh} + \frac{1}{2} ; \sin^2 \left( \frac{v}{2} \right) \right)\,.
    \label{eq:PeriodicBlock_FiniteV}
\end{align}
The integral can be performed analytically case by case for perturbative setups and numerically for non-integer dimensions.
Outside the domain of convergence, this expression is defined by analytic continuation in $p$.
Note that the limit $q \to 1$ correctly reproduces the infinite-volume periodic blocks~\eqref{eq:PeriodicBlocks_InfiniteV}.
The full correlator is again given by
\begin{equation}
   g(\tau)
   =
   \frac{1}{\beta^{2\Dh_{\smash{\hphi}}}}
   \sum_{\Oh} a_{\Oh}(q)\, F_{\Oh}(\tau)
   +
   \tilde{g}(\tau)\,.
   \label{eq:PeriodicOPE_FiniteV}
\end{equation}
The main difference with the infinite-volume case is that the periodic blocks in this setup are not given simply by the sum over images of the thermal blocks.\footnote{Nothing prevents such a construction to be valid, but it would move terms with vanishing OPE-channel discontinuity from $\tilde{g}(\tau)$ into the blocks.
Said differently, the periodic blocks associated to classical double-twist operators would not vanish and lead to $\tilde{g}(\tau)$ being unnecessarily more intricate.
See Section~\ref{subsubsec:PropertiesOfPeriodicBlocks} for more details regarding the latter point.}

It is instructive to consider the example of the antipodal configuration in free theories.
The exact correlator of two fields placed on the same defect is given by
\begin{equation}
   g(\tau)
   =
   \sum_{m=-\infty}^\infty \left[ 2 R \sinh \left( \frac{|\tau + m\beta|}{2R} \right) \right]^{\smash{-2\Dh_\hphi}}\,.
   \label{eq:FiniteV_GFF}
\end{equation}
Anticipating the discussion in Section~\ref{subsubsec:PropertiesOfPeriodicBlocks} in which we show that periodic blocks vanish for classical double-twist operators, the periodic OPE gives
\begin{equation}
    g(\tau)
    =
    \frac{F_{\hat{\mathds{1}}} (\tau)}{\beta^{2\Dh_{\hphi}}} + \tilde{g} (\tau)\,.
    \label{eq:FiniteV_GFF_PeriodicOPE}
\end{equation}
To reconstruct~\eqref{eq:FiniteV_GFF} from~\eqref{eq:FiniteV_GFF_PeriodicOPE}, it suffices to notice that the latter does not possess the correct analytic structure.
The role of $\tilde{g}(\tau)$ is thus purely to repair this mismatch.
In Appendix~\ref{app:FreeTheoryForTheAntipodalConfiguration}, we provide the explicit example of the free scalar field in four dimensions.
In fact, we argue in the following section that there exists a single function, up to a constant, that completes~\eqref{eq:FiniteV_GFF_PeriodicOPE} such that the correct analytic structure is restored and the properties listed in Section~\ref{subsubsec:AdditionalConstraints} are fulfilled.

\subsubsection{Properties of periodic blocks}
\label{subsubsec:PropertiesOfPeriodicBlocks}

We now discuss important properties of the expansion in periodic blocks.
We first show that the formulation~\eqref{eq:PeriodicOPE_InfiniteV}-\eqref{eq:PeriodicOPE_FiniteV} establishes a well-posed bootstrap problem in spite of the presence of $\tilde{g}(\tau)$.
We then highlight an important property of periodic blocks, which is that their contribution for classical double-twist operators vanishes both for the infinite and finite volume cases.

\paragraph{Uniqueness of $\tilde{g}(\tau)$.}
We first address the role of the function $\tilde{g}(\tau)$ appearing in~\eqref{eq:PeriodicOPE_InfiniteV} and~\eqref{eq:PeriodicOPE_FiniteV}.
Suppose first that the expansion in periodic blocks preserves all defining properties of the exact correlator: it is periodic, has exactly the same poles and discontinuities, and satisfies the required boundedness condition as $i\tau \to \pm\infty$ within a fundamental thermal strip.
Then $\tilde{g}(\tau)$ is entire, periodic, and bounded.
Under the map $w=e^{2\pi i\tau/\beta}$, it defines a holomorphic function on $\mathbb{C}^\ast$ that is bounded both as $w\to0$ and $w\to\infty$.
These two points are therefore removable singularities, so the function extends holomorphically to the Riemann sphere and must be constant:
\begin{equation}
   \tilde{g}(\tau)
   =
   \frac{c_{\smash{\hphi}}}{\beta^{\smash{2\Dh_\hphi}}}\,.
   \label{eq:ArcsAreConstant}
\end{equation}
In this case, $\tilde{g}(\tau)$ is therefore fixed up to an additive constant.

More generally, the raw periodic-block expansion may fail to reproduce one of the defining properties listed above, for example by generating spurious singularities, as is familiar from bouncing singularities in holographic correlators or generically for correlators on $S_\beta^1 \times S_R^{d-1}$.
The role of $\tilde{g} (\tau)$ is then to restore the correct analytic structure and asymptotic behavior.
Once the discontinuities and pole principal parts of the completed correlator are fixed, the difference between any two admissible completions is entire, periodic, and bounded, and must therefore be constant by the same argument.
Under these assumptions, $\tilde{g} (\tau)$ is unique up to an additive constant.

The remaining constant may be fixed by imposing an additional physical condition, such as the impurity susceptibility~\eqref{eq:DefectSusceptibility} (if known).
This is convenient, because one can show that the periodic blocks have vanishing regularized contributions in the integral:
\begin{equation}
    \int_0^\beta \mathrm{d}\tau\, F_\Oh (\tau) = 0\,.
    \label{eq:IntegratedPeriodicBlocksAreZero}
\end{equation}
Integrating over the thermal circle selects the zero Matsubara mode of the periodic block, which the prescriptions~\eqref{eq:PeriodicBlocks_InfiniteV},~\eqref{eq:PeriodicBlock_FiniteV} remove.
The integral~\eqref{eq:IntegratedPeriodicBlocksAreZero} should be understood as the analytic continuation of $\operatorname{Re} (p) > - 1/2$ to all values.
Relating the impurity susceptibility to the periodic OPE leads to
\begin{equation}
    \chi_{\smash{\hphi}} = \beta\, \left( \frac{c_{\smash{\hphi}}}{\beta^{\smash{2 \Dh_\hphi}}} - \vev{\hphi}^2 \right)\,.
\end{equation}
More generally, if the remainder function is not a constant, we have
\begin{equation}
    \chi_{\smash{\hphi}} = \int_0^\beta \mathrm{d}\tau\, \tilde{g}(\tau) - \beta \vev{\hphi}^2\,.
\end{equation}
For the tilt operator, the Ward identity fixes the susceptibility (see~\eqref{eq:TiltSusceptibilityIsZero}) while the disconnected part vanishes.
The constant is therefore zero:
\begin{equation}
    c_{t} = 0\,,
    \label{eq:TiltcIsZero}
\end{equation}
provided the expansion in periodic blocks respects the analytic structure of the full correlator.
At finite volume, as mentioned in Footnote~\ref{fn:Susceptibility_FiniteV}, this is only true if $g(\tau)$ encodes the contribution of North-North, South-South, and North-South two-point functions.

\paragraph{Double-twist operators.}
Consider classical double-twist operators $[\hphi\hphi]_n$, with dimensions
\begin{equation}
    \Dh_{\smash{[\hphi\hphi]_n}}
    =
    2\Dh_{\smash{\hphi}} + 2n\,,
    \qquad n\in\mathbb{Z}_{\geq 0}\,.
    \label{eq:DoubleTwist_ScalingDims}
\end{equation}
With our dispersive prescription, the corresponding periodic blocks vanish in both thermal geometries,
\begin{equation}
    F_{\smash{[\hphi\hphi]_n}} (\tau)
    =
    0\,.
    \label{eq:DoubleTwistPeriodicBlocksAreZero}
\end{equation}
This property is reminiscent of the Polyakov bootstrap~\cite{Polyakov:1974gs,Mazac:2018ycv,Guo:2026xyl}: classical double-twist operators do not contribute as independent terms to the periodic OPE.
When present, their contributions must instead be reconstructed from the analytic terms generated by the remaining periodic blocks together with the remainder, as discussed in the next subsection.
Although generic interacting theories need not contain operators at these exact dimensions, classical double-twist operators play an important role in perturbative regimes, including the $\veps$- and large-$N$ expansions.

\paragraph{Free theory.}
It is instructive to consider a free massless bulk scalar restricted to a trivial line defect, so that $\Dh_{\hphi}=(d-2)/2$ and with vanishing one-point function.
Its ordinary thermal OPE contains the identity and classical double-twist operators:
\begin{equation}
    g(\tau)
    =
    \frac{1}{\beta^{2\Dh_{\hphi}}}
    \left[
        f_{\hat{\mathds{1}}}(\tau)
        +
        \sum_{n=0}^{\infty}
        a_{[\hphi\hphi]_n}\,
        f_{[\hphi\hphi]_n}(\tau)
    \right].
\end{equation}
Promoting the ordinary blocks to periodic blocks eliminates the explicit double-twist contributions, since we showed above that $F_{[\hphi\hphi]_n}(\tau)=0$.
The identity periodic block alone reproduces the thermal image sum, so the remainder vanishes:
\begin{equation}
    g(\tau)
    =
    \frac{F_{\hat{\mathds{1}}}(\tau)}{\beta^{2\Dh_{\hphi}}}
    =
    \frac{1}{\beta^{2\Dh_{\hphi}}}
    \left[
        \zeta_H\left(2\Dh_{\hphi},\frac{\tau}{\beta}\right)
        +
        \zeta_H\left(2\Dh_{\hphi},1-\frac{\tau}{\beta}\right)
    \right].
    \label{eq:FreeTheory_PeriodicOPE_InfiniteV}
\end{equation}
The double-twist data have therefore not disappeared: they are encoded in the regular expansion of the identity periodic block around $\tau=0$.
Their OPE coefficients are given by:
\begin{equation}
    \tilde{a}^{(\hat{\mathds{1}})}_{{[\hphi\hphi]_n}}
    =
    2\zeta(2\Dh_{\smash{\hphi}} + 2n)
    \frac{\Gamma(1-2\Dh_{\smash{\hphi}})}
    {(2n)!\Gamma(1-2\Dh_{\smash{\hphi}}-2n)}\,.
    \label{eq:GFFFromAsymptotics_InfiniteV}
\end{equation}
At finite spatial volume, a non-trivial completion is instead required, as illustrated in Appendix~\ref{app:FreeTheoryForTheAntipodalConfiguration}.

\paragraph{Special case.}
At infinite volume, the periodic block has a simple pole when $\Dh_{\Oh}=2\Dh_{\hphi}-1$, corresponding to $p=-1/2$.
Indeed, the Laurent expansion of the Hurwitz-$\zeta$ function gives
\begin{equation}
    F_{\Oh}(\tau)
    =
    - \frac{2}{2p+1}
    +
    \text{finite terms}\,.
\end{equation}
The pole is independent of $\tau$ and can therefore be absorbed into $\tilde{g} (\tau)$, defining the block at the resonant dimension by its finite
part.
In perturbation theory, however, an approach to this dimension with $2p+1=\mathrm{O}(g)$ enhances the constant contribution of
$a_{\Oh} F_{\Oh} (\tau)$ by one inverse power of $g$.
This mechanism is reminiscent of weak-coupling "anomalies" in block expansions known at zero temperature~\cite{Cavaglia:2022qpg,Sakkas:2024dvm,Gabai:2025zcs,Belton:2025hbu,Drukker:2025dfm}.
We encounter an example of this mechanism in Section~\ref{subsec:AnalyticBootstrap}.

\subsubsection{Asymptotic OPE coefficients}
\label{subsubsec:AsymptoticOPECoefficients}

A manifestly periodic OPE expansion allows us to extract asymptotic information about the OPE coefficients of heavy operators~\cite{Barrat:2025nvu}.
The key observation is that, in both thermal geometries, each periodic block can be re-expanded in the ordinary OPE channel as
\begin{equation}
    a_\Oh\, F_\Oh (\tau)
    =
    a_\Oh\, f_\Oh (\tau)
    +
    \sum_{n=0}^\infty \tilde{a}^{(\Oh)}_{{[\hphi\hphi]_n}} f_{\smash{[\hphi\hphi]_n}} (\tau)\,,
    \label{eq:AnalyticContinuation}
\end{equation}
with $0 \leq q \leq 1$.
In other words, each periodic block associated with a defect operator $\Oh$ can be written as the original block together with an infinite tower of classical double-twist blocks with dimensions~\eqref{eq:DoubleTwist_ScalingDims}.
This follows because the periodic block reproduces the full local singular part of the original block, so their difference is even and analytic near $\tau=0$ and can therefore be expanded in classical double-twist blocks.
The coefficients $\tilde{a}^{(\Oh)}_{\smash{[\hphi\hphi]_n}}$ are determined by the OPE coefficient $a_{\Oh}$ of the light operator and provide an asymptotic estimate for the contribution of heavy operators.
The tilde emphasizes that these double-twist operators should be understood as asymptotic contributions and need not coincide with physical operators at finite dimension.
In the following, we use~\eqref{eq:AnalyticContinuation} to determine the asymptotic OPE coefficients associated with the classical double-twist families $[\hphi\hphi]_n$.

\paragraph{Infinite volume.}
At infinite volume, the asymptotic OPE coefficients appearing in~\eqref{eq:AnalyticContinuation} are given by
\begin{equation}
    \tilde{a}^{(\Oh)}_{{[\hphi\hphi]_n}}
    =
    a_\Oh\, 2\zeta(2n-2p) \frac{\Gamma(1+2p)}{(2n)!\Gamma(1+2p-2n)}\,.
    \label{eq:Asymptotics_InfiniteV}
\end{equation}
For $\Oh = \hat{\mathds{1}}$, one has $p = - \Dh_{\smash{\hphi}}$, and we recover~\eqref{eq:GFFFromAsymptotics_InfiniteV}

\paragraph{Finite volume.}
One can proceed similarly at finite volume.
The only subtlety is related to the presence of an automatically non-trivial completion $\tilde{g}(\tau)$ since the periodic blocks $F_\Oh(\tau)$ do not respect the analytic structure of the full correlator.
Provided that the completion is regular at the nearest KMS singularities and introduces no equally near singularity in the relevant expansion variable, its contribution is expected to be exponentially suppressed relative to that of the periodic OPE.
Ignoring these corrections, the asymptotic OPE coefficients in~\eqref{eq:AnalyticContinuation} can be written as
\begin{align}
    \tilde{a}^{(\Oh)}_{{[\hphi\hphi]_n}}
    &=
    a_{\Oh}(q)\,
    \frac{4q^{1/4}}{1+\sqrt q}
    \left[
        -\frac{(1+\sqrt q)\log (q)}{4(1-\sqrt q)}
    \right]^{2\Dh_{\smash{\hphi}}+2n}
    \left[
        -\frac{1-q}{\sqrt q\log (q)}
    \right]^{\Dh_{\Oh}}
    \frac{(2\Dh_{\smash{\hphi}}+2n)^{-2p-1}}{\Gamma(-2p)}
    \notag \\
    &\phantom{=\ } \times
    \left[
        1+
        \frac{
            8(2p+1)(\Dh_{\smash{\Oh}}-p)(1+q)
            -(8p+3)(1-\sqrt q)^2
        }{16\sqrt q\,(2\Dh_{\smash{\hphi}}+2n)}
        + \mathrm{O} \bigl((2\Dh_{\smash{\hphi}}+2n)^{-2}\bigr)
    \right]\,.
\end{align}
Arbitrarily many power corrections can be extracted from~\eqref{eq:PeriodicBlock_FiniteV}.
The infinite-volume asymptotic expression is recovered (term by term) in the limit $q \to 1$.

%% file: sections/EpsExpansionAndInterpolationTo3d.tex
\section{$\veps$-expansion and interpolation to $d=3$}
\label{sec:epsExpansionAndInterpolationTo3d}

In this section, we specialize our study of line defect correlators to the pinning-field line defect in the context of the $\veps$-expansion.
We first present an analytic bootstrap method that captures the one-loop correlator up to a constant.
The results are validated at infinite and finite volume through direct diagrammatic calculation.
We then study the thermal one-point of the operator $\hat{\phi}^1$ for the $3d$ Ising case (i.e., $N=1$) and interpolate its value to $d=3$ using the two-dimensional (free) theory.
We conjecture that the correlator $\vev{\hphi^1(\tau) \hphi^1(0)}$ can be well-approximated by truncating the periodic OPE at its first non-trivial operator, using the interpolation of $b_{\smash{\hphi^1}}$ as sole input.

\bigskip

We use the following conventions.
For $N>1$, we study the two-point function of tilt operators,
\begin{equation}
    \vev{t^a (\tau) t^b (0)}
    =
    \delta^{ab} g_{tt} (\tau)\,.
    \label{eq:gtt_Definition}
\end{equation}
As mentioned before, the tilt operators do not exist at $N=1$.
We thus also consider the correlator of two longitudinal singlet operators,
\begin{equation}
    \vev{\hphi^1 (\tau) \hphi^1 (0)}
    =
    g_{\smash{\hphi^1 \hphi^1}} (\tau)\,,
    \label{eq:ghphi1_Definition}
\end{equation}
which is the two-point function of the lowest-lying scalars in the absence of the tilt.
The remaining conventions follow Section~\ref{subsubsec:ThePinningFieldLineDefect}.

\subsection{Analytic bootstrap}
\label{subsec:AnalyticBootstrap}

The goal of this section is to determine the one-loop correlators $g_{tt} (\tau)$ and $g_{\smash{\hphi^1 \hphi^1}} (\tau)$ using the periodic OPE~\eqref{eq:PeriodicBlocks_InfiniteV} and~\eqref{eq:PeriodicBlock_FiniteV}.
We focus on the two-point function of tilt operators and give the result for $\vev{\hphi^1 \hphi^1}$, mentioning differences only when they are conceptual.
For most of this section, we focus on the infinite-volume case.
The starting point is to promote the OPE to its periodic version, which immediately fixes the correlator up to a finite number of constants.
After fixing these constants, we extract the OPE data and compare the results to direct Feynman diagram calculations.
Finally, we extend the calculation to the antipodal configuration at finite volume.

\subsubsection{Expansion in periodic blocks}
\label{subsubsec:ExpansionInPeriodicBlocks}

We use the tools of Section~\ref{subsec:PeriodicOPEAndAsymptotics}.
The first step is to expand the correlator in periodic blocks.

\paragraph{Expansion in thermal blocks.}
We first write the regular thermal OPE relevant for $g_{tt} (\tau)$.
The low-lying exchanged spectrum consists of
\begin{equation}
    \left. t \times t\ \right|_{s=0\,, \mathrm{O}(N-1) \text{ singlets}}
    \longrightarrow
    \hat{\mathds{1}}
    +
    \hphi^1
    +
    s_-
    +
    s_+
    +
    \ldots\,.
    \label{eq:tt_Spectrum}
\end{equation}
We have restricted the right-hand side to operators which are allowed to appear in thermal two-point functions, i.e., operators with transverse spin zero and uncharged under $\mathrm{O}(N-1)$ internal symmetry.
The operators are defined in Section~\ref{subsubsec:ThePinningFieldLineDefect}.

At zero coupling, only operators with even integer dimension appear, suggesting to organize the spectrum into multi-twist trajectories.
By naively counting $\veps$ powers, one sees that only double-twist operators have non-vanishing contributions at $\veps=0$.
In fact, further counting shows that the combinations $a_{\Oh}^{(0)} \gamma_{\Oh}^{(1)}$ appearing at $\mathrm{O} (\veps)$ can only consist of the operators $s_\pm$.
For other operators, either three-point functions or anomalous dimensions push this combination to appear only at higher orders of $\veps$.
For instance, the contribution of $\hphi^1$ starts at order $\mathrm{O}(\veps^{3/2})$.
This analysis is similar in spirit to what was done for bulk and defect correlators at zero temperature~\cite{Bissi:2019kkx,Gimenez-Grau:2022ebb},\footnote{See also~\cite{Barrat:2021yvp} for a similar counting in the strong-coupling limit of holographic theories.} as well as thermal correlators in bulk theories~\cite{Barrat:2025nvu}.
An important difference is that the constant part of the correlator may receive an order $\mathrm{O}(\sqrt{\veps})$ contribution due to IR resummation, as is well-known in the bulk theory~\cite{Dolan:1973qd,Sachdev:1996ed}.

All in all, the thermal OPE takes the form
\begin{align}
    \beta^2\, g_{tt} (\tau)
    &=
    f_{\hat{\mathds{1}}} (\tau)
    +
    \sum_{n=0}^\infty
    \left(
    \vev{a_{{\Dh_\mathrm{cl} = 2+2n}}^{(0)}}
    +
    \veps\,
    \vev{a_{{\Dh_\mathrm{cl} = 2+2n}}^{(1)}}
    \right)
    f_{\smash{\Dh_\mathrm{cl} = 2+2n}} (\tau) \notag \\
    &\phantom{=\ }
    +
    \veps\, \vev{a_{{\Dh_\mathrm{cl} = 2}}^{(0)} \gamma_{{\Dh_\mathrm{cl} = 2}}^{(1)}}\,
    \left. \pd_{\smash{\Dh}} f_{\smash{\Dh}} (\tau) \right|_{\Dh = 2}
    +
    \sqrt{\veps}
    \vev{a_{\smash{s_\pm}}^{(1/2)}}
    + \mathrm{O} (\veps^{3/2})\,.
    \label{eq:gtt_OPE}
\end{align}
Here the functions $f_{\Oh} (\tau)$ are the thermal blocks associated to an operator $\Oh$ and defined in~\eqref{eq:ThermalBlocks_InfiniteV}.
We define the degenerate OPE coefficients as
\begin{equation}
    \vev{ a_\Oh }
    =
    \sum_{\Oh',\ \Dh_{\Oh'} = \Dh_\Oh} a_{\Oh'}\,,
    \label{eq:veva_Definition}
\end{equation}
with
\begin{equation}
    a_\Oh
    =
    \sum_{k=0}^\infty \veps^{k/2}\, a_\Oh^{(k/2)}\,.
    \label{eq:a_PerturbativeExpansion}
\end{equation}
A similar expression should be used when anomalous dimensions are involved.
For instance,
\begin{equation}
    \vev{a_{{\Dh_\mathrm{cl} = 2}}^{(0)} \gamma_{{\Dh_\mathrm{cl} = 2}}^{(1)}}
    =
    f_{\smash{tts_-}}^{(0)} b_{\smash{s_-}}^{(0)} \gamma_{\smash{s_-}}^{(1)}
    +
    f_{\smash{tts_+}}^{(0)} b_{\smash{s_+}}^{(0)} (2 + \gamma_{\smash{s_+}}^{(1)})\,,
    \label{eq:vevagamma_Expansion}
\end{equation}
where on the right-hand side we have inserted the explicit definition~\eqref{eq:a_Definition}.
The additional $2$ in the second term comes from defining $s_+$ as the double-twist of $\hphi^1$ (see~\eqref{eq:sp_AnomalousDim}).

Let us discuss the expansion~\eqref{eq:gtt_OPE}.
The first term simply corresponds to the identity exchange, setting the dimension of the external operator to $\Dh_t = 1$.
The sum over $n$ in the second term encodes the contribution of the double-twist operators $[tt]_n$ and $[\hat{\phi}^1 \hat{\phi}^1]_n$ up to order $\mathrm{O}(\veps)$.\footnote{These families recombine and become degenerate at zero coupling.
There may be other multi-twist operators mixing with them at $\mathrm{O}(\veps)$.
Here we do not disentangle their individual contributions, and thus identifying them with double-twist operators does not affect the outcome of the calculation.}
The contributions $\vev{a_{\smash{\Dh_\mathrm{cl} = 2+2n}}^{(0)}}$ are tree level and easy to obtain, while $\vev{a_{\smash{\Dh_\mathrm{cl} = 2+2n}}^{(1)}}$ are associated to one-loop corrections and may not be directly accessible.
The latter contributions are the main goal of this analytic-bootstrap calculation, and using periodic blocks is a natural tool since we showed in Section~\ref{subsubsec:PropertiesOfPeriodicBlocks} that their contribution to classical double-twist operators vanish and are reconstructed through the periodic blocks of other operators.
Finally, the first term in the second line describes the single anomalous contribution relevant at one loop, while the constant term $\vev{a_{\smash{s_\pm}}^{(1/2)}}$ needs IR resummation as mentioned above.

\paragraph{Promoting to periodic blocks.}
We now promote the thermal blocks to periodic blocks.
It is convenient to split the correlator into three contributions:
\begin{equation}
    g_{tt} (\tau)
    =
    \frac{1}{\beta^2} F_{\hat{\mathds{1}}} (\tau)
    +
    g^\text{(int)}_{tt} (\tau)
    +
    \tilde{g}_{tt} (\tau)\,.
    \label{eq:g_Splitting}
\end{equation}
Here $F_{\hat{\mathds{1}}} (\tau)$ is simply the periodized contribution of the identity to the full correlator.
The second term, $g_{tt}^\text{(int)} (\tau)$, contains the non-trivial information of the periodic blocks.
The third term, $\tilde{g}_{tt} (\tau)$, is described by the action of commuting the OPE sum and the Cauchy integral, as explained in Section~\ref{subsubsec:PeriodicBlocks}.
We assume in the following that it can contribute at most a constant\footnote{One may be worried that the clustering property is lost since we show later that the periodic OPE truncates, and the latter diverges at $\tau \to i\infty$.
This should however be understood as a perturbative effect.
Put differently, the clustering property and perturbative expansion do not commute in general.
Our assumption that $\tilde{g}_{tt} (\tau) = 0$ is therefore a non-perturbative assumption performed \textit{before} expanding the correlator in $\veps$.} and, since the susceptibility of the tilt operator vanishes (see~\eqref{eq:TiltcIsZero}), we set
\begin{equation}
    \tilde{g}_{tt} (\tau)
    =
    0\,.
\end{equation}

We now consider each of the other terms separately.
The periodized contribution of the identity is simply given by
\begin{equation}
    F_{\hat{\mathds{1}}} (\tau)
    =
    \sum_{m=-\infty}^\infty
    \left( \frac{\tau + m\beta}{\beta} \right)^{-2}
    =
    \pi^2 \csc^2 \left( \frac{\pi\tau}{\beta} \right)\,.
    \label{eq:EpsExp_gGFF}
\end{equation}
This expression is valid across spacetime dimensions, as the scaling dimension of the tilt operator is protected.

To determine the interacting part $g_{tt}^\text{(int)} (\tau)$, we note that the periodic blocks associated with classical double-twist operators vanish whenever $n>0$.
The coefficients $\vev{a_{\smash{\Dh_\mathrm{cl} = 2n}}^{(1)}}$ multiply in~\eqref{eq:gtt_OPE} blocks precisely of this type.
Therefore, their contribution to the periodized expansion drops and we find
\begin{align}
    g^\text{(int)}_{tt} (\tau)
    &=
    \frac{1}{\beta^2}
    \biggl(
    - 2 a_{{\hphi^1}}^{(3/2)} \sqrt{\veps}
    - 2 a_{{\hphi^1}}^{(2)} \veps
    +
    \veps\,
    \vev{a_{{\Dh_\mathrm{cl} = 2}}^{(0)} \gamma_{{\Dh_\mathrm{cl} = 2}}^{(1)}}
    \left. \pd_{\smash{\Dh}} F_{\smash{\Dh}} (\tau) \right|_{\Dh = 2}
    +
    \mathrm{O} (\veps^{3/2})
    \biggr)\,.
    \label{eq:gInt_InPeriodicBlocks}
\end{align}
The constant terms in this expression stem from the fact that the OPE coefficient $a_{\smash{\hphi^1}}$, defined perturbatively as
\begin{equation}
    a_{\smash{\hphi^1}}
    =
    a_{{\hphi^1}}^{(3/2)} \veps^{3/2}
    +
    a_{{\hphi^1}}^{(2)} \veps^{2}
    +
    \ldots\,,
\end{equation}
multiplies the periodic block $F_{\smash{\hphi^1}}$, which has a divergent constant term when expanded in $\veps$:
\begin{equation}
    F_{\smash{\hphi^1}} (\tau)
    =
    - \frac{2}{\veps} + \text{finite}\,.
\end{equation}
This follows the discussion in Section~\ref{subsubsec:PropertiesOfPeriodicBlocks}.
Interestingly, it implies that the low-order constant $\vev{a_{\smash{s_\pm}}^{(1/2)}}$ is fixed by $a_{{\hphi^1}}^{(3/2)}$.

Equation~\eqref{eq:gInt_InPeriodicBlocks} is a drastic simplification compared to~\eqref{eq:gtt_OPE}.
In particular, we note that the $\tau$-dependent part of the correlator is completely fixed up to the overall multiplicative constant~\eqref{eq:vevagamma_Expansion}, which consists of zero-temperature anomalous dimensions and zero-coupling finite-temperature OPE coefficients.
We emphasize that the double-twist operators of dimension $\Dh=2+2n$ have not disappeared.
Indeed, their OPE coefficients are encoded in the expression~\eqref{eq:gInt_InPeriodicBlocks} and, after expanding it in the ordinary thermal blocks~\eqref{eq:ThermalBlocks_InfiniteV}, they re-emerge as output of this analytic bootstrap method.

The contribution~\eqref{eq:gInt_InPeriodicBlocks} can be written perhaps more explicitly as
\begin{align}
    \left. \pd_{\smash{\Dh}} F_{\smash{\Dh}} (\tau) \right|_{\Dh = 2}
    &=
    -
    \left[
    \zeta_{H}^{(1,0)} \left( 0, \frac{\tau}{\beta} \right) + \zeta_{H}^{(1,0)} \left( 0, 1- \frac{\tau}{\beta} \right)
    \right] \notag \\
    &=
    - \frac{1}{2} \log \left[ \frac{1}{4} \csc^2 \left( \frac{\pi \tau}{\beta} \right)\right]
    \,,
    \label{eq:gtt_BootstrapResult}
\end{align}
where the superscripts in the first line refer to derivatives with respect to the arguments of the Hurwitz-$\zeta$ function.

\paragraph{Fixing the constants.}
Equation~\eqref{eq:gInt_InPeriodicBlocks} fixes the full correlator up to the three constants $a_{{\hphi^1}}^{(3/2)}$, $a_{{\hphi^1}}^{(2)}$, and $\vev{a_{{\Dh_\mathrm{cl} = 2}}^{(0)} \gamma_{{\Dh_\mathrm{cl} = 2}}^{(1)}}$.
We now determine these constants through Feynman diagrams to provide an expression for the full correlator, using the Feynman rules provided in Appendix~\ref{subsec:FeynmanRules}.

In order to compute $\vev{a_{{\Dh_\mathrm{cl} = 2}}^{(0)} \gamma_{{\Dh_\mathrm{cl} = 2}}^{(1)}}$, we need to determine the zero-temperature three-point functions of the operators $s_\pm$ as well as their thermal one-point coefficients.
The zero-temperature three-point functions are easy to compute at tree level and they read
\begin{equation}
    f_{\smash{t t s_\pm}}^{(0)}
    =
    \frac{1}{n_{\smash{t}}^{(0)} \sqrt{n_{\smash{s_\pm}}}}
    \frac{A_\pm}{8 \pi^4}
    =
    \frac{\sqrt{2} A_\pm}{\sqrt{N A_\pm^2 - 2 A_\pm + 1}}\,.
    \label{eq:fttspm}
\end{equation}
The thermal one-point functions are also tree-level correlators and can be expressed as
\begin{equation}
    b_{\smash{s_\pm}}^{(0)}
    =
    \zeta_2 \sqrt{2} \frac{N\, A_\pm - 1}{\sqrt{N\, A_\pm^2 - 2 A_\pm + 1}}\,.
    \label{eq:bspm}
\end{equation}
Putting everything together, including the anomalous dimensions given in~\eqref{eq:sm_AnomalousDim}-\eqref{eq:sp_AnomalousDim}, we obtain the expression
\begin{equation}
    \vev{a_{{\Dh_\mathrm{cl} = 2}}^{(0)} \gamma_{{\Dh_\mathrm{cl} = 2}}^{(1)}}
    =
    2 \zeta_2 \frac{N+2}{N+8}\,.
    \label{eq:Input_OPECoefficient}
\end{equation}
The expression~\eqref{eq:Input_OPECoefficient} is remarkably simple, especially given the form of the anomalous dimensions~\eqref{eq:sm_AnomalousDim}-\eqref{eq:sp_AnomalousDim} and of the OPE coefficients separately.
Incidentally, the factor $\frac{N+2}{N+8}$ also corresponds to the anomalous dimension of the operator $\phi^2$ in the \textit{bulk} theory.
We will see when calculating Feynman diagrams that this coincidence is forced upon us by the fact that a single tadpole diagram contributes to the interacting expression.
This diagram does not involve defect interactions and thus it takes the same value as the bulk computation.
This also indicates that this effect will be lifted at a higher order in perturbation theory.

We now fix the constants $a_{{\hphi^1}}^{(3/2)}$ and $a_{{\hphi^1}}^{(2)}$.
The associated zero-temperature three-point function has been determined in~\cite{Gimenez-Grau:2022czc} to be
\begin{equation}
    f_{\smash{tt\hphi^1}}
    =
    \frac{\pi}{\sqrt{N+8}} \veps + \mathrm{O}(\veps^2)\,.
    \label{eq:ttPhi}
\end{equation}
The thermal one-point function $b_{\smash{\hphi^1}}$ can however not be determined through naive Feynman diagrams.
To understand this and eventually obtain $b_{\smash{\hphi^1}}^{(1/2)}$, consider the diagrammatic expansion
\begin{equation}
    \vev{\hphi^1 (0)}
    =
    \frac{1}{\sqrt{n_{\smash{\hphi^1}}}} Z_{\smash{\hphi^1}}^{-1}
    \left( \DiagPhiOne\ + \ldots \right)
    \sim
    - h \sum_{m} \int_0^\beta \mathrm{d}\tau\, G(\tau + m\beta, 0)\,.
    \label{eq:bPhi_Diagram}
\end{equation}
Although the leading diagram is scaleless in dimensional regularization, its perturbative expansion with thermal self-energy insertions is infrared divergent and requires resummation.
In fact, all diagrams contributing to~\eqref{eq:bPhi_Diagram} would require such a resummation.

The key observation to solve this issue is to note that the scalar propagator acquires a thermal mass $m_\text{th}^2 \sim \lambda$ in the bulk that screens IR divergences.
This effect is well-studied in bulk theories~\cite{Dolan:1973qd,Sachdev:1996ed}.
A standard way of regulating these divergences and extracting the correct behavior is to replace the massless propagator with a massive one and expand it around $m_\text{th} \sim 0$.

The exact resummed bulk propagator in momentum space has the form
\begin{equation}
    \tilde{G}_\text{massive} (\omega_n, k)
    =
    \frac{1}{\omega_n^2 + k^2 + \Pi(\omega_n, k)}\,.
    \label{eq:bspm_Gmassive}
\end{equation}
Non-analytic behavior arises in the soft region
\begin{equation}
    k \sim m_\text{th}\,,
    \label{eq:bspm_IRBehavior}
\end{equation}
and to extract the leading IR behavior it suffices to consider the leading term in
\begin{equation}
    \Pi (\omega,k)
    =
    m_\text{th}^2 + \ldots\,.
    \label{eq:bspm_Step4}
\end{equation}
The coincident renormalized bulk propagator can then be expanded as
\begin{equation}
    G_\text{massive} (0,0)
    =
    \frac{1}{12 \beta^2}
    -
    \frac{m_\text{th}}{4 \pi \beta}
    +
    \ldots\,.
    \label{eq:bspm_Step5}
\end{equation}
Using the known value of the thermal mass,
\begin{align}
    m_\text{th}^2
    &=
    \frac{N+2}{6} \lambda \left( \frac{1}{12 \beta^2} - \frac{m_\text{th}}{4 \pi \beta} \right) + \mathrm{O} \left(\lambda^2 \log(\lambda) \right) \notag \\
    &=
    \frac{2 \pi^2 \veps}{3 \beta^2}
    \frac{N+2}{N+8}
    \left(
        1 - \sqrt{\frac{6(N+2)}{N+8}} \sqrt{\veps}
        +
        \mathrm{O} \left( \veps \log(\veps)
        \right)
    \right)\,,
    \label{eq:ThermalMass}
\end{align}
we find
\begin{equation}
    G_\text{massive} (0,0)
    =
    \frac{1}{\beta^2}
    \left[
        \frac{1}{12}
        -\frac{1}{2\sqrt{6}}
        \sqrt{\frac{N+2}{N+8}} \sqrt{\veps}
        +\mathrm{O} \left( \veps \log(\veps) \right)
    \right]\,.
    \label{eq:bspm_Step6}
\end{equation}
Putting everything together in~\eqref{eq:bPhi_Diagram}, the one-point function of $\hphi^1$ reads
\begin{align}
    \beta^{\smash{\Dh_{\hphi^1}}} \vev{\hphi^1 (0)}
    &=
    \frac{h \beta m_\text{th}}{2} + \mathrm{O}(\veps^{3/2} \log(\veps)) \notag \\
    &=
    \pi \sqrt{\frac{N+2}{6}} \sqrt{\veps}
    -
    \frac{\pi}{2} \frac{N+2}{\sqrt{N+8}} \veps
    +
    \mathrm{O}(\veps^{3/2} \log(\veps))\,.
    \label{eq:bPhi_FinalResult}
\end{align}
Note that the order $\mathrm{O} (\veps)$ can be safely extracted from this method, however the form used for the massive propagator breaks down at the next order and logarithmic contributions in $\veps$ may appear.
This leads to the OPE coefficient $a_{\smash{\hphi^1}}$ being given by
\begin{equation}
    a_{\smash{\hphi^1}}
    =
    \frac{\pi^2}{\sqrt{6}} \sqrt{\frac{N+2}{N+8}} \veps^{3/2}
    -
    \frac{\pi^2}{2} \frac{N+2}{N+8} \veps^2
    +
    \mathrm{O}(\veps^{5/2} \log(\veps))\,.
    \label{eq:aPhi_FinalResult}
\end{equation}
This fixes the correlator~\eqref{eq:gInt_InPeriodicBlocks} fully up to order $\mathrm{O}(\veps)$.

As a consistency check, we can now verify the relation between $a_{\smash{\hphi^1}}$ and $a_{\smash{s_\pm}}$ inferred from the periodic expansion.
Replacing the coincident contraction in~\eqref{eq:bspm} by its massive counterpart gives
\begin{equation}
    b_{\smash{s_\pm}}
    =
    \frac{\sqrt{2}\,\zeta_2(N A_\pm-1)}
         {\sqrt{N A_\pm^2 - 2 A_\pm+1}}
    \left[
        1 - \sqrt{\frac{6 (N+2)}{N+8}}\,\sqrt{\veps}
    \right]
    +\mathrm{O} (\veps\log(\veps))\,.
    \label{eq:bspm_ThermalExpansion}
\end{equation}
The contribution proportional to $\vev{\hphi^1}^2$ starts only at order $\veps$, while the zero-temperature three-point coefficients contain no $\sqrt{\veps}$ correction.
Using~\eqref{eq:fttspm}, we therefore obtain
\begin{equation}
    \vev{a_{s_\pm}^{(1/2)}}
    =
    - \sqrt{\frac{6 (N+2)}{N+8}}
    \left( a_{s_-}^{(0)} + a_{s_+}^{(0)} \right)
    =
    -\pi^2 \sqrt{\frac{2(N+2)}{3(N+8)}}
    =
    -2 a_{{\hphi^1}}^{(3/2)}\,,
    \label{eq:aPhi_spm_Consistency}
\end{equation}
in agreement with the constant term generated by the $\hphi^1$ periodic block.
The relation between $\hphi^1$ and the combination of $s_\pm$ allows us to predict:
\begin{align}
    a_{s_-}^{(1)} + a_{s_+}^{(1)}
    &=
    -2a_{{\hphi^1}}^{(2)}
    +
    \log(2\pi)\,
    \vev{a_{{\Dh_\mathrm{cl} = 2}}^{(0)} \gamma_{{\Dh_\mathrm{cl} = 2}}^{(1)}}
    \notag\\
    &=
    \pi^2 \frac{N+2}{N+8}
    \left(
        1 + \frac{\log(2\pi)}{3}
    \right).
    \label{eq:aPhi_spm_Prediction_OneLoop}
\end{align}

\paragraph{Longitudinal singlet correlator $\vev{\hphi^1 \hphi^1}$.}
We now consider the thermal two-point function of the lowest $\mathrm{O} (N-1)$ singlet.
The calculation is analogous to the correlator of tilts, up to the following differences:
\begin{itemize}
    \item The scaling dimension of $\hphi^1$ changes with $\veps$.
    Consequently, the identity contribution in the periodic OPE is
    \begin{equation}
        F_{\hat{\mathds{1}}} (\tau)
        =
        \zeta_H \left( 2 \left(1 + \veps + \mathrm{O}(\veps^2) \right), \frac{\tau}{\beta} \right)
        +
        \zeta_H \left( 2 \left(1 + \veps + \mathrm{O}(\veps^2) \right), 1 - \frac{\tau}{\beta} \right)\,;
    \end{equation}
    \item We assume the remainder $\tilde{g}_{\smash{\hphi^1\hphi^1}} (\tau)$ to be constant as for the tilt.
    The impurity susceptibility of $\hphi^1$ is however not zero and is not given by a Ward identity.
    We thus do not fix the additive constant in the periodic OPE:
    \begin{equation}
        \chi_{\smash{\hphi^1}} = \beta\, \left( \frac{c_{\smash{\hphi^1}}}{\beta^{\smash{2\Dh_{\smash{\hphi^1}}}}} - \vev{\hphi^1}^2 \right)\,;
    \end{equation}
    \item The exchanged operators relevant for the periodic OPE are the same as for the tilt.
    We simply need to convert their associated OPE coefficients by using
    \begin{align}
        f_{{\hphi^1\hphi^1\hphi^1}}^{(1)}
        &=
        3 f_{{tt \hphi^1}}^{(1)}\,, \\
        f_{{\hphi^1\hphi^1 s_\pm}}^{(0)}
        &=
        \frac{A_\pm - 1}{A_\pm} f_{{tt s_\pm}}^{(0)}\,.
    \end{align}
\end{itemize}
The final result is then
\begin{align}
    g_{\smash{\hphi^1\hphi^1}} (\tau)
    &=
    \frac{1}{\beta^{\smash{2\Dh_{\smash{\hphi^1}}}}}
    \biggl[
    F_{\hat{\mathds{1}}} (\tau)
    -
    \frac{\pi^2}{6} \frac{N+2}{N+8} \log \left[ \frac{1}{4} \csc^2 \left( \frac{\pi \tau}{\beta} \right) \right] \veps \notag \\
    &\phantom{=\ }
    +
    \pi^2 \sqrt{\frac{6 (N+2)}{N+8}} \left( \sqrt{\veps} - \sqrt{\frac{3 (N+2)}{2 (N+8)}}\, \veps  \right)
    +
    \mathrm{O}(\veps^{3/2} \log(\veps))
    \biggr]
    + \frac{c_{\smash{\hphi^1}}}{\beta^{\smash{2\Dh_{\smash{\hphi^1}}}}}\,.
    \label{eq:g_hphi1_FinalResult}
\end{align}
At $N=1$, omit the contribution of $s_-$ and use the explicit formulas for the three-point functions since there is no tilt correlator to convert from.

\subsubsection{OPE data}
\label{subsubsec:OPEData}

We now extract the OPE data from the correlators~\eqref{eq:g_Splitting} and~\eqref{eq:g_hphi1_FinalResult}.
To do so, we simply need to rewrite~\eqref{eq:gtt_BootstrapResult} in the form
\begin{equation}
    \left. \pd_{\smash{\Dh}} F_{\smash{\Dh}} (\tau) \right|_{\Dh = 2}
    =
    \log \left( \frac{\tau}{\beta} \right)
    +
    \log(2\pi)
    -
    \sum_{n=1}^\infty \frac{\zeta_{2n}}{n} \left( \frac{\tau}{\beta} \right)^{2n}\,.
    \label{eq:ExpandBootstrapIntoBlocks}
\end{equation}
We use this result to extract the OPE data for the two-point functions of tilt and singlet operators.

\paragraph{OPE data for tilt two-point function.}
Comparing~\eqref{eq:gtt_OPE} and~\eqref{eq:ExpandBootstrapIntoBlocks}, we readily obtain the following expressions for the OPE data associated to the correlator of two tilt operators:
\begin{align}
    \vev{a_{\smash{\Dh_\text{cl} = 2}}}
    &=
    2\zeta_2
    -\pi^2\sqrt{\frac{2(N+2)}{3(N+8)}}\,\sqrt{\veps}
    \notag\\
    &\phantom{=\ }
    +
    \frac{N+2}{N+8}
    \left[\pi^2+2\zeta_2\log(2\pi)\right] \veps
    +
    \mathrm{O} \bigl( \veps^{3/2}\log(\veps) \bigr)\,, \\
    \vev{a_{\smash{\Dh_\text{cl} = 2 + 2n}}}_{n \geq 1}
    &=
    2(2n+1) \zeta_{2n+2}
    - \frac{2 \zeta_2 \zeta_{2n}}{n} \frac{N+2}{N+8} \veps
    + \mathrm{O} \bigl( \veps^{3/2}\log(\veps) \bigr)\,.
    \label{eq:OPEDataTilt}
\end{align}

\paragraph{OPE data for singlet two-point function.}
For the correlator of singlet operators, we must take into consideration the additional contributions coming from the identity periodic block:
\begin{align}
    \vev{a_{\smash{\Dh_\text{cl} = 2}}}
    &=
    2\zeta_2
    +
    \pi^2\sqrt{\frac{6(N+2)}{N+8}}\,\sqrt{\veps}
    +
    \left[
        4\zeta'(2)
        +\frac{N+2}{N+8}
        \left(-3\pi^2+2\zeta_2\log(2\pi)\right)
    \right] \veps
    \notag \\
    &\phantom{=\ }
    +
    c_{\hphi^1}
    +
    \mathrm{O} \bigl( \veps^{3/2}\log (\veps) \bigr)\,, \\
    \vev{a_{\smash{\Dh_\text{cl} = 2 + 2n}}}_{n \geq 1}
    &=
    2(2n+1) \zeta_{2n+2} \notag \\
    &\phantom{=\ }
    +
    \left(4 (2n+1) \left[ \zeta'_{2n+2} + (H_{2n+1} - 1) \zeta_{2n+2} \right]
    - \frac{2 \zeta_2 \zeta_{2n}}{n} \frac{N+2}{N+8} \right) \veps \notag \\
    &\phantom{=\ }
    + \mathrm{O} \bigl( \veps^{3/2}\log (\veps) \bigr)\,,
    \label{eq:OPEDataSinglet}
\end{align}
where the derivative of the $\zeta$ function should be understood as $\zeta'_s = \left. \pd_t \zeta_t \right|_{t=s}$.

\subsubsection{Comparison to Feynman diagrams}
\label{subsubsec:ComparisonToFeynmanDiagrams}

To validate our method, we compare the analytic bootstrap results above to direct diagrammatic calculations.
Computational details are collected in Appendix~\ref{app:DetailsOnTheFeynmanDiagrams}, such that we focus mostly on conclusive remarks here.

\paragraph{Diagrams.}
The diagrams contributing to $g_{tt} (\tau)$ up to order $\mathrm{O}(\veps)$ are
\begin{equation}
    g_{tt} (\tau)
    =\,
    \frac{1}{n_t} Z_t^{-2} \left( \DiagOneFiniteT\ +\ \DiagTwoFiniteT\ +\ \DiagThreeFiniteT\ + \ldots \right)\,.
    \label{eq:Diagrams}
\end{equation}
In this expression, we follow the convention in which we do not include the normalization and renormalization constants in the diagrams.
Note that the first two diagrams are already present in the zero-temperature correlator~\cite{Gimenez-Grau:2022czc}, while the scalar tadpole exists only at finite temperature and can be interpreted as non-vanishing due to the emergence of a thermal mass.

\paragraph{First and second diagrams.}
A simplification occurs for the computation of the first and second diagrams at finite temperature.
The first one is trivial and can be written as
\begin{align}
    \DiagOneFiniteT
    =
    \sum_{m=-\infty}^\infty
    \DiagOneZeroT\,,
    \label{eq:DiagramOne}
\end{align}
i.e., it is the sum of images of the zero-temperature propagator given in~\eqref{eq:ON_FreePropagator}.
As shown in Appendix~\ref{app:DetailsOnTheFeynmanDiagrams}, the second diagram can be manipulated such that it follows a similar structure:
\begin{align}
    \DiagTwoFiniteT
    =
    \sum_{m=-\infty}^\infty \left. \DiagTwoZeroT \right|_{\tau \to \tau+m \beta}\,.
    \label{eq:DiagramTwo}
\end{align}
The immediate consequence of~\eqref{eq:DiagramOne} and~\eqref{eq:DiagramTwo} is that the full correlator can be rewritten as
\begin{align}
    g_{tt} (\tau)
    =
    g^\text{(GFF)}_{tt} (\tau)
    +
    \frac{1}{n_t}\, \DiagThreeFiniteT\
    +
    \ldots\,,
    \label{eq:ReductionToOneDiagram}
\end{align}
with
\begin{equation}
    g^\text{(GFF)}_{tt} (\tau)
    =
    \frac{F_{\hat{\mathds{1}}} (\tau)}{\beta^2}\,.
    \label{eq:gGFFFromDiagrams}
\end{equation}
Thus, only the tadpole diagram contributes to $g_{tt}^\text{(int)} (\tau)$ and is non-trivial from this perspective.

\paragraph{Third diagram and matching.}
We now compute the third and final diagram.\footnote{Here a similar IR resummation as the one performed in Section~\ref{subsubsec:ExpansionInPeriodicBlocks} should be used in order to extract the order $\mathrm{O}(\sqrt{\veps})$.
We do not include it explicitly in subsequent expressions, but it can be shown to agree with the bootstrap expectation.}
It can be shown to be given in arbitrary dimensions by~\cite{Barrat:2025nvu}
\begin{align}
    \DiagThreeFiniteT\
    &=
    -\frac{\lambda(N+2)\Gamma(\nu)\Gamma(\nu-1)\zeta(2\nu)}
    {192\pi^{2\nu+2}\beta^{4\nu-2}}
    \left[
    \zeta_H\!\left(2\nu-2,\frac{\tau}{\beta}\right)
    +
    \zeta_H\!\left(2\nu-2,1-\frac{\tau}{\beta}\right)
    \right]\,,
    \label{eq:ThirdDiagram_Step1}
\end{align}
with $\nu = d/2-1$.
Details for obtaining this expression are provided in Appendix~\ref{app:DetailsOnTheFeynmanDiagrams}.
The analytically continued expression is finite and thus we can set $\nu = 1$ ($d=4$).
We obtain
\begin{align}
    \DiagThreeFiniteT\
    &=
    -\frac{\lambda(N+2)}{1152\pi^2\beta^2}
    \log\left[
    \frac14\csc^2\!\left(\frac{\pi\tau}{\beta}\right)
    \right]\,.
    \label{eq:ThirdDiagram_FinalResult}
\end{align}
This agrees with the $\tau$-dependent part of~\eqref{eq:gInt_InPeriodicBlocks} and~\eqref{eq:gtt_BootstrapResult} once we include the normalization.

\subsubsection{Finite volume}
\label{subsubsec:FiniteVolume}

We now repeat the same calculations as above for the antipodal configuration at finite volume.
For brevity, we focus on the two-point function of tilt operators, but $\vev{\hphi^1 \hphi^1}$ can be evaluated using similar techniques.
We first perform an analytic bootstrap calculation in a manner similar to the infinite-volume case, although technical difficulties arise.
We then extract the OPE data and compare the bootstrap result to an explicit diagrammatic calculation.

\paragraph{Analytic bootstrap.}
We now compute the two-point function of tilt operators at finite volume.
Our focus will mostly be on essential differences with the infinite-volume case rather than technical aspects.
Similarly to~\eqref{eq:g_Splitting}, the correlator can be written as
\begin{equation}
    g_{tt} (\tau)
    =
    \frac{1}{\beta^2} F_{\hat{\mathds{1}}} (\tau)
    +
    g^\text{(int)}_{tt} (\tau)
    +
    \tilde{g}_{tt} (\tau)\,.
    \label{eq:g_Splitting_FiniteV}
\end{equation}
As a starting point, we ask ourselves whether the interacting piece $g^\text{(int)}_{tt} (\tau)$ can be given by\footnote{We discard terms with half-integer powers $\veps^{k/2}$ since the conformal curvature gap removes the infrared zero-mode singularity at fixed $\beta/R$.
This correlator thus admits an expansion in \textit{integer} powers of $\veps$.
In particular, this implies that the $\hphi^1$ exchange does not contribute at this order.}
\begin{align}
    g^\text{(int)}_{tt} (\tau)
    &\overset{?}{=}
    \frac{\veps}{\beta^2}
    \vev{a_{{\Dh_\mathrm{cl} = 2}}^{(0)} (q) \gamma_{{\Dh_\mathrm{cl} = 2}}^{(1)}}
    \left. \pd_{\smash{\Dh}} F_{\smash{\Dh}} (\tau) \right|_{\Dh = 2}
    +
    \mathrm{O} (\veps^2)\,,
    \label{eq:FiniteV_StartingPoint}
\end{align}
as is the case at infinite volume.
The answer turns out to be no.
The reason is that~\eqref{eq:FiniteV_StartingPoint} generates logarithmic contributions multiplying arbitrary even powers of $\tau$, instead of having a single $\tau^0 \log (\tau/\beta)$:\footnote{Having a single $\log$ term is expected from the diagrams~\eqref{eq:Diagrams} (or, better suited,~\eqref{eq:FiniteV_Diagrams}) and checked in~\eqref{eq:FiniteV_LastDiagram_Result}.
This amounts to saying that the first thermal self-energy correction (the tadpole) can be interpreted as a mass term and corresponds to the constant power in $\tau$.
Another formulation of the same statement is to say that the loop in the tadpole diagram is position-independent and thus can at most contribute $\tau^0 \log(\tau/\beta)$.}
\begin{align}
    \left. \pd_{\smash{\Dh}} F_{\smash{\Dh}} (\tau) \right|_{\Dh = 2}
    &=
    f_{\smash{\Dh=2}} (\tau) \log \left( \frac{\tau}{\beta} \right) + \text{terms analytic at } \tau=0\,. \notag \\
    &=
    \left( 1 - \frac{1}{10} \frac{\tau^2}{R^2} + \frac{1}{168} \frac{\tau^4}{R^4} + \ldots \right) \log \left( \frac{\tau}{\beta} \right) + \text{terms analytic at } \tau=0\,.
\end{align}
The consequence is that operators other than $s_\pm$ must contribute with an anomalous dimension at order $\mathrm{O}(\veps)$.\footnote{This may seem at odds with the $\veps$-counting performed at infinite volume.
At finite volume, operators which had vanishing one-point functions at infinite volume can acquire an expectation value and thus contribute with anomalous dimensions at one loop.}
Fortunately, their contributions are entirely fixed by demanding that the terms $\tau^k \log(\tau/\beta)$ cancel for $k>0$.
This leads immediately to the expression
\begin{align}
    g^\text{(int)}_{tt} (\tau)
    &=
    \frac{\veps}{\beta^2}
    \vev{a_{{\Dh_\mathrm{cl} = 2}}^{(0)} (q) \gamma_{{\Dh_\mathrm{cl} = 2}}^{(1)}}
    \sum_{n=0}^\infty
    \frac{2 [(2n+1)!]^2}{(4n+2)!} \left( \frac{\beta}{R} \right)^{2n}
    \left. \pd_{\smash{\Dh}} F_{\smash{\Dh}} (\tau) \right|_{\Dh = 2+2n}
    +
    \mathrm{O} (\veps^2) \notag \\
    &=
    \frac{\veps}{\beta^2}
    \vev{a_{{\Dh_\mathrm{cl}=2}}^{(0)} (q) \gamma_{{\Dh_\mathrm{cl}=2}}^{(1)}}
    \left(
    \sum_{m = -\infty}^\infty
    \log \left[ \frac{u_m}{\sqrt{u_m^2+\pi^2}} \right]
    +
    \frac{\pi^2 R}{\beta}
    \right)
    +
    \mathrm{O}(\veps^2)\,,
    \label{eq:FiniteV_gInt_Final}
\end{align}
where we defined for compactness
\begin{equation}
    u_m
    =
    \frac{|\tau + m\beta|}{R}\,.
    \label{eq:um_Definition}
\end{equation}
It is remarkable that all terms are fixed simply by demanding consistency with the structure of the Feynman diagrams.

The remaining task is to fix $\tilde{g}_{tt}(\tau)$.
The analytic structure of~\eqref{eq:FiniteV_gInt_Final} does not match the expected analytic structure displayed in Figure~\ref{fig:AnalyticStructure}.
Expanding the spectral representation~\eqref{eq:SpectralDecomposition} to first order in $\veps$, and using that the leading energy differences are integer multiples of $1/R$, gives
\begin{equation}
    g_{tt}^{(1)} (\tau)
    =
    \frac{1}{2}
    \left( g_{tt}^{(1)} (\tau + 2\pi i R) + g_{tt}^{(1)} (\tau - 2\pi i R) \right)\,.
    \label{eq:OneLoopConstraint}
\end{equation}
Together with the OPE, this fixes the discontinuity of $\tilde{g}_{tt} (\tau)$ along the whole imaginary axis at each $\operatorname{Re} (\tau) = m\beta$, except for the pole terms located at $\tau = m\beta + 2\pi i n R$.
To fix these, we now assume that the first-order correction does not contain double poles or higher at these locations.
This can be justified using the fact that the one-loop self-energy is position-independent; the resulting correction can therefore develop only logarithmic and simple-pole singularities at the physical pole locations.\footnote{A position-independent self-energy acts as a constant mass shift.
Expanding a cylinder mode around its free frequency $k/R$ produces a correction proportional to $e^{-k \tau/R}(1/k + \tau/R)$.
Near a pole location, the two sums behave as $\sum_{ k \geq 1} e^{-k \operatorname{Re}(\tau)/R}/k \sim - \log \left( \operatorname{Re}(\tau) \right)$ and
$\sum_{k \geq 1} e^{-k \operatorname{Re}(\tau)/R} \sim 1/\operatorname{Re}(\tau)$, respectively.
Thus only logarithmic and simple-pole singularities occur.}
A candidate completion yielding a correlator with these properties is
\begin{align}
    \tilde{g}_{tt} (\tau)
    &=
    \tilde{g}_{tt}^{(\text{free})} (\tau) \notag \\
    &\phantom{=\ }
    +
    \frac{\veps}{\beta^2}
    \vev{a_{{\Dh_\mathrm{cl} = 2}}^{(0)} (q) \gamma_{{\Dh_\mathrm{cl} = 2}}^{(1)}}
    \left[
    \sum_{m=-\infty}^\infty
    \left(
    \log \left[ ( 1 - e^{-u_m} ) \frac{\sqrt{\pi^2 + u_m^2}}{u_m} \right]
    -
    \frac{u_m}{e^{u_m} - 1}
    \right)
    -
    \frac{\pi^2 R}{\beta}
    \right] \notag \\
    &\phantom{=\ } + \frac{c_t (q)}{\beta^2} + \mathrm{O} \left( \veps^2 \right)\,.
    \label{eq:FiniteV_Completion}
\end{align}
The first term is simply the completion of the free part already discussed in Section~\ref{subsubsec:PeriodicBlocks}.
As argued in Section~\ref{subsubsec:PropertiesOfPeriodicBlocks}, the second term is the unique completion (up to a constant) that restores consistent properties of the full correlator, including~\eqref{eq:OneLoopConstraint}.
Under these assumptions, Equation~\eqref{eq:g_Splitting_FiniteV} with~\eqref{eq:FiniteV_gInt_Final} and~\eqref{eq:FiniteV_Completion} solves the finite-volume bootstrap problem and fixes the one-loop correlator up to a constant:\footnote{As explained in Footnote~\ref{fn:Susceptibility_FiniteV}, this constant is not equivalent to the tilt susceptibility.
One would need to calculate the North-South correlator as well to be able to relate them.}
\begin{align}
    g_{tt} (\tau) - g_{tt}^\text{(free)} (\tau)
    &=
    \frac{\veps}{\beta^2}
    \vev{a_{{\Dh_\mathrm{cl} = 2}}^{(0)} (q) \gamma_{{\Dh_\mathrm{cl} = 2}}^{(1)}}
    \sum_{m = -\infty}^\infty
    \left[
        \log(1-e^{-u_m})
        -\frac{u_m}{e^{u_m}-1}
    \right] \notag \\
    &\phantom{=\ }
    + \frac{c_t (q)}{\beta^2} + \mathrm{O} \left( \veps^2 \right)\,.
    \label{eq:FiniteV_FinalBootstrapResult}
\end{align}

\paragraph{OPE data.}
We now extract the OPE data from the tilt correlator at finite volume.
The results are more intricate than at infinite volume and we find closed expressions in recursive form.
These expressions are nevertheless efficient to evaluate at very high scaling dimension $2 + 2n$.
At tree level, the OPE coefficients read:
\begin{equation}
    \vev{a_{{\Dh_\text{cl} = 2 + 2n}}^{(0)} (q)}
    =
    \frac{2}{(2n)!} \left( \frac{\beta}{R} \right)^{2+2n} L_{2n+1} (q)
    -
    \sum_{k=0}^{n-1} \vev{a_{{\Dh_\text{cl} = 2 + 2k}}^{(0)} (q)}
    \left( \frac{\beta}{R} \right)^{2(n-k)} \Bm_{n - k} (2+2k)\,,
\end{equation}
where we have defined
\begin{align}
    L_j (q)
    &=
    \sum_{k=1}^{\infty} \frac{k^j q^k}{1 - q^k}
    =
    \frac{\psi_q^{(j)}(1)}{(\log q)^{j+1}}\,, \label{Lq_Definition} \\
    \Bm_j (h)
    &=
    - \frac{(h + 2j - 1)(h + 2j)}{j (2h + 2j-1)}
    \sum_{k=0}^{j - 1}
    \frac{\Bm_k (h)}{[2(j-k) + 2]!}\,,
    \quad
    \Bm_0 (h) = 1\,,
\end{align}
where we used the $q$-polygamma function $\psi_q^{(j)}(1)$ in~\eqref{Lq_Definition}.
The coefficients present in~\eqref{eq:FiniteV_gInt_Final} give rise to the combinations
\begin{equation}
    \vev{a_{{\Dh_\text{cl} = 2 + 2n}}^{(0)} (q) \gamma_{{\Dh_\text{cl} = 2 + 2n}}^{(1)}}
    =
    \frac{N+2}{N+8}
    \frac{4 [(2n+1)!]^2}{(4n+2)!}
    \left( \frac{\beta}{R} \right)^{2n+2}
    L_1 (q)\,.
\end{equation}
Finally, the one-loop OPE coefficients are given recursively by
\begin{align}
    \vev{a_{{\Dh_\text{cl} = 2 + 2n}}^{(1)} (q)}_{n \geq 1}
    &=
    \frac{2(N+2)}{(N+8)(2n)!}
    \left( \frac{\beta}{R} \right)^{2+2n}
    L_1(q) \notag \\
    &\phantom{=\ } \times
    \left[
        -\frac{2n-1}{2n}\, B_{2n}
        +2 (2n-1) L_{2n-1}(q)
        -2 \frac{\beta}{R}\, q \partial_q L_{2n-1}(q)
    \right] \notag \\
    &\phantom{=\ }
    -
    \sum_{k=0}^{n-1}
    \left( \frac{\beta}{R} \right)^{2(n-k)}
    \biggl[
        \vev{a_{{\Dh_\text{cl} = 2 + 2k}}^{(1)} (q)}
        \Bm_{n-k}(2+2k) \notag \\
        &\phantom{=\ }
        +
        \vev{a_{{\Dh_\text{cl} = 2 + 2k}}^{(0)} (q) \gamma_{{\Dh_\text{cl} = 2 + 2k}}^{(1)}}
        \left.\partial_h \Bm_{n-k}(h)\right|_{h=2+2k}
    \biggr]\,,
\end{align}
with the starting value
\begin{align}
    \vev{a_{{\Dh_\text{cl} = 2}}^{(1)} (q)}
    &=
    2\frac{N+2}{N+8} \log^2 (q)\,
    L_1(q)
    \left[
        \log(-\log (q)) - 1
        +2 \sum_{m=1}^{\infty} \log(1-q^m)
        +2 \log (q)\,
        L_1(q)
    \right] \notag \\
    &\phantom{=\ }
    +
    c_t^{(1)} (q)\,,
\end{align}
where we used the Bernoulli numbers $B_n$.
The tree-level constant $c_t^{(0)} (q)$ is already included in the tree-level coefficient.

\paragraph{Comparison to Feynman diagrams.}
We now repeat the computation of Section~\ref{subsubsec:ComparisonToFeynmanDiagrams} for the antipodal configuration.
In this case, the finite-temperature diagrams are given by
\begin{equation}
    g_{tt} (\tau)
    =
    \frac{Z_t^{-2}}{n_t}
    \left(\
    \DiagOneFiniteTFiniteV\
    +
    \DiagTwoFiniteTFiniteV\
    +
    \DiagThreeFiniteTFiniteV\
    +
    \DiagFourFiniteTFiniteV\
    +
    \DiagFiveFiniteTFiniteV\
    + \ldots
    \right)\,.
    \label{eq:FiniteV_Diagrams}
\end{equation}
All the diagrams have a zero-temperature counterpart (see~\eqref{eq:Diagrams_FiniteV_ZeroT}), except for the tadpole one.
Similarly to the infinite-volume case, we can recast~\eqref{eq:FiniteV_Diagrams} as
\begin{equation}
    g_{tt} (\tau) - g_{tt}^{(\text{GFF})} (\tau)
    =
    \frac{1}{n_t}\
    \DiagFiveFiniteTFiniteV\
    +
    \ldots\,.
\end{equation}
The calculation of this diagram is detailed in Appendix~\ref{app:DetailsOnTheFeynmanDiagrams}.
We find
\begin{align}
    \DiagFiveFiniteTFiniteV\
    &=
    \frac{\lambda(N+2) L_1(q)}{96\pi^4 R^2}
    \sum_{m = -\infty}^\infty
    \left[
        \log(1-e^{-u_m})
        -\frac{u_m}{e^{u_m}-1}
    \right]\,.
    \label{eq:FiniteV_LastDiagram_Result}
\end{align}
This agrees with the bootstrap result~\eqref{eq:g_Splitting_FiniteV} with~\eqref{eq:FiniteV_gInt_Final} and~\eqref{eq:FiniteV_Completion}.
In the conventions of~\eqref{eq:FiniteV_Completion}, direct matching with the diagram also fixes the remaining constant to vanish through this order,
\begin{equation}
    c_t^{(1)} (q)
    =
    0\,.
\end{equation}

\subsection{$\vev{\hphi^1}$ and interpolation to $d=3$}
\label{subsec:bPhi1AndInterpolationTo3d}

In this section, we shift our focus to the thermal one-point function $\vev{\hphi^1}$ for the $3d$ Ising ($N=1$).
We compute its value in the $2d$ theory ($\veps = 2$) and interpolate it to reach an estimate at $d=3$.\footnote{The identification of the $d \to 2$ limit of the Wilson--Fisher $\mathrm{O}(N)$ fixed point with the weakly coupled non-linear $\sigma$ model is subtle at finite $N>2$: recent work has identified a mismatch in protected operators that obstructs a straightforward analytic connection between the two fixed-point families~\cite{DeCesare:2025ukl,DeCesare:2026dwm}.}
Finally, we translate this expression to an OPE coefficient in the two-point function $\vev{\hphi^1 \hphi^1}$ and conjecture that the latter is well-estimated by truncating the periodic OPE.

\subsubsection{$\vev{\hphi^1}$ at $d=2$ and Padé interpolation}
\label{subsubsec:bPhi1At2dAndInterpolation}

\paragraph{Infinite volume.}
In $d=2$, the defect setup~\eqref{eq:ON_Action} with $N=1$ corresponds to the product of two copies (left and right) of the Ising CFT, as explained in Section~\ref{subsubsec:ThePinningFieldLineDefect}.
This setup was solved in~\cite{Cardy:1989ir,Ishibashi:1988kg} and the spectrum of boundary operators of a single CFT is just given by the Virasoro descendants of the identity.
The operator $\hphi^1$ corresponds to a combination of the left and right displacement operators and has dimension $\Dh_{\smash{\hphi^1}} = 2$ in the $2d$ theory (see~\eqref{eq:hphi1In2d} and Figure~\ref{fig:LargeNSpectrum} for the large $N$ counterpart).

We therefore need to determine the one-point function of the stress-tensor operator $T$.
To do so, we map the plane to a cylinder of circumference $\beta$ and obtain
\begin{equation}
    T(\tau) \to \left( \frac{\mathrm{d}z}{\mathrm{d}\tau} \right)^2 T(z) + \frac{c}{12} \lbrace z, \tau \rbrace\,,
    \qquad
    z = \tan \left( \frac{\pi \tau}{\beta} \right)\,,
    \qquad
    c = \frac{1}{2}\,,
    \label{eq:bPhi_2d_Step5}
\end{equation}
with $\lbrace z, \tau \rbrace$ the Schwarzian derivative of $z$.
Since $\vev{T(z)} = 0$ in the vacuum and
\begin{equation}
    \lbrace z,\tau \rbrace
    =
    \frac{2 \pi^2}{\beta^2}\,,
    \label{eq:bPhi_2d_Step6}
\end{equation}
one obtains
\begin{equation}
    \vev{T_L}
    =
    \vev{T_R}
    =
    \frac{\pi^2}{12 \beta^2}\,,
    \qquad
    \vev{D}_\beta
    =
    0\,.
    \label{eq:bPhi_2d_Step7}
\end{equation}
Putting everything together, we find
\begin{equation}
    \vev{\hphi^1}_{2d}
    =
    \sqrt{2}\, \left( \vev{T_L} + \vev{T_R} \right)
    =
    \frac{\pi^2}{3 \sqrt{2} \beta^2}
    \sim
    \frac{2.32629}{\beta^2}\,.
    \label{eq:bPhi_2d_FinalResult}
\end{equation}

\begin{figure}
    \centering
    \begin{minipage}{.5\textwidth}
        \centering
        \includegraphics[width=0.95\linewidth]{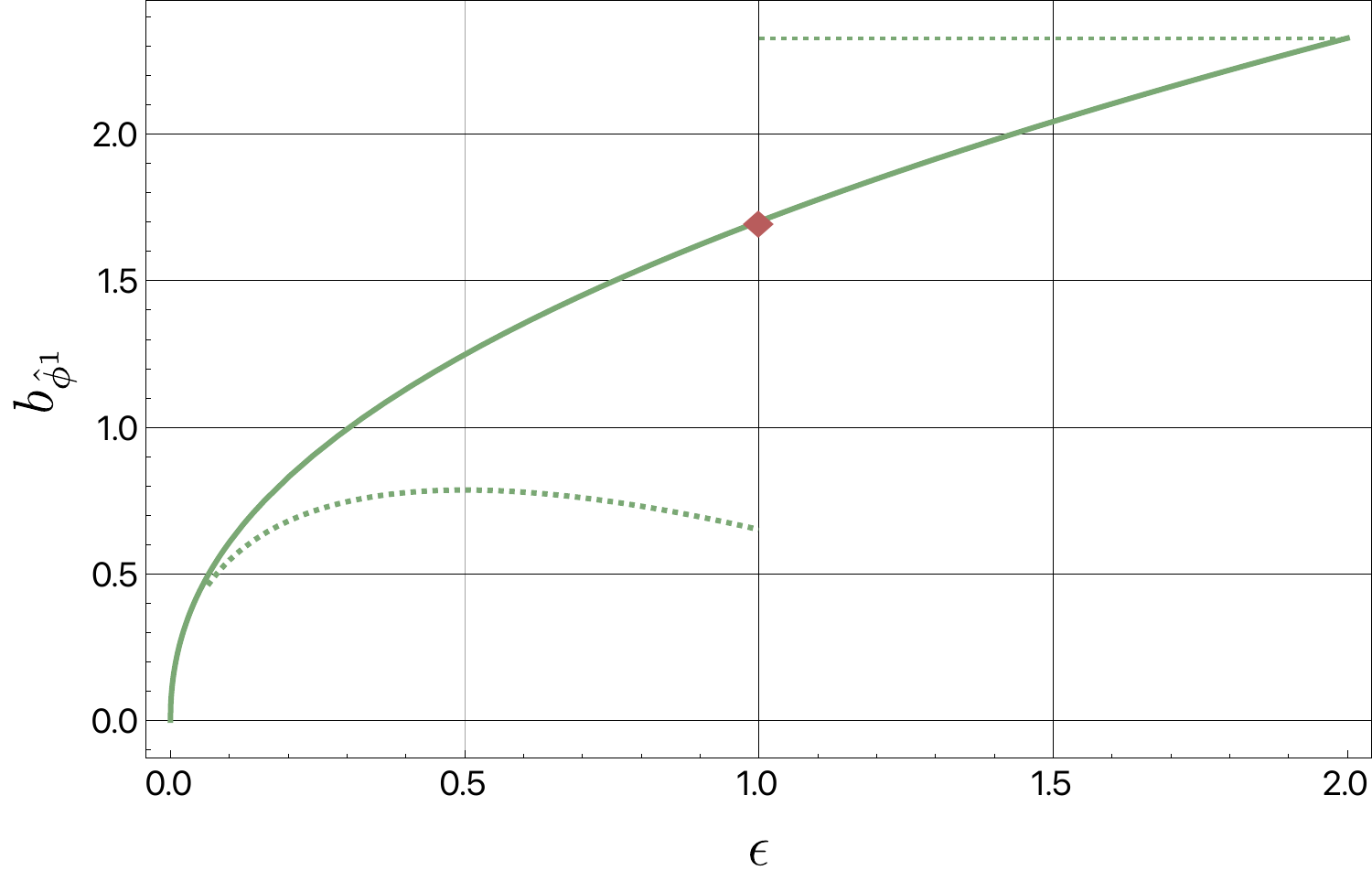}
    \end{minipage}%
    \begin{minipage}{0.5\textwidth}
        \centering
        \includegraphics[width=0.95\linewidth]{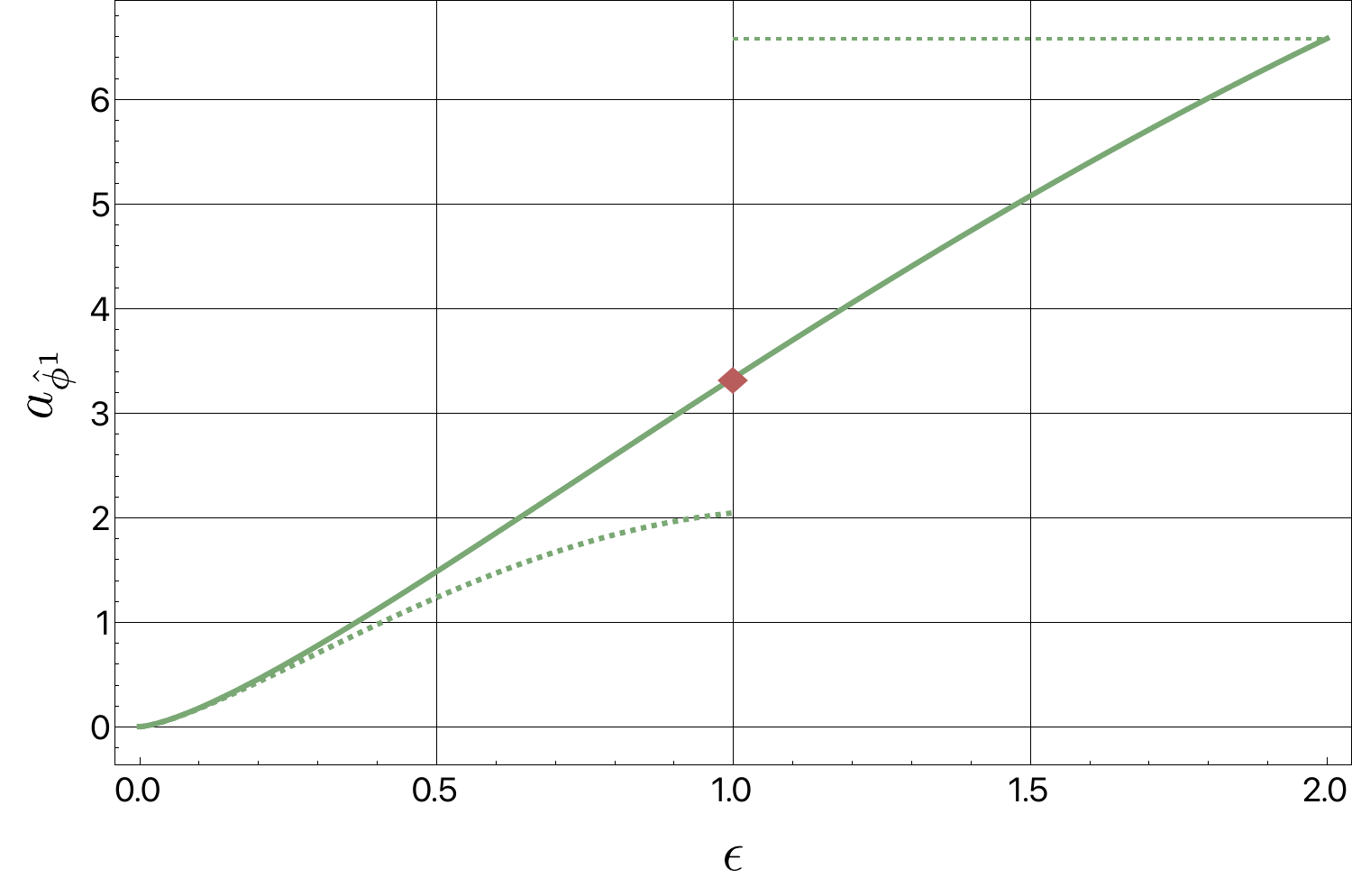}
    \end{minipage}
    \caption{
    \textbf{Left}: Padé approximant for $b_{\smash{\hphi^1}}$ (green solid curve) interpolating between the perturbative $4d$ value (at $\veps = 0$) and the $2d$ one (at $\veps=2$), shown here as green dashed curves.
    We obtain $b_{\smash{\hphi^1}} \sim 1.701$ at $d=3$ ($\veps = 1$), here shown as a red diamond.
    \textbf{Right}: Padé interpolation (green solid curve) between the values of the OPE coefficient $a_{\smash{\hphi^1}}$ obtained perturbatively near $d=4$ and in $2d$ (green dashed curves).
    At $d=3$, the interpolation predicts $a_{\smash{\hphi^1}} \sim 3.333$ (red diamond).
    }
    \label{fig:bPhi_Pade}
\end{figure}

We now perform a standard $[2/1]$ Padé interpolation between these two endpoints.
Combining~\eqref{eq:bPhi_2d_FinalResult} with~\eqref{eq:bPhi_FinalResult}, we obtain the function
\begin{equation}
    b^{(\text{Padé})}_{{\hphi^1}} (\veps)
    =
    \frac{
        \pi(3\sqrt{2}-\pi)\sqrt{\veps}
        +\frac{\pi}{\sqrt{2}}(2\pi-3\sqrt{2})\veps
    }{
        6-\sqrt{2}\pi+\pi\sqrt{\veps}
    }\,.
    \label{eq:ON_bPhi_Pade}
\end{equation}
At $d=3$, this gives
\begin{equation}
    b^{(\text{Padé})}_{{\hphi^1}} (1)
    \sim
    1.701\,.
    \label{eq:ON_bPhi_3d}
\end{equation}
The interpolating function with its $d=3$ point is shown in Figure~\ref{fig:bPhi_Pade}.
We similarly perform an interpolation for the OPE coefficient $a_{\smash{\hphi^1}}$:\footnote{Note importantly that this OPE coefficient is the one appearing in $\vev{\hphi^1 \hphi^1}$, not in $\vev{tt}$ as previously defined.
We use the same symbol in order not to burden the notation further.}
\begin{equation}
    a^{(\text{Padé})}_{{\hphi^1}} (\veps)
    =
    \frac{
        \pi^2\veps^{3/2}(2\sqrt{2}-\sqrt{\veps})
    }{
        4+\sqrt{2\veps}
    }\,.
    \label{eq:ON_aPhi_Pade}
\end{equation}
At $d=3$, this gives
\begin{equation}
    a^{(\text{Padé})}_{{\hphi^1}} (1)
    \sim
    3.333\,.
    \label{eq:ON_aPhi_3d}
\end{equation}
The interpolation is shown in Figure~\ref{fig:bPhi_Pade}.

Note that different interpolation schemes result in different $3d$ values.
We compare in Figure~\ref{fig:LargeN_AcrossDims} the results of this Padé scheme for large $N$ alongside the numerical solution of the equations of motion and observe a good agreement.
It should be possible to perform Monte Carlo simulations to obtain an independent estimate.
Additionally, the concepts developed in this work open the door to a numerical bootstrap analysis similar, e.g., to~\cite{Iliesiu:2018fao,Iliesiu:2018zlz,Barrat:2025wbi,Barrat:2025nvu}.

\paragraph{Finite volume.}
We repeat the same computation at finite volume.
The antipodal geometry splits into two fixed-boundary Ising strips of width $\pi R$.
Each strip has partition function $Z_{\text{strip}} = \chi_0(q)$, the Ising vacuum character.
This follows from the standard boundary-CFT strip construction~\cite{Cardy:1989ir}.
The Hilbert space of each strip is therefore the vacuum Virasoro module $\mathcal V_0$, and its Hamiltonian is
\begin{equation}
    H_{\mathrm{strip}}=\frac{1}{R}\left(L_0 - \frac{1}{48}\right)\,.
\end{equation}
Here, $L_0$ is the Virasoro zero-mode generator, whose eigenvalues on $\mathcal V_0$ are the conformal levels of the vacuum descendants.
We measure energies relative to the Casimir ground-state energy, i.e., we replace $\chi_0(q)$ by $q^{1/48} \chi_0(q)= \operatorname{tr}_{\Vm_0} q^{L_0}$ in the following.
The logarithmic derivative of the vacuum-energy-subtracted partition function gives the thermal expectation value of $L_0$, so that
\begin{equation}
    \vev{\hphi^1}
    =
    \frac{2\sqrt{2}}{R^2} \vev{L_0}
    =
    \frac{2\sqrt{2}}{R^2}\, q \partial_q \log \left( q^{1/48}\chi_0(q) \right)\,,
\end{equation}
from which we read
\begin{equation}
    \left. b_{\hphi^1}(q)\right|_{2d}
    =
    2\sqrt{2} \log^2(q)\,
    q \partial_q \log ( q^{1/48} \chi_0 (q) )\,.
    \label{eq:bPhi_FiniteV_2d}
\end{equation}
Near $d=4$, we determine the same correlator to be given by
\begin{equation}
    \left. b_{\smash{\hphi^1}} \right|_{d=4-\veps}
    =
    - \pi \log (q)\,L_1(q)\,\veps
    + \mathrm{O} (\veps^2)\,.
    \label{eq:bPhi_FiniteV_epsExp}
\end{equation}
We perform a [1/1] Padé interpolation, which results in
\begin{equation}
    b^{(\text{Padé})}_{{\hphi^1}} (\veps)
    =
    \frac{
        -\pi\log (q)\,L_1(q)\,\veps
    }{
        1-\frac{\veps}{2}
        -\frac{\pi L_1(q)}
        {2\sqrt{2}\log (q)\,q\partial_q\log (q^{1/48} \chi_0 (q))}\,\veps
    }\,.
    \label{eq:bPhi_FiniteV_Pade}
\end{equation}
In three dimensions, this gives
\begin{equation}
    b^{(\text{Padé})}_{{\hphi^1}} (1)
    =
    \frac{
        2\sqrt{2}\pi\log^2 (q)\,
        L_1(q)\,q\partial_q\log (q^{1/48} \chi_0 (q))
    }{
        \pi L_1(q)
        -\sqrt{2}\log (q)\,q\partial_q\log (q^{1/48} \chi_0 (q))
    }\,.
    \label{eq:bPhi_FiniteV_3d}
\end{equation}
One should be cautious when interpreting this result.
Equations~\eqref{eq:bPhi_FiniteV_Pade} and~\eqref{eq:bPhi_FiniteV_3d} are not expected to produce accurate results near the infinite-volume regime $q \to 1$.
The reason is that the IR resummation problem emerges in this limit, and the correct way to proceed is to include the $\mathrm{O}(\sqrt{\veps})$ term in the perturbative expansion.
The results of~\eqref{eq:bPhi_FiniteV_3d} are presented in Figure~\ref{fig:LargeN_bPhi_FiniteV}, alongside the large $N$ results.
The large-$N$ comparison suggests that the approximation may already be poor not far from $q \sim 0$, motivating a more elaborate interpolation that also incorporates the independently determined infinite-volume limit.
We reserve a better understanding of the OPE coefficients and interpolating schemes in the antipodal configuration to future work.

\subsubsection{Truncating the periodic OPE}
\label{subsubsec:TruncatingThePeriodicOPE}

\begin{figure}
    \centering
    \includegraphics[width=0.8\linewidth]{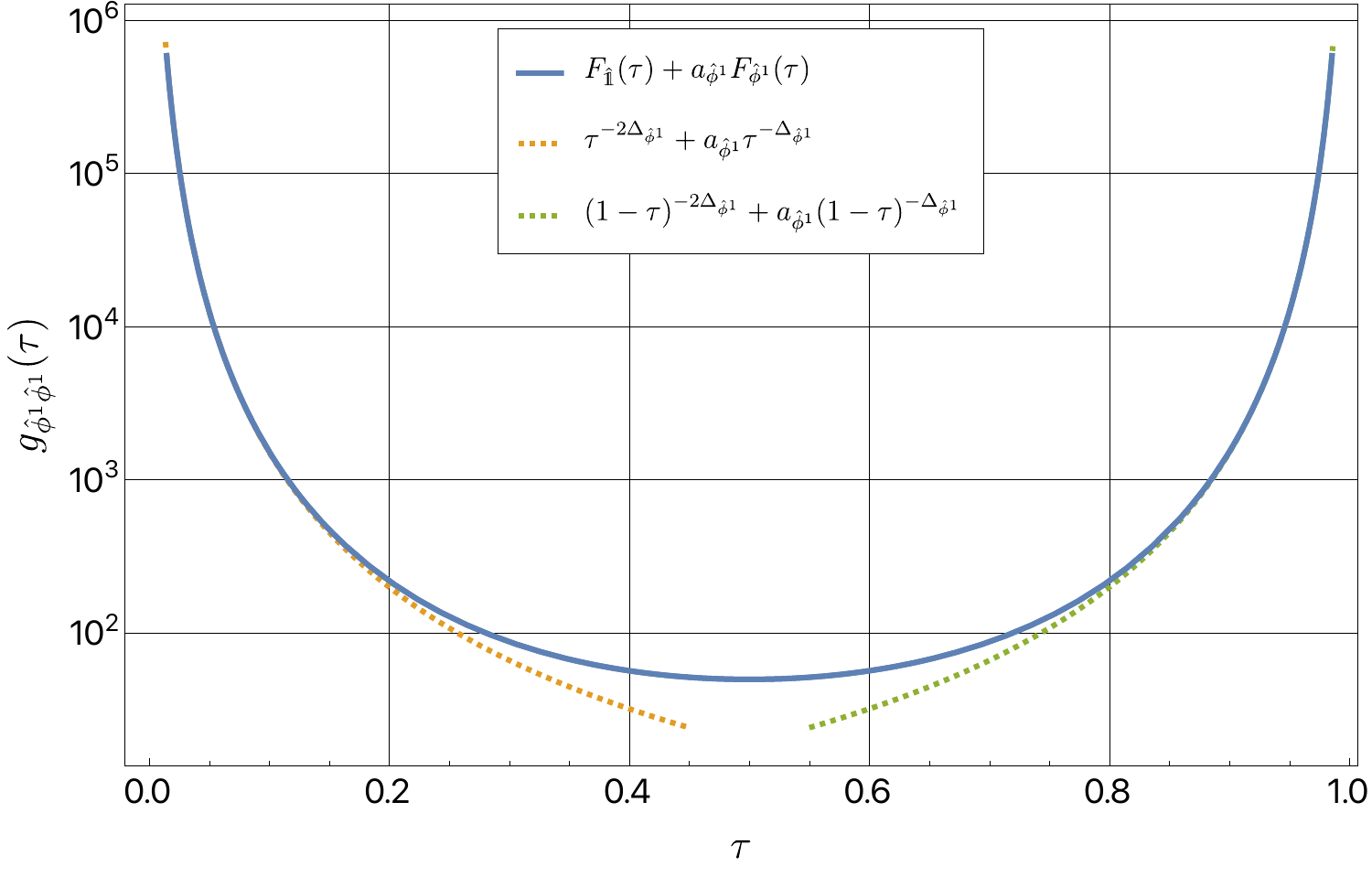}
    \caption{Logarithmic plot of the truncated periodic OPE $g_{\smash{\hphi^1\hphi^1}} (\tau) \sim F_{\hat{\mathds{1}}} + a_{\smash{\hphi^1}} F_{\smash{\hphi^1}} (\tau) + c_{\smash{\hphi^1}}$ for $c_{\smash{\hphi^1}}=0$ and $\beta=1$ (solid blue line).
    The expansions in ordinary thermal blocks for the $s$- and $t$-channels are shown as orange and green dashed lines.
    We conjecture that the truncated OPE interpolates well between these channels, up to a possible constant.}
    \label{fig:gPhiPhi_Ising_3d}
\end{figure}

We use the Padé estimate for $a_{\smash{\hphi^1}}$ to propose an approximation for the Ising magnetic-impurity correlator,
\begin{equation}
    g_{\smash{\hphi^1\hphi^1}}(\tau)
    \sim
    \frac{1}{\beta^{\smash{2\Dh_{\smash{\hphi^1}}}}}
    \left[
        F_{\hat{\mathds{1}}} (\tau)
        +
        a_{\smash{\hphi^1}}
        F_{\smash{\hphi^1}} (\tau)
        +
        c_{\smash{\hphi^1}}
    \right]\,,
    \label{eq:Ising_TruncatedPeriodicOPE}
\end{equation}
where we use the dimension determined by a similar [2/1] Padé interpolation in~\cite{Cuomo:2021kfm}, $\Dh_{\smash{\hphi^1}} \sim 1.57$.
This truncation incorporates the identity and lowest-singlet OPE singularities at both ends of the thermal circle while satisfying KMS symmetry exactly.
The dimensionless constant $c_{\smash{\hphi^1}}$ is not fixed by these inputs.
Figure~\ref{fig:gPhiPhi_Ising_3d} shows the approximation setting $c_{\smash{\hphi^1}} = 0$, together with its two-term ordinary OPE in each channel.

In order to test the accuracy of this proposal, we compute in the following section $g_{tt}(\tau)$ and $g_{\smash{\hphi^1\hphi^1}}(\tau)$ numerically at large $N$ and compare the output to the truncated OPE.
Anticipating the results presented in Section~\ref{subsec:LargeNDefectCorrelators}, we find strikingly good agreement between "exact" results and the truncated OPE.
This suggests a possible interpretation in which the heavy thermal OPE contributions approach free behavior, at least in an averaged sense, with the dominant scalar double-twist families becoming effectively classical at large $\Dh$.
A concrete illustration is provided by the large-$N$ trajectories $[tt]_{n,\pm}$: their anomalous dimensions, given in~\eqref{eq:LargeN_AnomalousDims}, vanish as $n\to\infty$, so that both families approach the classical dimensions $2+2n$.
Such a \textit{liberation} mechanism is understood for bulk theories in the large spin limit~\cite{Fitzpatrick:2012yx,Komargodski:2012ek}, however in the present situation the parameter is $\Dh$ and all the operators are scalars.
This picture is reminiscent of analytic bootstrap results in one-dimensional CFTs, where crossing symmetry constrains the averaged large-dimension OPE data, with generalized-free theories providing natural reference spectra~\cite{Mazac:2018mdx,Qiao:2017xif}.
Whether the combined thermal OPE coefficients of double-twist operators approach free values remains to be established; the agreement of the truncated correlators alone does not prove this stronger statement.
Finally, Monte Carlo simulations of the three-dimensional Ising model with a pinning-field line defect would provide a valuable independent test
of our predictions and could determine the remaining additive constant.

%% file: sections/LargeN.tex
\section{Large $N$}
\label{sec:LargeN}

In this section, we consider the magnetic impurity in the $\mathrm{O}(N)$ model at large $N$ in thermal geometries.
In order to derive defect OPE data and correlators, we first solve numerically the appropriate system of equations for the bulk profiles $\vev{s(x)}$ and $\vev{\phi^1 (x)}$ in presence of the defect.
We obtain a numerical estimation of the defect thermal one-point functions $\vev{\hphi^1}$ and $\vev{s_-}$ for $2 < d < 4$, and adapt the method to the finite-volume geometry $S_\beta^1 \times S_R^{d-1}$.
Finally, we obtain numerically the defect correlators $g_{tt} (\tau)$ and $g_{\smash{\hphi^1 \hphi^1}} (\tau)$ and compare them to the truncated periodic OPE.

\subsection{Bulk profiles and thermal OPE data}
\label{subsec:BulkProfilesAndOPEData}

Our first step consists in determining the bulk profiles $\vev{s(x)}$ and $\vev{\phi^1 (x)}$ in thermal geometries, extending the zero-temperature results of~\cite{Cuomo:2021kfm,Gimenez-Grau:2022czc}.
We describe the saddle at finite temperature and infinite volume, calculate numerically the profiles and extract the OPE data for the low-lying defect operators $\hphi^1 (\tau)$ and $s_- (\tau)$.
We then extend the results to the finite-volume geometry.

\subsubsection{Finite-temperature saddle at infinite volume}
\label{subsubsec:FiniteTemperatureSaddleAtInfiniteVolume}

In this section, we discuss how the system described by~\eqref{eq:ZeroT_LargeN_Action} behaves when the temperature is non-vanishing.

\paragraph{Thermal bulk profiles and transverse propagator.}
At zero temperature, the bulk profiles $\vev{s(x)}$ and $\vev{\phi^1 (x)}$ acquire a non-trivial dependence on the transverse distance $r = |\bm{x}_\perp|$ between the bulk operator and the defect.
We express the profiles as\footnote{Note that the large $N$ definition of $F_\phi(z)$ absorbs a factor $1/\sqrt{N}$.}
\begin{align}
    \vev{s(x)}
    &=
    \frac{F_s(z)}{\beta^2}\,,
    \label{eq:Fs_Definition} \\
    \frac{\vev{\phi^1(x)}}{\sqrt{N}}
    &=
    \frac{F_\phi(z)}{\beta^{\Delta_\phi}}\,,
    \label{eq:Fphi_Definition} \\
    \vev{\phi^a(x)}
    &=
    0\,,
    \label{eq:LargeN_phiAIsZero}
\end{align}
where we defined the dimensionless variable
\begin{equation}
    z
    =
    \frac{r}{\beta}\,.
    \label{eq:z_Definition}
\end{equation}
The transverse propagator,
\begin{equation}
    \vev{\phi^a (x) \phi^b (x')}
    =
    \delta^{ab} G_\perp (x,x')\,,
    \label{eq:Gperp_Definition}
\end{equation}
satisfies the equation of motion
\begin{equation}
    \left[
    - \partial_\tau^2 - \nabla_{\bm{x}_\perp}^2 + \frac{F_s (z)}{\beta^2}
    \right]
    G_\perp(x,x')
    =
    \delta (\tau - \tau')
    \delta^{(d-1)}(\bm{x}_\perp - \bm{x}'_\perp)\,,
    \label{eq:LargeN_GreenEquation}
\end{equation}
which relates $G_\perp (x,x')$ to $F_s (z)$.
Here the $\delta$-function involving the time coordinate is understood in the periodic sense:
\begin{equation}
    \delta (\tau)
    =
    \sum_{m= - \infty}^\infty \delta(\tau+m\beta)\,.
    \label{eq:deltaTemporal_Definition}
\end{equation}

\paragraph{Mean field and gap equations.}
At finite temperature, the mean-field and gap equations~\eqref{eq:LargeN_Fphi} and~\eqref{eq:LargeN_Gap} become
\begin{align}
    F_\phi''(z)
    +\frac{d-2}{z}F_\phi'(z)
    -F_s(z)F_\phi(z)
    &=0\,,
    \label{eq:LargeN_MeanFieldEquation}
    \\
    F_\phi(z)^2
    +\beta^{d-2} G_\perp (x,x)
    &=0\,.
    \label{eq:LargeN_GapEquation}
\end{align}
Here the first equation relates $F_s$ and $F_\phi$, while the second one fixes $F_\phi$ in terms of $G_\perp$ at coincidence.
Thus, the relations~\eqref{eq:LargeN_GreenEquation},~\eqref{eq:LargeN_MeanFieldEquation}, and~\eqref{eq:LargeN_GapEquation} fix the problem uniquely once appropriate boundary conditions are provided.

Note that the transverse bulk propagator diverges at coincidence.
To make it finite, we define it formally as
\begin{equation}
    G_\perp (x,x)
    =
    \lim_{x'\to x}
    \left[
        G_\perp(x,x')
        -
        G_{\mathrm{bulk}}(x,x')
    \right]\,,
    \label{eq:LargeN_ThermalSubtraction}
\end{equation}
with
\begin{equation}
    G_{\mathrm{bulk}} (x,x')
    =
    \frac{1}{\beta}
    \sum_{n\in\mathbb{Z}}
    \int\frac{\mathrm{d}^{d-1}k}{(2\pi)^{d-1}}
    \frac{
        e^{i\omega_n(\tau-\tau')
        +i\mathbf{k} \cdot (\bm{x}_\perp-\bm{x}'_\perp)}
    }{
        \mathbf{k}^2+\omega_n^2+m_\text{th}^2
    }\,,
    \qquad
    \omega_n=\frac{2\pi n}{\beta}\,.
    \label{eq:LargeN_BulkPropagator}
\end{equation}
This simply amounts to subtracting the divergence that already appears in the bulk propagator without defect.

\paragraph{Near-defect boundary conditions.}
We now give the boundary conditions necessary for solving Equations~\eqref{eq:LargeN_GreenEquation},~\eqref{eq:LargeN_MeanFieldEquation}, and~\eqref{eq:LargeN_GapEquation} numerically.
To that effect, we need the thermal mass $m_\mathrm{th}$ (as a function of $d$), which we express as a dimensionless quantity $\mu_d$ through
\begin{equation}
    \mu_d = m_\mathrm{th} (d) \beta\,.
    \label{eq:mud_Definition}
\end{equation}
One can obtain an exact relation for $\mu_d$, involving a momentum-space integral~\cite{Chubukov:1993aau}:
\begin{equation}
    \int\frac{\mathrm{d}^{d-1}k}{(2\pi)^{d-1}}
    \left[
        \frac{
            \coth\!\left(\frac12\sqrt{k^2+\mu_d^2}\right)
        }{
            2\sqrt{k^2+\mu_d^2}
        }
        -\frac{1}{2k}
    \right]
    =
    0\,.
    \label{eq:LargeN_ThermalMass}
\end{equation}
This integral is finite and can be computed for arbitrary choices of $2 < d < 4$.
At $d=3$, it famously yields
\begin{equation}
    \mu_{d=3}
    =
    2\log\!\left(\frac{1 + \sqrt{5}}{2}\right)
    \sim
    0.9624236501\,.
    \label{eq:LargeN_ThermalMass_3d}
\end{equation}
The boundary conditions can then be summarized as
\begin{align}
    &z\to 0:
    &F_s(z)
    &\sim\frac{(d-2)(4-d)}{4z^2}
    &F_\phi(z)
    &\sim
    \frac{\mu_{\phi\hat{\mathds{1}}}}{\sqrt{N}} z^{-(d-2)/2}\,,
    \label{eq:LargeN_UVBoundaryConditions}
    \\
    &z\to\infty:
    &F_s(z)
    &\longrightarrow\mu_d^2,
    &
    F_\phi(z)
    &\sim
    c_{\mathrm{IR}}\,
    z^{-(d-3)/2}
    K_{(d-3)/2}(\mu_d z)\,.
    \label{eq:LargeN_IRBoundaryConditions}
\end{align}
The first line follows from~\eqref{eq:ZeroT_LargeN_ConformalSymmetry}.
The large-$z$ behavior is determined from thermal bulk physics.
The Bessel function follows by replacing $F_s(z)$ in Equation~\eqref{eq:LargeN_MeanFieldEquation} by $\mu_d^2$, in which case it can be solved explicitly up to the multiplicative factor $c_{\mathrm{IR}}$.
For $d=3$, these boundary conditions become
\begin{align}
    &z\to 0:
    &F_s(z) &\sim \frac{1}{4z^2},
    &F_\phi(z) &\sim - \frac{0.210744}{\sqrt z}\,,
    \\
    &z\to\infty:
    &F_s(z)&\longrightarrow\mu_3^2,
    &F_\phi(z)&\sim c_{\mathrm{IR}}K_0(\mu_3 z)\,.
    \label{eq:LargeN_BoundaryConditions3d}
\end{align}

\subsubsection{Numerical solution and OPE extraction}
\label{subsubsec:NumericalSolutionAndOPEExtraction}

We now discuss how to solve Equations~\eqref{eq:LargeN_GreenEquation},~\eqref{eq:LargeN_MeanFieldEquation}, and~\eqref{eq:LargeN_GapEquation}, together with their boundary conditions~\eqref{eq:LargeN_UVBoundaryConditions}-\eqref{eq:LargeN_IRBoundaryConditions}, numerically.
Our goal is to obtain the bulk profiles $F_s (z)$ and $F_\phi (z)$ to high accuracy, such that we can extract the OPE coefficients associated to the defect operators $\hphi^1 (\tau)$ and $s_- (\tau)$.

\begin{figure}
    \centering
    \begin{minipage}{.5\textwidth}
        \centering
        \includegraphics[width=0.95\linewidth]{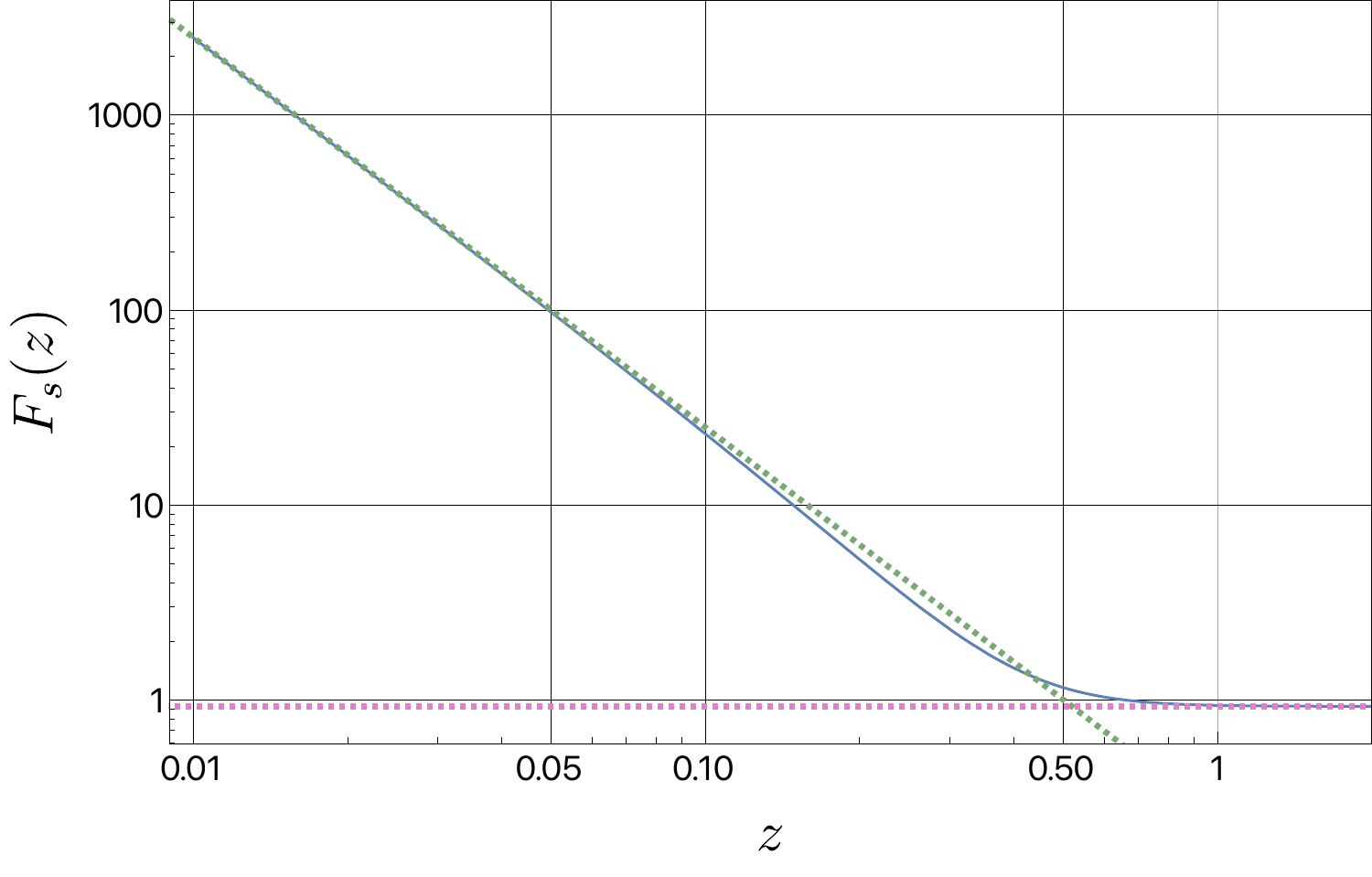}
    \end{minipage}%
    \begin{minipage}{0.5\textwidth}
        \centering
        \includegraphics[width=0.95\linewidth]{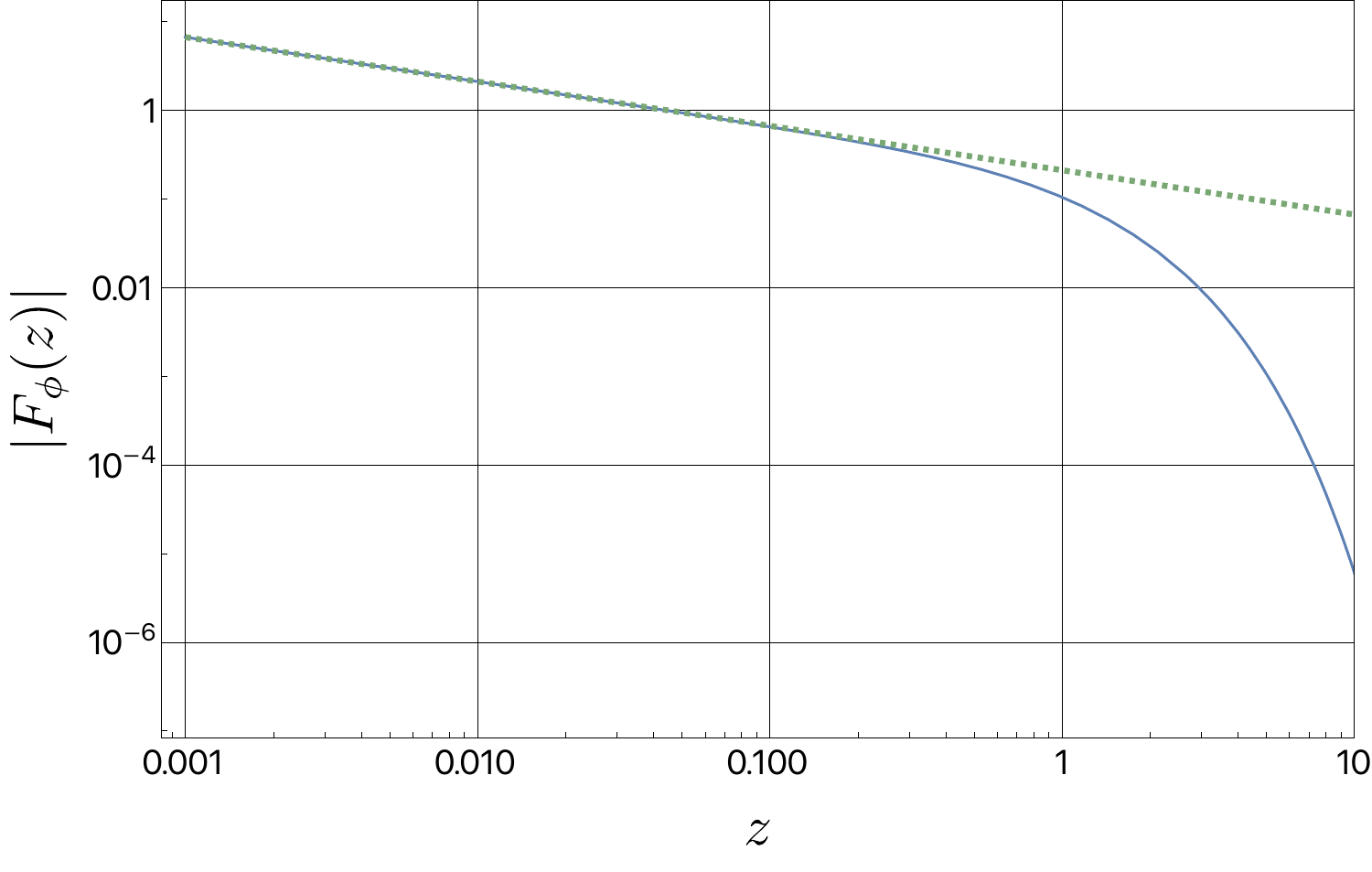}
    \end{minipage}
    \caption{
    Logarithmic plot of the bulk profiles $F_s(z)$ (\textbf{left}) and $|F_\phi(z)|$ (\textbf{right}) at $d=3$.
    The solid lines represent the numerical solution to the system of equations formed by~\eqref{eq:LargeN_GreenEquation},~\eqref{eq:LargeN_MeanFieldEquation}, and~\eqref{eq:LargeN_GapEquation}, while the dashed lines show the asymptotic behaviors~\eqref{eq:LargeN_UVBoundaryConditions}-\eqref{eq:LargeN_IRBoundaryConditions} that were used as boundary conditions by the solver.
    }
    \label{fig:LargeN_BulkProfiles}
\end{figure}

\paragraph{Near-defect expansion and numerical strategy.}
Our strategy will be to extend the boundary conditions~\eqref{eq:LargeN_UVBoundaryConditions} such that the UV behavior of the bulk profiles is fixed to be
\begin{align}
    z^2 F_s(z)
    &=
    \frac{(d-2)(4-d)}{4}
    +
    \mu_{\smash{s \hphi^1}} b_{\smash{\hphi^1}} z^{\smash{\Dh_{\smash{\hphi^1}}}}
    +
    \mu_{\smash{s\, s_-}} b_{\smash{s_-}} z^{\smash{\Dh_{\smash{s_-}}}}
    +
    \mu_{\smash{s\, s_+}} b_{\smash{s_+}} z^{\smash{\Dh_{\smash{s_+}}}}
    +
    \ldots\,,
    \label{eq:LargeN_ProfileOPEs}
    \\
    z^{\Delta_\phi} \sqrt{N} F_\phi(z)
    &=
    \mu_{\smash{\phi\hat{\mathds{1}}}}
    +
    \mu_{\smash{\phi \hphi^1}} b_{\smash{\hphi^1}} z^{\smash{\Dh_{\smash{\hphi^1}}}}
    +
    \mu_{\smash{\phi\, s_-}} b_{\smash{s_-}} z^{\smash{\Dh_{\smash{s_-}}}}
    +
    \mu_{\smash{\phi\, s_+}} b_{\smash{s_+}} z^{\smash{\Dh_{\smash{s_+}}}}
    +
    \ldots\,,
    \label{eq:LargeN_ProfileOPEphi}
\end{align}
Equation~\eqref{eq:LargeN_MeanFieldEquation} readily fixes the coefficients of~\eqref{eq:LargeN_ProfileOPEphi} in terms of the coefficients appearing in~\eqref{eq:LargeN_ProfileOPEs}:
\begin{align}
    \mu_{\smash{\phi\, \Oh}}
    &=
    \frac{\mu_{\smash{\phi\hat{\mathds{1}}}} \mu_{\smash{s\, \Oh}}}{\Dh_\Oh (\Dh_\Oh - 1)}\,,
    \qquad
    \text{with } \Oh\in \lbrace \hphi^1,s_- \rbrace\,,
    \label{eq:LargeN_LinearProfileRelation}
    \\
    \mu_{\smash{\phi\, s_+}} b_{\smash{s_+}}
    &=
    \frac{\mu_{\smash{\phi\hat{\mathds{1}}}}}
    {\Dh_{s_+}(\Dh_{s_+}-1)}
    \left(
        \mu_{\smash{s\, s_+}} b_{\smash{s_+}}
        +\frac{\mu_{\smash{s\, \hphi^1}}^2 b_{\smash{\hphi^1}}^2}
        {\Dh_{\smash{\hphi^1}}(\Dh_{\smash{\hphi^1}}-1)}
    \right)\,,
    \label{eq:LargeN_QuadraticProfileRelation} \\
    &\, \vdots \notag
\end{align}
Furthermore, by identifying $s_+$ as the classical double-twist $[\hphi^1\hphi^1]_0$ at strict large $N$, we can replace its dimension and one-point function in~\eqref{eq:LargeN_ProfileOPEs}-\eqref{eq:LargeN_ProfileOPEphi} by
\begin{equation}
    \Dh_{s_+}
    =
    2\Dh_{\smash{\hphi^1}}\,,
    \qquad
    b_{s_+}
    =
    \frac{b_{\smash{\hphi^1}}^2}{\sqrt2}\,.
    \label{eq:LargeN_ThermalFactorization}
\end{equation}

We now wish to turn Green's equation~\eqref{eq:LargeN_GreenEquation} into a one-dimensional problem.
We expand the transverse bulk propagator in Matsubara modes:
\begin{equation}
    G_\perp(x,x')
    =
    \frac{1}{\beta^{d-2}}
    \sum_{n= - \infty}^\infty
    e^{2\pi i n(\tau-\tau')/\beta}
    \sum_{\ell=0}^{\infty}\sum_m
    Y_{\ell m}(\Omega)\,
    Y_{\ell m}^{*}(\Omega')\,
    \gamma_{n\ell}(z,z')\,,
    \label{eq:LargeN_MatsubaraDecomposition}
\end{equation}
with $\Omega,\Omega'\in S^{d-2}$, and
\begin{equation}
    \int_{S^{d-2}}\!\mathrm d\Omega\,
    Y_{\ell m}^{*}(\Omega)Y_{\ell' m'}(\Omega)
    =\delta_{\ell\ell'}\delta_{mm'}\,.
    \label{eq:SphericalHarmonics}
\end{equation}
Equation~\eqref{eq:LargeN_GreenEquation} then simplifies to
\begin{equation}
    \left[
        -\partial_z^2-\frac{d-2}{z}\partial_z
        +(2\pi n)^2
        +\frac{\ell(\ell+d-3)}{z^2}
        +F_s(z)
    \right]\gamma_{n\ell}(z,z')
    =
    \frac{\delta(z-z')}{z^{d-2}}\,,
    \label{eq:LargeN_RadialGreenEquation}
\end{equation}
at the price of having introduced the indices $n$, $\ell$.
This procedure however allows us to introduce a cutoff for these indices and approximate the large-$n,\ell$ modes using a WKB approximation, as we describe in Appendix~\ref{app:LargeN_Numerics}.

\begin{table}[t]
    \centering
    \caption{
        Thermal one-point coefficients extracted from the large-$N$ saddle at infinite spatial volume for $2 < d < 4$.
        The analytically known cases $d=2,4$ are not included.
        The quoted uncertainties are empirical numerical sensitivity allowances, not rigorous error bounds.
    }
    \begin{tabular}{c|cc}
        \hline
        $d$
        & $b_{\hphi^1}/\sqrt N$
        & $b_{s_-}/\sqrt N$
        \\
        \hline
        $2.2$ & $2.19 \pm 0.05$ & $-1.18 \pm 0.06$ \\
        $2.4$ & $1.82 \pm 0.02$ & $-0.78 \pm 0.06$ \\
        $2.6$ & $1.544 \pm 0.009$ & $-0.39 \pm 0.05$ \\
        $2.8$ & $1.319 \pm 0.004$ & $\phantom{-}0.00 \pm 0.05$ \\
        $3.0$ & $1.126 \pm 0.003$ & $\phantom{-}0.32 \pm 0.05$ \\
        $3.2$ & $0.951 \pm 0.002$ & $\phantom{-}0.57 \pm 0.05$ \\
        $3.4$ & $0.7864 \pm 0.0010$ & $\phantom{-}0.75 \pm 0.05$ \\
        $3.6$ & $0.6204 \pm 0.0006$ & $\phantom{-}0.91 \pm 0.05$ \\
        $3.8$ & $0.4347 \pm 0.0004$ & $\phantom{-}1.11 \pm 0.05$ \\
        \hline
    \end{tabular}
    \label{tab:LargeN_ThermalCoefficientsAcrossDimensions}
\end{table}

Note that, in this notation, the coincident transverse propagator takes the form
\begin{equation}
    G_\perp (x,x)
    =
    \frac{1}{\beta^{d-2} \operatorname{Vol}(S^{d-2})}
    \sum_{n\in\mathbb Z} \sum_{\ell=0}^{\infty}
    D_\ell
    \left[
        \gamma_{n\ell}(z,z)
        -z^{3-d}I_{\nu_\ell}(k_nz)K_{\nu_\ell}(k_nz)
    \right]\,,
    \label{eq:LargeN_ModeSubtraction}
\end{equation}
where we used the angular degeneracy $D_\ell$ defined in~\eqref{eq:ZeroT_LargeN_Definitions}, and
\begin{equation}
    \nu_\ell=\ell+\frac{d-3}{2}\,,
    \qquad
    k_n^2=(2\pi n)^2+\mu_d^2\,.
    \label{eq:LargeN_SomeDefintions}
\end{equation}

\paragraph{Bulk profiles and OPE data in three dimensions.}
We start by solving Equations~\eqref{eq:LargeN_GreenEquation},~\eqref{eq:LargeN_MeanFieldEquation}, and~\eqref{eq:LargeN_GapEquation} numerically for the case $d=3$.
The dimensions of the lowest-lying operators in the OPE Ansatz~\eqref{eq:LargeN_ProfileOPEs} are
\begin{align}
    \Dh_{\smash{\hphi^1}} &\sim 1.5417281\,, \label{eq:LargeN_Dh_hphi1_3d} \\
    \Dh_{\smash{s_-}} &\sim 2.5564691\,, \label{eq:LargeN_Dh_s-_3d} \\
    \Dh_{\smash{s_+}} &\sim 3.0834562\,. \label{eq:LargeN_Dh_s+_3d}
\end{align}
Details about the numerical solver are given in Appendix~\ref{app:LargeN_Numerics}.
The resulting bulk profiles are presented in Figure~\ref{fig:LargeN_BulkProfiles}.

We now extract the OPE data for the low-lying operators.
In order to do so, we need the associated bulk-defect two-point functions $\mu_{\smash{s \Oh}}$, which can be determined from~\cite{Cuomo:2021kfm} to be

\begin{equation}
    \left. \mu_{\smash{s\, \hphi^1}} \right|_{d=3}
    \sim -\frac{0.6475485}{\sqrt{N}}\,,
    \qquad
    \left. \mu_{\smash{s\, s_-}} \right|_{d=3}
    \sim \frac{1.268576}{\sqrt{N}}\,.
    \label{eq:LargeN_BulkDefectCoefficients3d}
\end{equation}
The numerical profiles yield the following values for the thermal one-point functions;
\begin{align}
    \left. b_{\smash{\hphi^1}} \right|_{d=3}
    &=
    (1.126 \pm 0.003) \sqrt{N}\,, \label{eq:LargeN_b_hphi_3d} \\
    \left. b_{\smash{s_-}} \right|_{d=3}
    &=
    (0.32 \pm 0.05) \sqrt{N}\,. \label{eq:LargeN_b_s-_3d}
\end{align}
The uncertainties provided here are not rigorous and estimated by varying the parameters (see Appendix~\ref{app:LargeN_Numerics} for more details).
The factorization formula~\eqref{eq:LargeN_ThermalFactorization} for the operator $s_+$ gives
\begin{equation}
    \left. b_{\smash{s_+}} \right|_{d=3}
    =
    (0.896 \pm 0.005) N.
    \label{eq:LargeN_DoubleTraceCoefficient_3d}
\end{equation}

\begin{figure}
    \centering
    \begin{minipage}{.5\textwidth}
        \centering
        \includegraphics[width=0.95\linewidth]{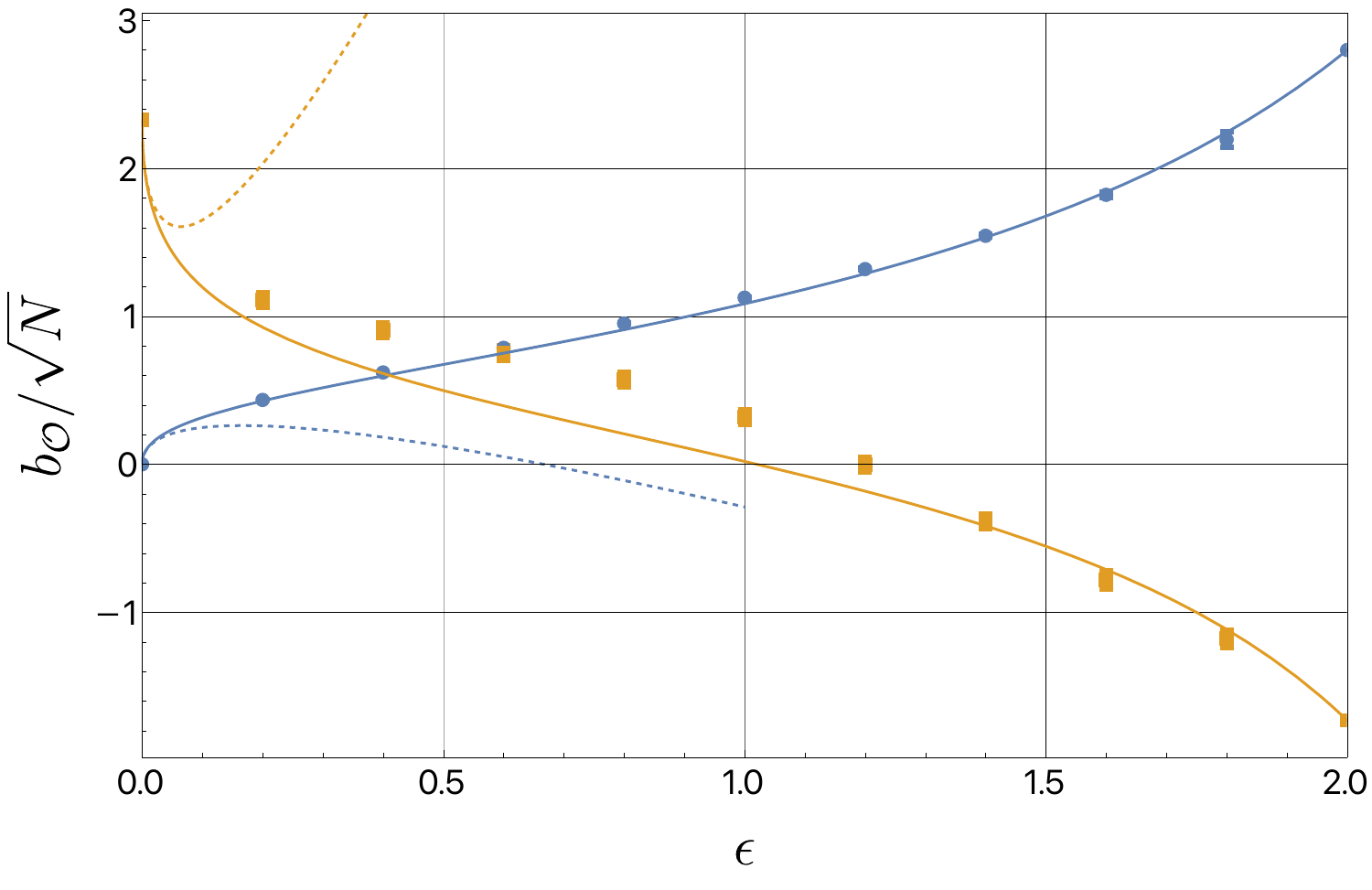}
    \end{minipage}%
    \begin{minipage}{0.5\textwidth}
        \centering
        \includegraphics[width=0.95\linewidth]{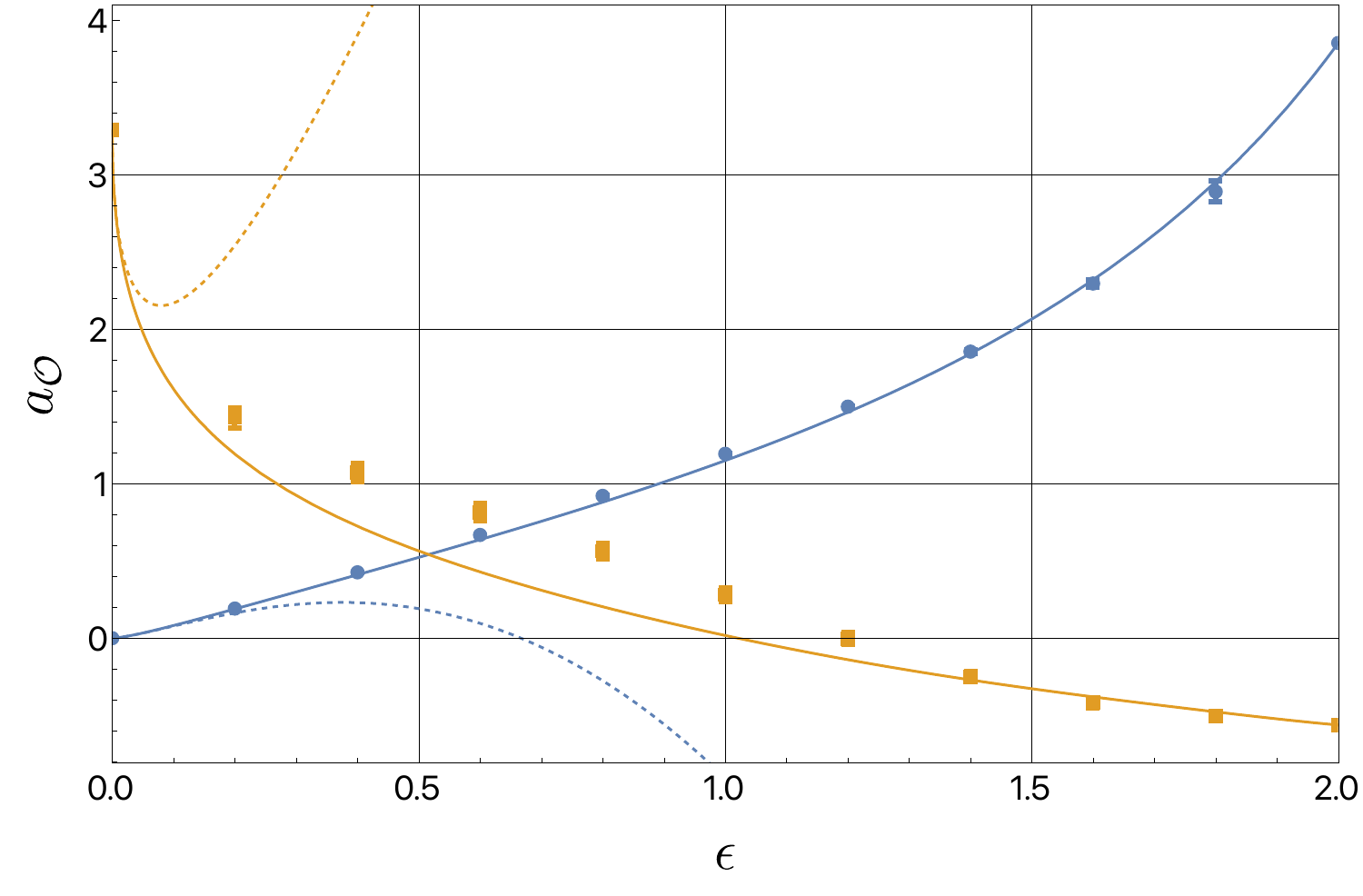}
    \end{minipage}
    \caption{\textbf{Left}: Plot of the thermal one-point functions $b_{\smash{\hphi^1}}/\sqrt{N}$ (blue) and $b_{\smash{s_-}}/\sqrt{N}$ (orange) at large $N$ for $2 \leq d = 4-\veps \leq 4$ in steps of $0.2$.
    The dots correspond to numerical solutions of the system~\eqref{eq:LargeN_GreenEquation},~\eqref{eq:LargeN_MeanFieldEquation}, and~\eqref{eq:LargeN_GapEquation} with OPE Ansatz~\eqref{eq:LargeN_ProfileOPEs}.
    The errors are not rigorous and estimated by varying the parameters.
    The dashed lines represent the $\veps$-expansion calculations provided in~\eqref{eq:bPhi_FinalResult}, \eqref{eq:bspm_ThermalExpansion}, and~\eqref{eq:aPhi_spm_Prediction_OneLoop}, while the solid lines are Padé interpolations between the $\veps$-expansion result and the $d=2$ theory.
    Note that the coefficient $b_{s_-}^{(1)}$ is determined from~\eqref{eq:aPhi_spm_Prediction_OneLoop} by using the large $N$ factorization~\eqref{eq:LargeN_ThermalFactorization} for $b_{s_+}$.
    The Padé interpolation performs well for $b_{\smash{\hphi^1}}$ but seems less accurate for $b_{\smash{s_-}}$.
    \textbf{Right}: Same plot but for the OPE coefficients $a_{\smash{\hphi^1}}$ (blue) and $a_{\smash{s_-}}$ (orange) after accounting for the large $N$ three-point functions, which are calculated in Appendix~\ref{app:LargeN_WittenDiagrams}.
    The same observations hold for the accuracy of the Padé interpolation.}
    \label{fig:LargeN_AcrossDims}
\end{figure}

\paragraph{OPE data in arbitrary dimensions.}
The results above can be extended by solving~\eqref{eq:LargeN_GreenEquation}, \eqref{eq:LargeN_MeanFieldEquation}, and~\eqref{eq:LargeN_GapEquation} numerically at arbitrary $2 < d < 4$.
The obtained values for the thermal one-point functions $b_{\smash{\hphi^1}}$ and $b_{\smash{s_-}}$ are presented in Table~\ref{tab:LargeN_ThermalCoefficientsAcrossDimensions} and plotted against $\veps$-expansion analytic predictions in Figure~\ref{fig:LargeN_AcrossDims}.

\subsubsection{Antipodal configuration}
\label{subsubsec:AntipodalConfiguration}

We now extend the results of the previous sections to the finite-volume geometry $S_\beta^1 \times S_R^{d-1}$.
Our focus is to provide a numerical evaluation of the thermal one-point function coefficient $b_{\smash{\hphi^1}}$ for different values of $\beta/R$ at $d=3$.
The method presented here can easily be generalized to arbitrary $d$.

\paragraph{Geometry and saddle equations.}
At finite volume, the metric is given by
\begin{equation}
    \mathrm{d}s^2
    =
    \mathrm{d}\tau^2
    +
    R^2\left(
        \mathrm{d}\theta^2
        +
        \sin^2(\theta)\,\mathrm{d}\Omega_{d-2}^2
    \right)\,.
    \label{eq:LargeN_FiniteV_Metric}
\end{equation}
The antipodal configuration corresponds to
\begin{equation}
    S_{\mathrm{def}}
    =
    \sqrt{N}\, h \int_0^\beta\mathrm d\tau\,
    \left[
        \phi^1(\tau, N)
        +\phi^1(\tau, S)
    \right]\,,
    \label{eq:LargeN_FiniteV_Action}
\end{equation}
where $N$ refers to $\theta=0$ (North) and $S$ to $\theta=\pi$ (South).
We then define the bulk profiles through
\begin{align}
    \vev{s(\theta)}
    &=
    \frac{F_s(\theta)}{\beta^2}\,, \\
    \frac{\vev{\phi^1(\theta)}}{\sqrt N}
    &=
    \frac{F_\phi(\theta)}{\beta^{\Delta_\phi}}\,.
    \label{eq:LargeN_FiniteV_Profiles}
\end{align}
Equations~\eqref{eq:LargeN_GreenEquation},~\eqref{eq:LargeN_MeanFieldEquation}, and~\eqref{eq:LargeN_GapEquation} become
\begin{align}
    \left[
        \partial_\theta^2
        +(d-2)\cot(\theta)\,\partial_\theta
        -\frac{(d-2)^2}{4}
        -\frac{R^2}{\beta^2}F_s(\theta)
    \right]F_\phi(\theta)
    &=0\,,
    \label{eq:LargeN_FiniteV_MeanField} \\
    \left[
        -\partial_\tau^2
        -\frac{1}{R^2}\nabla_{S^{d-1}}^2
        +\frac{(d-2)^2}{4R^2}
        +\frac{F_s(\theta)}{\beta^2}
    \right]G_\perp(x,x')
    &=
    \frac{\delta (\tau-\tau') \delta (\theta-\theta')\,
    \delta (\Omega,\Omega')}{R^{d-1}\sin^{d-2}(\theta)}\,,
    \label{eq:LargeN_FiniteV_GreenEquation} \\
    F_\phi(\theta)^2
    +\beta^{d-2}G_\perp (x,x)
    &=0\,.
    \label{eq:LargeN_FiniteV_GapEquation}
\end{align}

\begin{table}
    \centering
    \caption{
        Thermal one-point coefficient $b_{\smash{\hphi^1}}(q)/\sqrt{N}$ at large $N$ and $d=3$ in the antipodal configuration, with $q$ defined in~\eqref{eq:q_Definition}.
        The quoted uncertainties are empirical numerical sensitivity allowances, not rigorous error bounds.
    }
    \begin{tabular}{c c}
        \hline
        $\beta/R$ & $b_{\smash{\hphi^1}}(q)/\sqrt{N}$ \\
        \hline
        $0.25$ & $1.126 \pm 0.003$ \\
        $0.50$ & $1.1197 \pm 0.0010$ \\
        $1$    & $0.974 \pm 0.002$ \\
        $2$    & $0.591 \pm 0.002$ \\
        $3$    & $0.331 \pm 0.003$ \\
        $4$    & $0.174 \pm 0.005$ \\
        $6$    & $0.042 \pm 0.002$ \\
        $8$    & $0.009 \pm 0.002$ \\
        $10$   & $0.002 \pm 0.002$ \\
        \hline
    \end{tabular}
    \label{tab:LargeN_bPhi_FiniteV}
\end{table}

\paragraph{Matsubara decomposition.}
At finite volume, the expansion in Matsubara modes turns into
\begin{equation}
    G_\perp(x,x')
    =
    \frac{1}{\beta R^{d-3}}
    \sum_{n\in\mathbb Z}
    e^{2\pi i n(\tau-\tau')/\beta}
    \sum_{\ell,m}
    Y_{\ell m}(\Omega)Y_{\ell m}^{*}(\Omega')
    \gamma_{n\ell}(\theta,\theta')\,,
    \label{eq:LargeN_FiniteV_Matsubara}
\end{equation}
and the counterpart of Equation~\eqref{eq:LargeN_RadialGreenEquation} for the modes $\gamma_{n\ell}(\theta,\theta')$ is
\begin{equation}
    \left[
        -\partial_\theta^2
        -(d-2)\cot(\theta)\,\partial_\theta
        +\frac{\ell(\ell+d-3)}{\sin^2(\theta)}
        +\left(\frac{2\pi nR}{\beta}\right)^2
        +\frac{(d-2)^2}{4}
        +\frac{R^2}{\beta^2}F_s(\theta)
    \right]\gamma_{n\ell}(\theta,\theta')
    =
    \frac{\delta(\theta-\theta')}{\sin^{d-2}(\theta)}\,.
    \label{eq:LargeN_FiniteV_RadialGreenEquation}
\end{equation}

\begin{figure}
    \centering
    \includegraphics[width=0.65\linewidth]{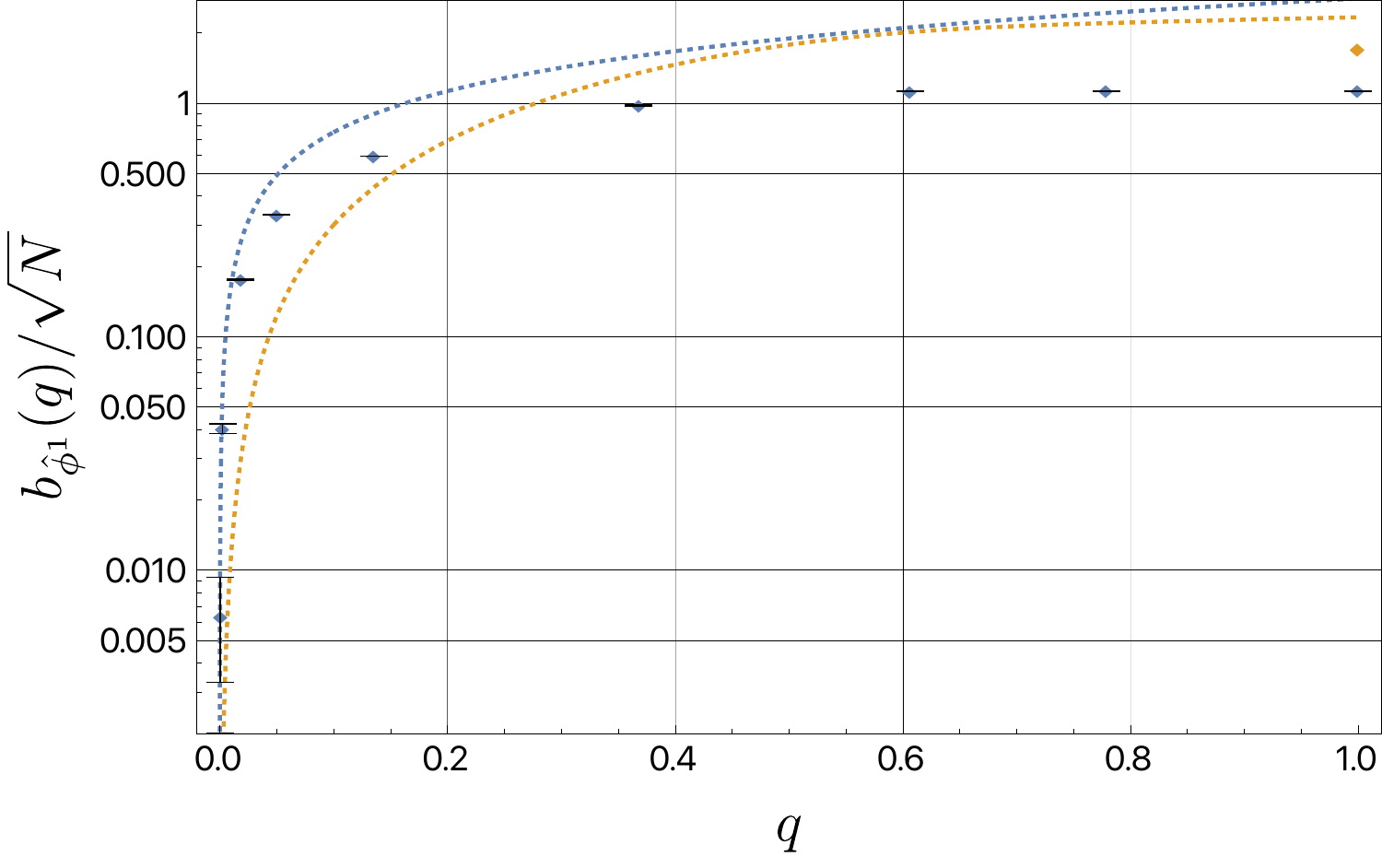}
    \caption{Logarithmic plot of the thermal one-point function $b_{\smash{\hphi^1}} (q)/\sqrt{N}$ for $N=1$ (orange) and large $N$ (blue) for the antipodal configuration at $d=3$.
    The dashed curves show Padé interpolations between $d=4-\veps$ and the $2d$ theory.
    The orange diamond shows the infinite-volume Padé interpolation, while the blue diamonds present the numerical results for large $N$, determined by solving the finite-volume equations~\eqref{eq:LargeN_FiniteV_MeanField}-\eqref{eq:LargeN_FiniteV_GapEquation} with the bulk profile Ansatz~\eqref{eq:LargeN_FiniteV_OPEs}.
    The error bars refer to numerical uncertainty in solving the equations and propagating in the coefficient.
    At large $N$, we observe that the Padé interpolation performs poorly away from the $q \sim 0$ regime (zero temperature, finite volume).
    This may reflect the need to incorporate IR resummation results into the interpolation scheme, and thus the current finite-volume Padé functions should be viewed as qualitative instead of quantitative.}
    \label{fig:LargeN_bPhi_FiniteV}
\end{figure}

\paragraph{Near-defect expansion and OPE extraction.}
We follow the same strategy as in the infinite volume case to solve the equations~\eqref{eq:LargeN_FiniteV_MeanField}-\eqref{eq:LargeN_FiniteV_GapEquation} numerically.
Using~\eqref{eq:ThermalBulkOnePointFunctions_FiniteVolume}, we consider the following OPE Ansatz for the bulk profile $F_s (\theta)$:
\begin{align}
    \left(\frac{R\sin(\theta)}{\beta}\right)^2
    F_s(\theta)
    &=
    \frac{(d-2)(4-d)}{4} \notag \\
    &\phantom{=\ }
    +
    \sum_{\Oh \ne \hat{\mathds{1}}}
    \mu_{s\Oh}\,
    b_{\Oh}(q)
    \left(\frac{R\sin(\theta)}{\beta}\right)^{
        \Dh_{\Oh}}
    {}_2F_1\left(
        \frac{\Dh_{\Oh}}{2},
        \frac{\Dh_{\Oh}}{2};
        \Dh_{\Oh}+\frac{1}{2};
        \sin^2(\theta)
    \right)\,,
    \label{eq:LargeN_FiniteV_OPEs}
\end{align}
and similarly for $F_\phi(\theta)$.
The boundary conditions~\eqref{eq:LargeN_UVBoundaryConditions}-\eqref{eq:LargeN_IRBoundaryConditions} can be adapted to the finite-volume case by replacing the large-$z$ behavior with the conditions
\begin{equation}
    \partial_\theta F_s(\pi/2)
    =
    \partial_\theta F_\phi(\pi/2)
    =
    0\,.
    \label{eq:LargeN_FiniteV_BoundaryConditions}
\end{equation}
Following the same procedure as in Section~\ref{subsubsec:NumericalSolutionAndOPEExtraction}, we find the values for the OPE coefficient $b_{\smash{\hphi^1}}$ shown in Table~\ref{tab:LargeN_bPhi_FiniteV} and plotted in Figure~\ref{fig:LargeN_bPhi_FiniteV}.

\subsection{Defect correlators}
\label{subsec:LargeNDefectCorrelators}

We now extract the defect correlators $g_{tt} (\tau)$ and $g_{\smash{\hphi^1 \hphi^1}} (\tau)$ numerically and compare the results to the truncated periodic OPE.
We focus on the infinite-volume case for clarity, although the method can also be used at finite volume.

\subsubsection{Tilt correlator $g_{tt} (\tau)$}
\label{subsubsec:TiltCorrelator}

\begin{figure}
    \centering
    \begin{minipage}{.5\textwidth}
        \centering
        \includegraphics[width=0.95\linewidth]{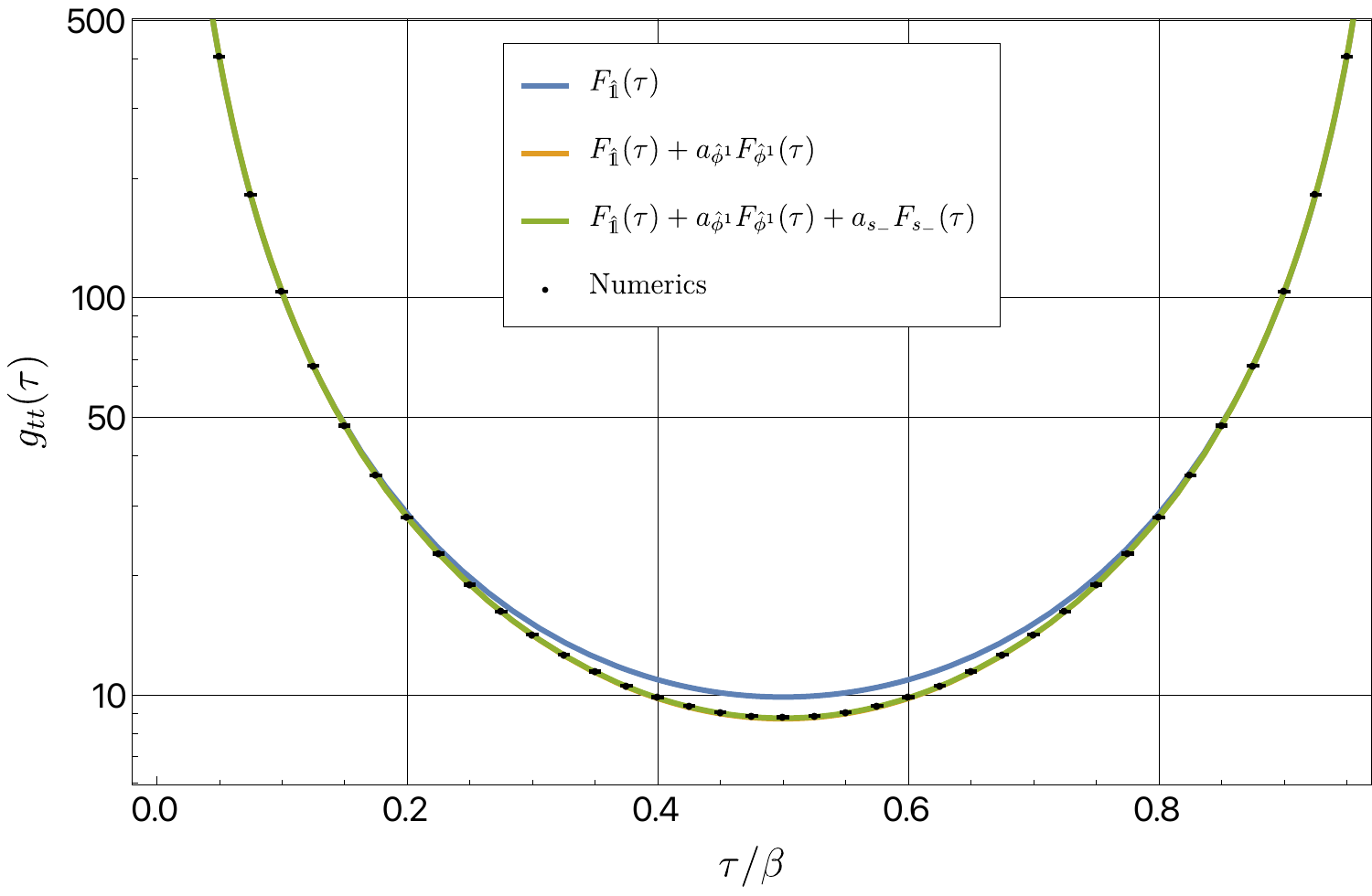}
    \end{minipage}%
    \begin{minipage}{0.5\textwidth}
        \centering
        \includegraphics[width=0.95\linewidth]{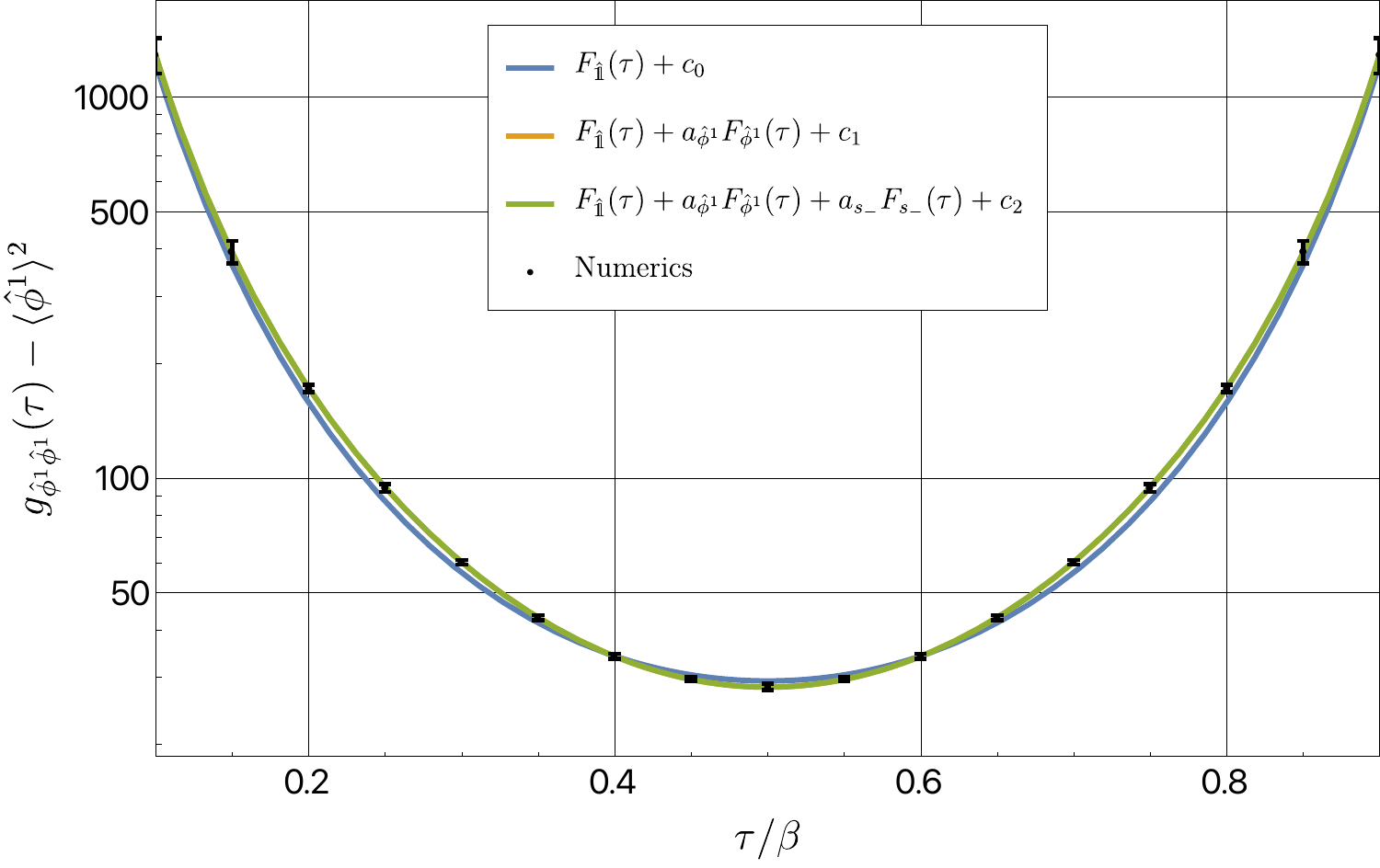}
    \end{minipage}
    \caption{\textbf{Left}: Comparison of the numerical results for $g_{tt} (\tau)$ (dots with error bars) with the truncated periodic OPE~\eqref{eq:LargeN_gtt0_Definition}--\eqref{eq:LargeN_gtt2_Definition} (solid lines).
    We observe a very good agreement once the contribution of $\hphi^1$ is incorporated.
    The curves with and without $s_-$ are indistinguishable as the improvement is minimal.
    \textbf{Right}: Same plot for the connected part of the correlator $g_{\smash{\hphi^1\hphi^1}} (\tau)$, after fitting the constants $c_j$ as prescribed in~\eqref{eq:LargeN_ghphi10_Definition}--\eqref{eq:LargeN_ghphi12_Definition} with the results~\eqref{eq:LargeN_hphi1_MaxDiscrepancy_0}-\eqref{eq:LargeN_hphi1_MaxDiscrepancy_2}.
    For this correlator, the dots are midpoints of intervals obtained from numerical uncertainties, while the error bars are the associated half-widths.
    We observe again a very good agreement once the contribution of $\hphi^1$ is taken into account.
    Note that considering the identity alone may seem more accurate than for $g_{tt} (\tau)$, however this is an artifact of fitting $c_0$; the shapes show a clear disagreement which we estimated to be of the same magnitude.}
    \label{fig:LargeN_DefectCorrelators}
\end{figure}

\paragraph{Extraction from the transverse propagator.}
We first determine the correlator of tilt operators $g_{tt} (\tau)$, defined using the conventions~\eqref{eq:gtt_Definition}.
It can be immediately extracted from the transverse propagator $G_\perp (x,x')$ by making use of the bulk-to-defect OPE~\eqref{eq:ZeroT_BulkDefectOPE} and keeping only the leading contribution:

\begin{equation}
    g_{tt} (\tau)
    =
    \frac{1}{\mu_{\phi t}^2}
    \lim_{r,r'\to0}
    (r r')^{\Delta_\phi-1}
    G_\perp (x,x')\,.
    \label{eq:LargeN_TiltBoundaryLimit}
\end{equation}
Note in particular that there is no disconnected term since thermal one-point functions $\vev{t^a}$ vanish.
The zero-temperature bulk-defect two-point function $\mu_{\phi t}$ can be determined from~\cite{Cuomo:2021kfm} to be
\begin{equation}
    \mu_{\phi t}^2
    =
    \frac{1}{\pi\operatorname{Vol}(S^{d-2})}\,.
    \label{eq:LargeN_mu_phit}
\end{equation}
The result for $g_{tt} (\tau)$ is shown in Figure~\ref{fig:LargeN_DefectCorrelators} for the case $d=3$.

\paragraph{Comparison with the periodic OPE.}
We now wish to compare the numerical $g_{tt} (\tau)$ to the truncated periodic OPE, following the insights of Section~\ref{subsec:PeriodicOPEAndAsymptotics} and with the goal of justifying the $N=1$ approximation provided in Equation~\eqref{eq:Ising_TruncatedPeriodicOPE}.
We define the following truncations:
\begin{align}
    g_{tt}^{[0]} (\tau) &= \frac{1}{\beta^2} F_{\hat{\mathds{1}}} (\tau)\,, \label{eq:LargeN_gtt0_Definition} \\
    g_{tt}^{[1]} (\tau) &= \frac{1}{\beta^2} \left( F_{\hat{\mathds{1}}} (\tau) + a_{\smash{\hphi^1}} F_{\smash{\hphi^1}} (\tau) \right)\,, \label{eq:LargeN_gtt1_Definition} \\
    g_{tt}^{[2]} (\tau) &= \frac{1}{\beta^2} \left( F_{\hat{\mathds{1}}} (\tau) + a_{\smash{\hphi^1}} F_{\smash{\hphi^1}} (\tau) + a_{\smash{s_-}} F_{\smash{s_-}} (\tau) \right)\,. \label{eq:LargeN_gtt2_Definition}
\end{align}
Using the results of Section~\ref{subsubsec:NumericalSolutionAndOPEExtraction} for the thermal one-point functions combined with the three-point functions determined in Appendix~\ref{app:LargeN_WittenDiagrams}, we find that the OPE coefficients in the truncated OPE are
\begin{equation}
    \left.a_{\hphi^1}\right|_{d=3}
    =
    1.193 \pm 0.003,
    \qquad
    \left.a_{s_-}\right|_{d=3}
    =
    0.28 \pm 0.05.
    \label{eq:LargeN_TiltOPECoefficients_3d}
\end{equation}
The curves are plotted against the numerical correlator in Figure~\ref{fig:LargeN_DefectCorrelators}.
For each case, the maximum discrepancy can be evaluated in the meaningful window for $\tau/\beta$ to be
\begin{equation}
    \max_{0.05\leq \tau/\beta \leq 0.95}
    \left|
        \frac{g_{tt}^{[j]}(\tau) - g_{tt}^{\text{(num)}}(\tau)}
             {g_{tt}^{\text{(num)}}(\tau)}
    \right|
    \sim
    \begin{cases}
        12.45 \%, & j=0,\\
        0.86 \%, & j=1,\\
        0.47 \%, & j=2.
    \end{cases}
    \label{eq:LargeN_Tilt_PeriodicOPEComparison}
\end{equation}
We observe a spectacular improvement of the agreement from $j=0$ to $j=1$, which corresponds to adding the contribution of the operator $\hphi^1 (\tau)$ on top of the identity.
Although it is not clear how this agreement continues down to $N=1$, it provides support for Equation~\eqref{eq:Ising_TruncatedPeriodicOPE} being a good approximation of the correlator.

\subsubsection{Longitudinal singlet correlator $g_{\smash{\hphi^1\hphi^1}} (\tau)$}
\label{subsubsec:LongitudinalSingletCorrelator}

\paragraph{Extraction from the longitudinal propagator.}
For the sake of completeness, we now consider the correlator of two $\hphi^1 (\tau)$ operators.
As explained in~\cite{Cuomo:2021kfm} for the zero-temperature case, this involves understanding the longitudinal degrees of freedom associated to the bulk operator $\phi^1 (x)$.
We define the fluctuations of $\phi^1 (x)$ as
\begin{equation}
    \delta\phi^1(x)
    =
    \phi^1(x)-\vev{\phi^1(x)}\,.
    \label{eq:LargeN_FluctuationsOfphi1}
\end{equation}
We also define the following shorthand notation for the longitudinal propagator:
\begin{equation}
    G_\parallel (x,x')
    =
    \vev{\delta \phi^1 (x) \delta \phi^1 (x')}\,.
    \label{eq:Gparallel_Definition}
\end{equation}
The defect correlator is then obtained through the bulk-to-defect OPE~\eqref{eq:ZeroT_BulkDefectOPE} and is given by
\begin{equation}
    g_{\smash{\hphi^1\hphi^1}} (\tau)
    =
    \frac{1}{\mu_{\smash{\phi\hphi^1}}^2}
    \lim_{r,r'\to0}
    (r r')^{\Delta_\phi - \Dh_{\hphi^1}}
    G_\parallel (x,x')
    +
    \vev{\hphi^1}^2\,.
    \label{eq:LargeN_ghphi1FromGparallel}
\end{equation}
Our goal is to obtain the longitudinal propagator $G_\parallel (x,x')$ at sufficiently high accuracy in order to take this limit.
Note that this disconnected part $\vev{\hphi^1}^2$ is non-vanishing, as opposed to the tilt correlator studied in Section~\ref{subsubsec:TiltCorrelator}.
The disconnected contribution is determined at leading order in large $N$ by the one-point coefficient~\eqref{eq:LargeN_b_hphi_3d}.

We define the fluctuations of $s(x)$ through
\begin{equation}
    \delta s(x)
    =
    s(x) - \vev{s (x)}\,,
    \label{eq:LargeN_FluctuationsOfs}
\end{equation}
and expand the large $N$ action to second order after integrating out the transverse fields, yielding
\begin{align}
    S^{(2)}
    &=
    \frac{1}{2} \int_0^\beta \mathrm{d}\tau \int \mathrm{d}^{d-1} \bm{x}_\perp\,
    \delta\phi^1 (x)
    \bigl[ -\partial_x^2 + \vev{s (x)} \bigr]
    \delta\phi^1 (x)
    \notag\\
    &\quad
    +
    \int_0^\beta \mathrm{d}\tau \int \mathrm{d}^{d-1} \bm{x}_\perp\,
    \vev{\phi^1 (x)}\,
    \delta\phi^1(x) \delta s(x)
    \notag\\
    &\quad
    -\frac{N}{4}
    \int_0^\beta \mathrm{d}\tau \int \mathrm{d}^{d-1} \bm{x}_\perp \int_0^\beta \mathrm{d}\tau' \int \mathrm{d}^{d-1} \bm{x}_\perp'\,
    G_\perp(x,x')^2\,
    \delta s(x)\delta s(x')\,.
    \label{eq:LargeN_MixedQuadraticAction}
\end{align}
The longitudinal propagator is then related to the fluctuations $\delta s(x)$ through
\begin{align}
    G_\parallel(x,x')
    &=
    G_\perp(x,x')
    +
    \int_0^\beta \mathrm{d} \tau_1 \int \mathrm{d}^{d-1} \bm{x}_{\perp,1}
    \int_0^\beta \mathrm{d} \tau_2 \int \mathrm{d}^{d-1} \bm{x}_{\perp,2} \notag \\
    &\phantom{=\ } \times
    G_\perp(x,x_1)
    \vev{\phi^1(x_1)}
    G_{\delta s\delta s}(x_1,x_2)
    \vev{\phi^1(x_2)}
    G_\perp(x_2,x')\,,
    \label{eq:LargeN_LongitudinalPropagator}
\end{align}
where we defined the propagator $G_{\delta s\delta s}(x,x')$ as
\begin{equation}
    G_{\delta s\delta s}(x,x')
    =
    \vev{\delta s(x)\delta s(x')}\,.
    \label{eq:LargeN_deltasPropagator}
\end{equation}
Defining the inverse propagator $K_{\delta s\delta s}$ via
\begin{equation}
    G_{\delta s\delta s} (x,x')
    =
    \frac{1}{N} K_{\delta s\delta s}^{-1} (x,x')\,,
    \label{eq:LargeN_ThermalAuxiliaryKernel}
\end{equation}
the equation to solve can be obtained from the action and reads
\begin{equation}
    K_{\delta s\delta s} (x,x')
    =
    -\frac12G_\perp(x,x')^2
    -
    \frac{\vev{\phi^1(x)}}{\sqrt N}
    G_\perp(x,x')
    \frac{\vev{\phi^1(x')}}{\sqrt N}\,,
    \label{eq:LargeN_EquationForK}
\end{equation}
with $\vev{\phi^1(x)}$ given by~\eqref{eq:Fphi_Definition} and shown in Figure~\ref{fig:LargeN_BulkProfiles}.

The numerical problem is now posed entirely.
It consists of three steps: after solving~\eqref{eq:LargeN_EquationForK}, we need to perform the inversion
\begin{equation}
    \int_0^\beta \mathrm{d}\tau_1 \int \mathrm{d}^{d-1} \bm{x}_{\perp,1}\,
    K_{\delta s\delta s}(x,x_1)
    G_{\delta s\delta s}(x_1,x')
    =
    \frac{1}{N} \delta (\tau - \tau') \delta^{(d-1)}(\bm{x}_\perp-\bm{x}'_\perp)\,.
    \label{eq:LargeN_AuxiliaryPropagatorEquation}
\end{equation}
We then take the limit in which the auxiliary-field propagator approaches the defect:
\begin{align}
    g_{\smash{\hphi^1\hphi^1}} (\tau)
    =
    \frac1{\mu_{\smash{s\hphi^1}}^2}
    \lim_{r,r'\to0}
    (r r')^{2-\Dh_{\hphi^1}}
    G_{\delta s\delta s} (x,x')
    +
    \frac{b_{\smash{\hphi^1}}^2}{\beta^{\smash{2\Dh_{\hphi^1}}}}\,.
    \label{eq:LargeN_LongitudinalFullCorrelator}
\end{align}
It is harder than it may seem to obtain an accurate numerical solution.
The reason is that, while it is sufficient to know $G_{\delta s\delta s} (x,x')$ at small $r,r'$, it involves knowing $K_{\delta s\delta s}(x,x')$ sufficiently accurately for \textit{all} $x,x'$ since Equation~\eqref{eq:LargeN_AuxiliaryPropagatorEquation} is non-local.
The results are displayed in Figure~\ref{fig:LargeN_DefectCorrelators}.

\paragraph{Comparison with the periodic OPE.}
We now wish to compare the numerical results to the truncated periodic OPE.
Using the three-point functions determined in Appendix~\ref{app:LargeN_WittenDiagrams}, we can convert the OPE coefficients~\eqref{eq:LargeN_TiltOPECoefficients_3d} to the ones appearing in $g_{\smash{\hphi^1\hphi^1}} (\tau)$:
\begin{align}
    \left.
    a_{\smash{\hphi^1}}
    \right|_{d=3}
    &= 2.04 \pm 0.04\,, \\
    \left.
    a_{\smash{s_-}}
    \right|_{d=3}
    &= 0.12 \pm 0.05\,.
\end{align}
As for the tilt correlator, we consider the truncations
\begin{align}
    g_{{\hphi^1\hphi^1}}^{[0]} (\tau) &= \frac{1}{\beta^{\smash{2\Dh_{\hphi^1}}}} \left( F_{\hat{\mathds{1}}} (\tau) + c_0 + b_{{\hphi^1}}^2 \right) \,, \label{eq:LargeN_ghphi10_Definition} \\
    g_{{\hphi^1\hphi^1}}^{[1]} (\tau) &= \frac{1}{\beta^{\smash{2\Dh_{\hphi^1}}}} \left( F_{\hat{\mathds{1}}} (\tau) + a_{\smash{\hphi^1}} F_{\smash{\hphi^1}} (\tau)+ c_1  + b_{{\hphi^1}}^2 \right)\,, \label{eq:LargeN_ghphi11_Definition} \\
    g_{{\hphi^1\hphi^1}}^{[2]} (\tau) &= \frac{1}{\beta^{\smash{2\Dh_{\hphi^1}}}} \left( F_{\hat{\mathds{1}}} (\tau) + a_{\smash{\hphi^1}} F_{\smash{\hphi^1}} (\tau) + a_{\smash{s_-}} F_{\smash{s_-}} (\tau)+ c_2  + b_{{\hphi^1}}^2 \right)\,. \label{eq:LargeN_ghphi12_Definition}
\end{align}
where we included $b_{\smash{\hphi^1}}^2$ to account for the disconnected part and the undetermined truncation-dependent constants $c_j$ to account for a possible shift not captured by the periodic blocks, as explained in Section~\ref{subsubsec:PropertiesOfPeriodicBlocks}.
Fitting the constant against the numerical result, we find the following results and maximum discrepancies for the connected correlator:
\begin{align}
    j&=0: \quad c_0&= 11.6 \pm 0.3\,, \quad \text{max. discrepancy on } 0.2 \leq \tau/\beta \leq 0.8&: 8.25\%\,, \label{eq:LargeN_hphi1_MaxDiscrepancy_0} \\
    j&=1: \quad c_1&= -8.6 \pm 0.7\,, \quad \text{max. discrepancy on } 0.2 \leq \tau/\beta \leq 0.8&: 0.36\%\,, \label{eq:LargeN_hphi1_MaxDiscrepancy_1} \\
    j&=2: \quad c_2&= -8.5 \pm 0.8\,, \quad \text{max. discrepancy on } 0.2 \leq \tau/\beta \leq 0.8&: 0.40\%\,.
    \label{eq:LargeN_hphi1_MaxDiscrepancy_2}
\end{align}
The range is chosen to be tighter than~\eqref{eq:LargeN_Tilt_PeriodicOPEComparison} because the error is larger at the edges.\footnote{This is due to the non-locality of Equation~\eqref{eq:LargeN_AuxiliaryPropagatorEquation}, which reduces the numerical precision of the solver.}
We observe again a good match between numerical solution and truncated periodic OPE once the contribution of $\hphi^1$ is included (see also Figure~\ref{fig:LargeN_DefectCorrelators}).
It is not possible at present to determine the effect of including $s_-$, as the numerical uncertainty is bigger than the discrepancy.

As an additional check, we attempt to derive the constant $c_j$ independently.
First, it is straightforwardly related to the impurity susceptibility~\eqref{eq:DefectSusceptibility} since
\begin{equation}
    \int_0^\beta \mathrm{d}\tau\, F_\Oh (\tau) = 0\,,
    \label{eq:LargeN_PeriodicBlocksDoNotContributeInSusceptibility}
\end{equation}
as discussed in Section~\ref{subsubsec:PropertiesOfPeriodicBlocks}.
We therefore have
\begin{equation}
    \chi_{\smash{\hphi^1}}
    =
    \frac{c_\text{exact}}{\beta^{\smash{2\Dh_{\smash{\hphi^1}} - 1}}}
    \sim
    \frac{c_j}{\beta^{\smash{2\Dh_{\smash{\hphi^1}} - 1}}}\,.
    \label{eq:LargeN_RelationChiAndcj}
\end{equation}
Since each periodic block vanishes in this integral, the left-hand side is independent of $j$.\footnote{In~\eqref{eq:LargeN_hphi1_MaxDiscrepancy_0}-\eqref{eq:LargeN_hphi1_MaxDiscrepancy_2}, the constant $c_j$ is evaluated from fitting the periodic OPE and thus it does depend on $j$, presumably weakly once $j \geq 1$.}
The susceptibility can be computed independently by adding a source term coupled to $\hphi^1$ to the action (see Section~\ref{subsubsec:AdditionalConstraints}).
We solve the linearized saddle equations with this static source, for which the near-defect response takes the form
\begin{align}
    \delta\!\left[z^2 F_s(z)\right]
    &=
    \alpha z^{\smash{1-\Dh_{\smash{\hphi^1}}}}+\cdots + B z^{\smash{\Dh_{\smash{\hphi^1}}}} + \ldots\,, \\
    \chi_{\smash{\hphi^1}}
    &=
    \frac{\sqrt{\pi}\,\Gamma(\Dh_{\smash{\hphi^1}}-1/2)}{\Gamma(\Dh_{\smash{\hphi^1}}) \beta^{\smash{2\Dh_{\smash{\hphi^1}} - 1}}}
    \frac{B}{\alpha}\,,
\end{align}
where $\delta$ denotes the linear change of the saddle-point profile induced by the small static source.
The normalization follows by integrating the zero-temperature bulk-defect two-point function.
Extracting the renormalized finite-part static response jointly from the two bulk fields gives at $d=3$ the independent prediction
\begin{equation}
    \chi_{\smash{\hphi^1}}
    =
    - \frac{8.7 \pm 1.0}{\beta^{\smash{2\Dh_{\smash{\hphi^1}} - 1}}}\,.
    \label{eq:LargeN_IndependentPrediction}
\end{equation}
As usual, the uncertainty is an empirical estimate from cutoff and extraction tests rather than a rigorous bound.
We observe that the fit values listed in~\eqref{eq:LargeN_hphi1_MaxDiscrepancy_0}-\eqref{eq:LargeN_hphi1_MaxDiscrepancy_2} are compatible with this value once the contribution of $\hphi^1$ is included.

We conclude this section by mentioning a caveat specific to this correlator.
The fact that the correlator contains a disconnected piece of order $\mathrm{O}(N)$ allows for the emergence of the anomalous dimension of $s_+$.
The periodic block associated to $s_+$ would thus take the form
\begin{equation}
    a_{\smash{s_+}} F_{\smash{s_+}} (\tau)
    =
    a^{(0)}_{\smash{s_+}} \gamma^{(1)}_{\smash{s_+}}\, \pd_{\Dh} \left. F_{\Oh} (\tau) \right|_{\Dh = 2\Dh_{\smash{\hphi^1}}} + \mathrm{O} (1/N)\,,
    \label{eq:LargeN_Caveat}
\end{equation}
where the superscripts in the OPE coefficient and anomalous dimension denote the order of the expansion in $N$.
There is a priori no reason to discard this term, as it is of the same order in $N$ as the connected part.
Unfortunately, the anomalous dimension $\gamma^{(1)}_{\smash{s_+}}$ is not known to the best of our knowledge and determining it would require calculating a zero-temperature four-point function up to order $\mathrm{O}(1/N)$.
We therefore do not evaluate its magnitude at present and simply note that the truncations~\eqref{eq:LargeN_ghphi10_Definition}-\eqref{eq:LargeN_ghphi12_Definition} seem to produce a result consistent with~\eqref{eq:LargeN_Caveat} being a small contribution.
We reserve a deeper analysis to future work.

%% file: sections/Discussion.tex
\section{Discussion and outlook}
\label{sec:DiscussionAndOutlook}

\paragraph{Summary.}
In this work, we studied thermal correlators of the pinning-field line defect in the critical $\mathrm{O}(N)$ model.
We developed a manifestly periodic OPE for defect two-point functions and related asymptotic heavy-operator contributions to low-lying thermal data.
We applied this framework in the $\veps$-expansion and at large $N$, determining thermal one-point coefficients and correlators, and explored its extension to the finite-volume antipodal geometry.
Combining perturbative results with the two-dimensional Ising case also provided Padé estimates for the three-dimensional magnetic impurity at $N=1$.
For the lowest-lying operator $\hphi^1$, we obtain
\begin{equation}
    \left. \vev{\hphi^1} \right|_{3d\,, \text{ Padé}}
    \sim
    \frac{1.701}{\beta^{\smash{\Dh_{\smash{\hphi^1}}}}}\,.
    \label{eq:Discussion_3dEstimate}
\end{equation}
We conjecture that the correlator $\vev{\hphi^1 (\tau) \hphi^1 (0)}$ is well-approximated by truncating the periodic OPE,
\begin{equation}
    g_{\smash{\hphi^1\hphi^1}} (\tau)
    \sim
    \frac{1}{\beta^{\smash{2\Dh_{\smash{\hphi^1}}}}}
    \left(
    F_{\hat{\mathds{1}}}(\tau)
    +
    a_{\smash{\hphi^1}} F_{\smash{\hphi^1}}(\tau)
    \right)
    +
    \frac{c_{\smash{\hphi^1}}}{\beta^{\smash{2\Dh_{\smash{\hphi^1}}}}}\,,
    \label{eq:Discussion_TruncatedPeriodicOPE}
\end{equation}
with $c_{\smash{\hphi^1}}$ a constant undetermined by the framework.
Our large-$N$ comparisons support the usefulness of low-lying truncations of the periodic OPE.
At $d=3$, we find that the two-point function of tilt operators solved numerically has a maximum discrepancy with the truncated periodic OPE of $\sim 0.86\%$ over the interval $0.05 \leq \tau/\beta \leq 0.95$.
This agreement is surprising, as it suggests that the heavy contributions to the thermal OPE may be effectively organized into families of \textit{scalar} double-twist operators at large $\Dh$.
This picture is reminiscent of analytic results in one-dimensional CFTs, where crossing constrains the asymptotic OPE density and generalized-free spectra provide natural reference distributions~\cite{Qiao:2017xif,Mazac:2018ycv}.
The understanding of the systematic accuracy of the truncated periodic OPE and of the determination of thermal data from bootstrap constraints alone remain important open questions.

\bigskip

There are several interesting directions worth exploring following this work:
\begin{itemize}
    \item \textbf{Thermal correlators on supersymmetric Wilson lines.}
    It is natural to extend the application of the bootstrap methods presented here to other line defect setups for which the spectrum is well-understood.
    For the supersymmetric Wilson line in $\Nm=4$ Super Yang-Mills (SYM), many states are known from integrability~\cite{Cavaglia:2021bnz,Cavaglia:2022qpg,Julius:2025dce} as well as perturbative methods~\cite{Giombi:2017cqn,Liendo:2018ukf,Kiryu:2018phb,Ferrero:2021bsb,Barrat:2021tpn,Barrat:2022eim,Ferrero:2023znz,Ferrero:2023gnu,Artico:2024wut} and provide strong constraints on the finite-temperature correlator.
    Several regimes can be investigated.
    At strong coupling, the effective action for the two-dimensional black hole background in asymptotic AdS$_2$ was derived in~\cite{Giombi:2026kdz}, where the wave equation was solved for two-point functions of tilt and displacement operators, revealing the presence of the bouncing singularity well-known from bulk theories~\cite{Fidkowski:2003nf,Ceplak:2024bja,Dodelson:2025jff,Grozdanov:2026cut,Komargodski:2026oob}.
    Interestingly, the correlator of tilt operators is special in that it does not exhibit the bouncing singularity and is simply given by the identity contribution in the periodic OPE:
    \begin{equation}
        g_{tt} (\tau)
        =
        \frac{F_{\hat{\mathds{1}}} (\tau)}{\beta^2}
        =
        \frac{\pi^2}{\beta^2} \csc^2 \left( \frac{\pi \tau}{\beta} \right)\,.
    \end{equation}
    It would be instructive to determine the first correction in $1/\sqrt{\lambda}$ to this correlator, either through analytic bootstrap methods or direct calculation.
    Moreover,~\cite{Giombi:2026kdz} focused on retarded correlators.
    Reconstructing the Euclidean correlator as in~\cite{Barrat:2025twb} would also give access to OPE contributions that are invisible to the discontinuity.
    A valuable additional direction is to study the thermal two-point function of tilt operators on the supersymmetric Wilson line numerically~\cite{WilsonLineThermalNumericsWIP}.
    \item \textbf{Defect probes of confinement.}
    The extension of the periodic OPE to finite-volume defect systems opens interesting perspectives.
    It is well-known that large $N$ $\Nm=4$ SYM exhibits a confinement-deconfinement phase transition when put on the geometry $S_\beta^1 \times S_R^3$~\cite{Witten:1998zw,Aharony:2003sx}.
    In the confining regime, correlators of bulk single-trace operators turn out to take a very simple form at large $N$, known as inheritance principle~\cite{Brigante:2005bq} and valid at all orders in perturbation theory.
    For two-point functions, it takes the form
    \begin{equation}
        \vev{\Om (\tau, \theta) \Om (0,0)}
        =
        \sum_{m=-\infty}^\infty \vev{\Om (\tau + m\beta, \theta) \Om (0,0)}_{T=0}\,.
    \end{equation}
    It is natural to wonder what happens to correlators of defect operators in the antipodal configuration.
    How do defect correlators change across the bulk confinement-deconfinement transition, and does the inheritance principle extend to the antipodal defect configuration?
    \item \textbf{Finite-volume bootstrap.}
    Another direction is to adapt the finite-volume framework developed here to bulk theories.
    One could for instance try to determine the thermal one-point functions of the low-lying operators of the $3d$ Ising as functions of $\beta/R$, and in particular try to improve the curve for $b_T$ shown in Figure~13 of~\cite{Iliesiu:2018fao}.
    The success of the thermal bootstrap at infinite volume for the $3d$ Ising model~\cite{Iliesiu:2018zlz,Barrat:2025wbi} is an encouraging starting point.
    A related direction would be to obtain bounds on the critical temperature mentioned above in $\Nm=4$ SYM and other holographic theories sharing similar features.
    \item \textbf{UV $\leftrightarrow$ IR.}
    Recent works have connected UV OPE data to IR data, namely quasinormal modes, in bulk theories~\cite{Barrat:2026jfg,Arnaudo:2026axe}.
    Such relations are expected to persist in defect theories.
    Since the line defect correlators are one-dimensional, it may be easier to learn something about UV physics from the periodic blocks.
    Moreover, at finite volume, one can relate the OPE data to the spectral representation.
    It would be worthwhile seeing whether new zero-temperature OPE data can be extracted from this perspective.
    \item \textbf{Analytic bootstrap for higher-dimensional defects.}
    Recent work was done on higher-dimensional defects at finite temperature~\cite{Li:2026udh,Artico:2026rye}, completing and extending the results of~\cite{Barrat:2024aoa}.
    These systems are more difficult to constrain than line defects wrapping the thermal circle as they break more symmetry.
    In particular, the current constraints involve a mix of bulk and defect operators, making them difficult to use effectively.
    It would be interesting to understand whether there exists a generalization of the periodic OPE for higher-dimensional defects.
    \item \textbf{Defect eigenstate thermalization.}
    It would be worthwhile to investigate whether correlators of light defect operators in typical heavy defect states approach thermal correlators in the antipodal geometry.
    Through the defect state-operator correspondence, this would relate heavy-heavy-light three-point coefficients to the thermal one-point data studied here.
    Explicit heavy-light correlators on the supersymmetric Wilson line~\cite{Giombi:2021zfb,Giombi:2022anm} and studies of eigenstate thermalization in conformal quantum mechanics~\cite{Nayak:2019khe} provide useful starting points, although the special large-charge states considered in the former need not exhibit thermal behavior.
    A complementary approach would be to adapt methods relating thermal one-point functions to averaged heavy-heavy-light coefficients~\cite{Buric:2025uqt}, distinguishing such averaged statements from thermalization in individual eigenstates.
\end{itemize}

%% file: appendices/FreeTheory.tex
\section{Free theory for the antipodal configuration}
\label{app:FreeTheoryForTheAntipodalConfiguration}

In this appendix, we show how to reconstruct the free scalar correlator in the finite-volume antipodal geometry for the case $d=4$.
The lowest defect scalar, which we denote $\hphi$, has scaling dimension $\Dh_{\hphi} = 1$.
The exact two-point function can be evaluated as the sum of images of the free propagator on $\mathbb{R} \times S^3_R$:
\begin{equation}
    g (\tau)
    =
    \sum_{m=-\infty}^\infty
    \left[2R \sinh \left( \frac{|\tau + m\beta|}{2R} \right)\right]^{-2}
    =
    \frac{1}{\beta^2}
    \left[
    \psi_q^{(1)} \left( \frac{\tau}{\beta} \right)
    +
    \psi_q^{(1)} \left( 1-\frac{\tau}{\beta} \right)
    \right]\,,
    \label{eq:FreeTheory_Exact}
\end{equation}
with we selected the strip $0 < \tau < \beta$ in the second equality.
In the following, we reconstruct this expression from the periodic block expansion derived in Section~\ref{subsec:PeriodicOPEAndAsymptotics}.

\subsection{Expansion in periodic blocks}
\label{subsec:ExpansionInPeriodicBlocks}

We begin by writing the regular thermal block expansion.
The exchanged spectrum consists of the identity $\hat{\mathds{1}}$ and of classical double-twist operators $[\hphi \hphi]_n$:
\begin{equation}
    g(\tau)
    =
    \frac{1}{\beta^2}
    \left(
    f_{\hat{\mathds{1}}} (\tau)
    +
    \sum_{n=0}^\infty a_{[\hphi \hphi]_n} (q)\, f_{[\hphi \hphi]_n} (\tau)
    \right)\,.
\end{equation}
The next step is to promote the thermal blocks to periodic blocks.
The correlator simply becomes
\begin{equation}
    g (\tau)
    =
    \frac{F_{\hat{\mathds{1}}} (\tau)}{\beta^2}
    +
    \tilde{g} (\tau)\,,
\end{equation}
since periodic blocks of $[\hphi \hphi]_n$ vanish, as explained in Section~\ref{subsubsec:PropertiesOfPeriodicBlocks}.
The identity contribution can be determined from the expression~\eqref{eq:PeriodicBlock_FiniteV} to be
\begin{equation}
    F_{\hat{\mathds{1}}} (\tau)
    =
    \pi^2 \csc^2
    \left(
    \frac{\pi \tau}{\beta}
    \right)\,.
    \label{eq:FreeTheory_Identity}
\end{equation}
Note that it is independent of $R$ and does not reproduce~\eqref{eq:FreeTheory_Exact}; thus, the completion $\tilde{g} (\tau)$ must be non-trivial.

\subsection{Completion}
\label{subsec:Completion}

The key observation is that $F_{\hat{\mathds{1}}} (\tau)$ has poles only at $\tau = m \beta$.
We need to restore the expected poles at $\tau = m \beta + 2 \pi i n R$ (see Figure~\ref{fig:AnalyticStructure}).
A special property of the free conformally coupled scalar in $d=4$ is that its normal-mode frequencies on $S^3_R$ are $\omega_\ell = (\ell+1)/R$.
Its thermal two-point function therefore satisfies
\begin{equation}
    g(\tau+ 2\pi i R)
    =
    g(\tau)\,.
\end{equation}
We thus write the following candidate for $\tilde{g}(\tau)$:
\begin{equation}
    \tilde{g}(\tau)
    =
    \sum_{n \neq 0}
    \frac{\pi^2}{\beta^2}
    \csc^2 \left( \frac{\pi(\tau+2\pi n i R)}{\beta} \right)
    +
    \frac{c_{\hphi} (q)}{\beta^2}\,,
\end{equation}
and invoke the uniqueness argument of Section~\ref{subsubsec:PropertiesOfPeriodicBlocks} to argue that it is the correct answer.
Combining this expression for $\tilde{g} (\tau)$ with~\eqref{eq:FreeTheory_Identity} resums to~\eqref{eq:FreeTheory_Exact}, up to the constant which we identify to be given by 
\begin{equation}
    c_{\hphi} (q)
    =
    - \frac{\beta}{R}
    =
    \log (q)\,.
\end{equation}

%% file: appendices/FeynmanDiagrams.tex
\section{Details on Feynman diagrams}
\label{app:DetailsOnTheFeynmanDiagrams}

This appendix provides details on the perturbative computations performed in Section~\ref{sec:epsExpansionAndInterpolationTo3d}.

\subsection{Feynman rules}
\label{subsec:FeynmanRules}

We first list the Feynman rules associated to the action~\eqref{eq:ON_Action}.

\paragraph{Infinite volume.}
The $d$-dimensional scalar propagator on $\mathbb{R}^d$ is defined as
\begin{equation}
    G(\tau, \bm{x}_\perp)
    =
    \frac{\Gamma(d/2-1)}{4 \pi^{d/2}}
    \frac{1}{(\tau^2 + \bm{x}_\perp^2)^{d/2-1}}\,.
    \label{eq:ON_FreePropagator}
\end{equation}
The action~\eqref{eq:ON_Action} contains a four-scalar bulk vertex, which reads
\begin{equation}
    \VertexFourScalars\
    =
    - \frac{\lambda}{3} (\delta^{IJ} \delta^{KL} + \delta^{IK} \delta^{JL} + \delta^{IL} \delta^{JK}) \int_0^\beta \mathrm{d}\tau \int \mathrm{d}^{d-1} \bm{x}_\perp\,.
    \label{eq:ON_BulkVertex}
\end{equation}
The defect vertex reads
\begin{equation}
    \VertexDefect\
    =
    - h \delta^{I1} \int_0^\beta \mathrm{d}\tau\,.
    \label{eq:ON_DefectVertex}
\end{equation}
For multiple vertices, symmetry factors would need to be taken into account.

\paragraph{Finite volume.}
On $\mathbb R\times S_R^{d-1}$, the zero-temperature scalar propagator is
\begin{equation}
    G(\tau,\theta)
    =
    \frac{\Gamma(d/2-1)}{4\pi^{d/2}}
    \frac{1}{
        \left[2R^2\bigl(\cosh(\tau/R)-\cos\theta\bigr)\right]^{d/2-1}
    },
\end{equation}
where $\theta$ is the angular separation on the sphere.
At finite temperature, each propagator is replaced by its sum over thermal images.

The bulk vertex retains the same coupling and index structure, with integration measure
\begin{equation}
    \VertexFourScalars\
    =
    - \frac{\lambda}{3} (\delta^{IJ} \delta^{KL} + \delta^{IK} \delta^{JL} + \delta^{IL} \delta^{JK})
    R^{d-1}
    \int_0^\beta \mathrm{d}\tau
    \int_{S^{d-1}} \mathrm{d}\Omega_{d-1}\,.
\end{equation}
For the antipodal configuration, each defect insertion is summed over the North and South poles, with the same vertex weight $-h\delta^{I1} \int_0^\beta\mathrm{d} \tau$ at either pole.
A bulk point at polar angle $\theta$ therefore connects to the north pole through $G_\beta(\tau,\theta)$ and to the south pole through $G_\beta(\tau,\pi-\theta)$.

\subsection{Reducing the first and second diagrams to the GFF correlator}
\label{subsec:ReducingTheFirstAndSecondDiagramsToTheGFFCorrelator}

We first show how to reach Equation~\eqref{eq:ReductionToOneDiagram} through~\eqref{eq:DiagramOne} and~\eqref{eq:DiagramTwo}.
We show explicitly the infinite-volume case.
The antipodal configuration can be treated in the very same way.

\paragraph{First diagram.}
The first diagram in~\eqref{eq:Diagrams} is simply given by free theory in $d$ dimensions, i.e.,
\begin{align}
    \DiagOneFiniteT
    &=
    \frac{\Gamma(d/2-1)}{4 \pi^{d/2}}
    \sum_{m=-\infty}^\infty
    \frac{1}{|\tau + m\beta|^{d-2}}\,.
    \label{eq:ON_FiniteV_DiagramOne}
\end{align}
We keep the sum of images explicit to be able to relate the computation to the zero-temperature correlator, which is the statement~\eqref{eq:DiagramOne} in the main text.

\paragraph{Second diagram.}
The second diagram can be expressed as
\begin{align}
    \DiagTwoFiniteT\
    &=
    -\frac{\lambda h^2}{6}\,
    \operatorname{Vol}(S^{d-2})
    \sum_{m_i \in \mathbb{Z}}
    \int_0^\beta \mathrm{d}\tau_1
    \int_0^\beta \mathrm{d}\tau_2
    \int_0^\beta \mathrm{d}\tau_3
    \int_0^\infty \mathrm{d}r\, r^{d-2}
    \notag\\
    &\phantom{=\ } \times
    G(\tau_3+m_1\beta,\bm{x}_\perp)
    G(\tau_3-\tau+m_2\beta,\bm{x}_\perp)
    G(\tau_{13}+m_3\beta,\bm{x}_\perp)
    G(\tau_{23}+m_4\beta,\bm{x}_\perp)\,.
\end{align}
We can easily rewrite it as a sum of images of a zero-temperature integral:
\begin{align}
    \DiagTwoFiniteT\
    &=
    -\frac{\lambda h^2}{6}\,
    \operatorname{Vol}(S^{d-2})
    \sum_{m\in\mathbb Z}
    \int_{-\infty}^{\infty}\mathrm{d}\tau_1
    \int_{-\infty}^{\infty}\mathrm{d}\tau_2
    \int_{-\infty}^{\infty}\mathrm{d}\tau_3
    \int_0^\infty \mathrm{d}r\, r^{d-2}
    \notag\\
    &\phantom{=\ } \times
    G(\tau_3,\bm{x}_\perp)
    G(\tau_3-\tau-m\beta,\bm{x}_\perp)
    G(\tau_{13},\bm{x}_\perp)
    G(\tau_{23},\bm{x}_\perp) \notag \\
    &=
    \sum_{m=-\infty}^\infty \left. \DiagTwoZeroT \right|_{\tau \to \tau+m \beta}\,,
\end{align}
This leads to~\eqref{eq:DiagramTwo} and thus to the conclusion given in Equation~\eqref{eq:ReductionToOneDiagram}.

\subsection{Computation of the third diagram}
\label{subsec:ComputationOfTheThirdDiagram}

We now detail the calculation of the tadpole diagram.
We first perform the infinite-volume calculation and then the finite-volume one.

\paragraph{Infinite volume.}
The tadpole diagram can be unfolded such as to obtain a single image sum of a zero-temperature integral:
\begin{equation}
    \DiagThreeFiniteT\
    =
    -\frac{\lambda(N+2)}6
    \frac{\Gamma(\nu)\zeta(2\nu)}{2\pi^{\nu+1}\beta^{2\nu}}
    \sum_{m = -\infty}^\infty
    \int_{-\infty}^\infty \mathrm{d}\tau' \int \mathrm{d}^{d-1}\bm{x}_\perp'\,
    G( \tau',\bm{x}_\perp' )\, G ( \tau'-(\tau+m\beta), \bm{x}_\perp' )\,.
\end{equation}
The remaining zero-temperature two-point integral is well-known and reads~\cite{tHooft:1978jhc}
\begin{equation}
    \int_{-\infty}^\infty \mathrm{d}\tau' \int \mathrm{d}^{d-1}\bm{x}_\perp'\,
    G( \tau',\bm{x}_\perp' )\, G ( \tau'-(\tau+m\beta), \bm{x}_\perp' )
    =
    \frac{\Gamma(\nu-1)}{16\pi^{\nu+1}\bigl[(\tau+m\beta)^2\bigr]^{\nu-1}}\,.
\end{equation}
Performing the image sum therefore gives, by analytic continuation in $d$,
\begin{equation}
    \DiagThreeFiniteT\
    =
    -\frac{\lambda(N+2)\Gamma(\nu)\Gamma(\nu-1)\zeta(2\nu)}{192\pi^{2\nu+2}\beta^{4\nu-2}}
    \left[
        \zeta_H\left(2\nu-2,\frac{\tau}{\beta}\right)
        +
        \zeta_H\left(2\nu-2,1-\frac{\tau}{\beta}\right)
    \right].
\end{equation}
As $\nu \to 1$ ($d=4$), the pole of $\Gamma(\nu-1)$ cancels the zero of the Hurwitz-$\zeta$ combination, yielding
\begin{equation}
    \DiagThreeFiniteT\
    =
    -\frac{\lambda(N+2)}{1152\pi^2\beta^2}
    \log\left[
        \frac{1}{4} \csc^2 \left(\frac{\pi\tau}{\beta} \right)
    \right].
\end{equation}
This expression is the one given in~\eqref{eq:ThirdDiagram_FinalResult} in the main text.
The non-analytic terms associated with thermal screening require the infrared resummation discussed in the main text.

\paragraph{Finite volume.}
We now consider the finite-volume case.
Step by step, we find:
\begin{align}
    \DiagFiveFiniteTFiniteV\
    &=
    - \frac{\lambda}{6} (N+2) \left( \sum_{n=-\infty}^\infty G(n\beta,0) \right)
    R^{d-1} \text{Vol}(S^{d-3})
    \sum_{m_i}
    \int_0^\beta \mathrm{d}\tau' \notag \\
    &\phantom{=\ } \times
    \int_0^\pi \mathrm{d}\theta'\, \sin^{d-2} (\theta') \int_0^\pi \mathrm{d}\varphi'\, \sin^{d-3} (\varphi')\,
    G(\tau' + m_1\beta,\theta') G(\tau'-\tau + m_2\beta,\theta') \notag \\
    &=
    - \frac{\lambda}{6} (N+2) \frac{2 \pi^{d/2-1} R^{d-1}}{\Gamma(d/2-1)} \left( \sum_{n=-\infty}^\infty G(n\beta,0) \right)
    \sum_{m=-\infty}^\infty
    \int_{-\infty}^\infty \mathrm{d}\tau' \notag \\
    &\phantom{=\ } \times
    \int_0^\pi \mathrm{d}\theta'\, \sin^{d-2} (\theta') \int_0^\pi \mathrm{d}\varphi'\, \sin^{d-3} (\varphi')
    G(\tau',\theta') G(\tau'-\tau + m\beta,\theta') \notag \\
    &=
    - \frac{\lambda (N+2)}{6} \frac{\Gamma^2(d/2-1) R^{3-d}}{2^{d+1} \pi^{(d+1)/2} \Gamma((d-1)/2)} \left( \sum_{n=-\infty}^\infty G(n\beta,0) \right)
    \sum_{m=-\infty}^\infty
    \int_{-\infty}^\infty \mathrm{d}\tau' \notag \\
    &\phantom{=\ } \times
    \int_0^\pi \frac{\mathrm{d}\theta'\, \sin^{d-2} (\theta')}{\left[ (\cosh (\tau'/R) - \cos(\theta') )(\cosh ((\tau' - \tau + m\beta)/R) - \cos(\theta') ) \right]^{d/2-1}}\,,
\end{align}
In the first line the propagators $G(\tau,\theta)$ are defined as zero-temperature propagators on $\mathbb{R} \times S^{d-1}$.
In particular, since the external points are aligned (i.e., superimposed on $S^{d-1}$), the integrand does not depend on the second angle $\varphi'$.
In the second line we have used the standard trick of trading one sum and the thermal $\tau'$-integral against a zero-temperature integral.
In the third line we unwrapped the integrand.
It can be shown to be equal to
\begin{align}
    \DiagFiveFiniteTFiniteV\
    &=
    - \lambda (N+2) \frac{\Gamma(\nu)}{24 \pi^{\nu+1} R^{2\nu-2}}
    \left( \sum_{n=-\infty}^\infty G(n\beta,0) \right)
    \sum_{m=-\infty}^\infty \sum_{\ell=0}^\infty
    \frac{e^{-(\nu + \ell)\frac{|\tau + m\beta|}{R}}}{2(\nu + \ell)} \notag \\
    &\phantom{=\ } \times
    \left(
    \frac{|\tau + m\beta|}{R} + \frac{1}{\nu + \ell}
    \right)
    C_\ell^{(\nu)} (1)\,,
\end{align}
with $\nu = d/2-1$.
At $d=4$, this expression is finite and can be recast in the form
\begin{align}
    \DiagFiveFiniteTFiniteV\
    &=
    \frac{\lambda(N+2) L_1(q)}{96\pi^4 R^2}
    \sum_{m = -\infty}^\infty
    \left[
        \log(1-e^{-u_m})
        -\frac{u_m}{e^{u_m}-1}
    \right]\,,
\end{align}
with $L_1(q)$ defined in~\eqref{Lq_Definition}.
This is the result provided in the first line of Equation~\eqref{eq:FiniteV_LastDiagram_Result}.

%% file: appendices/WittenDiagrams.tex
\section{Details on Witten diagrams for the large $N$ $\mathrm{O}(N)$ model}
\label{app:LargeN_WittenDiagrams}

This appendix is dedicated to the zero-temperature calculations used in Section~\ref{sec:LargeN} for translating the thermal one-point coefficients $b_{\Oh}$ into OPE coefficients $a_{\Oh}$.
The three-point functions are determined using AdS$_2 \times S^{d-2}$ formalism, following in particular~\cite{Cuomo:2021kfm}.
We first review this formulation, then proceed to the calculation of $f_{tt\Oh}$ and $f_{\smash{\hphi^1\hphi^1\Oh}}$ for the relevant operators $\Oh$.

\subsection{AdS$_2$ description}
\label{subsec:AdS2Description}

\paragraph{Geometry and fields.}
The large $N$ $\mathrm{O}(N)$ model can be formulated on $\mathrm{AdS}_2\times S^{d-2}$ using a Weyl transformation of the metric:
\begin{equation}
    \mathrm{d}s_{\mathbb R^d}^2
    =
    r^2\left[
    \frac{\mathrm{d}\tau^2+\mathrm{d}r^2}{r^2}
    +\mathrm{d}\Omega_{d-2}^2
    \right]\,.
\end{equation}
Both factors have unit radius, and the line defect lies at the
$\mathrm{AdS}_2$ boundary.
The Weyl-rescaled fields and the conformal saddle are
\begin{align}
    \tilde{\phi}^I
    &=
    r^{(d-2)/2}\phi^I\,, \\
    \sigma
    &=
    \sqrt N\,r^2(s-s_0)\,, \\
    r^2s_0
    &=
    \frac{(d-2)(4-d)}4\,, \\
    \vev{\tilde{\phi}^1}
    &=
    \sqrt N\,C_d\,.
\end{align}
The curvature coupling cancels the saddle contribution to the mass of the lowest transverse harmonic, which is therefore a massless $\mathrm{AdS}_2$ scalar.

\paragraph{External legs.}
All defect primaries have unit two-point normalization.
For an angle-independent singlet $\Oh$, define the bulk-to-boundary propagator as
\begin{equation}
    \vev{\sigma(X)\Oh(\tau_i)}
    =
    \sqrt N\,\mu_{s\Oh} K_{\hat\Delta_\Oh}(X,\tau_i),
\end{equation}
with
\begin{equation}
    K_\Delta(X,\tau_i)
    =
    \left[
    \frac{r}{r^2+(\tau-\tau_i)^2}
    \right]^\Delta\,,
\end{equation}
and $\mu_{s\Oh}$ the associated bulk-defect coefficient.
The corresponding tilt leg is
\begin{equation}
    \vev{\tilde\phi^a(X) t^b(\tau_i)}
    =
    \delta^{ab} \mu_{\phi t} K_1(X,\tau_i)\,,
\end{equation}
with the bulk-defect coefficient known to be
\begin{equation}
    \mu_{\phi t}^2
    =
    \frac{1}{\pi\,\operatorname{Vol}(S^{d-2})}\,.
\end{equation}
The singlet dimensions and residues are obtained from the zero-temperature quadratic kernel.
At $d=3$, our conventions give
\begin{equation}
    \begin{array}{c|cc}
        \Oh & \hat\Delta_\Oh & \sqrt N\,\mu_{s\Oh} \\ \hline
        \hphi^1 & 1.5417281 & -0.6475485 \\
        s_-     & 2.5564691 &  1.268576
    \end{array}.
\end{equation}

\subsection{Three-point functions $f_{tt\Oh}$}
\label{subsec:ThreePointFunctionsfttOh}

We now use the $\mathrm{AdS}_2\times S^{d-2}$ formulation above to compute three-point functions involving two tilt operators with an $\mathrm{O}(N-1)$ singlet in the form of Witten diagrams.

\paragraph{Contact diagram.}
The relevant interaction is
\begin{equation}
    S_{\sigma\phi\phi}
    =
    \frac{1}{2\sqrt N}
    \int_{\mathrm{AdS}_2\times S^{d-2}}
    \mathrm{d}^dX\sqrt g\,
    \sigma\,\tilde{\phi}^a \tilde{\phi}^a.
\end{equation}
This can be used to compute the three-point function of two tilt operators with a (single-particle) operator $\Oh$.
The two Wick contractions cancel the factor $1/2$, while the sphere volume cancels against the normalization of the two tilt legs, yielding
\begin{equation}
    \sqrt N\,f_{tt\Oh}
    =
    -\frac{\sqrt N\,\mu_{s\Oh}}{\pi}\,
    I(1,1,\hat\Delta_\Oh)\,.
    \label{eq:Witten_TiltCoefficient}
\end{equation}
The general contact integral is standard and can be expressed as~\cite{Freedman:1998tz}
\begin{equation}
    \int_{\mathrm{AdS}_2}
    \frac{\mathrm{d}\tau\,\mathrm{d}r}{r^2}
    \prod_{i=1}^3 K_{\Delta_i}(X,\tau_i)
    =
    \frac{I (\Delta_1,\Delta_2,\Delta_3)}
    {|\tau_{12}|^{\Delta_1+\Delta_2-\Delta_3}
     |\tau_{13}|^{\Delta_1+\Delta_3-\Delta_2}
     |\tau_{23}|^{\Delta_2+\Delta_3-\Delta_1}}\,,
\end{equation}
with the function $I(\Delta_1,\Delta_2,\Delta_3)$ defined as
\begin{equation}
    I(\Delta_1,\Delta_2,\Delta_3)
    =
    \frac{\sqrt\pi\,
    \Gamma\!\left(\frac{\Delta_1+\Delta_2+\Delta_3-1}{2}\right)}
    {2\Gamma(\Delta_1)\Gamma(\Delta_2)\Gamma(\Delta_3)}
    \prod_{i=1}^3
    \Gamma\!\left(
    \frac{\Delta_1+\Delta_2+\Delta_3}{2}-\Delta_i
    \right)\,.
\end{equation}
In particular, for two external tilts we have
\begin{equation}
    I(1,1,\Delta)
    =
    \frac{\pi^2\,2^{-\Delta}}{\sin(\pi\Delta/2)}\,,
\end{equation}
which leads to
\begin{equation}
    \sqrt{N}\,f_{tt\Oh}
    =
    -\frac{\pi\, \sqrt N\,\mu_{s\Oh} \,2^{-\Dh_\Oh}}
    {\sin(\pi\Dh_\Oh/2)}\,.
    \label{eq:Witten_TiltClosedForm}
\end{equation}
All dependence on the spacetime dimension enters through $\Dh_\Oh$ and $\mu_{s\Oh}$.

\paragraph{Finite part and results.}
For $\Delta>2$, the integral is defined by analytic continuation.
For $2<\Delta<4$, an equivalent subtraction, with boundary insertions at $0$, $1$, and infinity, is
\begin{equation}
    I(1,1,\Delta)
    =
    \lim_{L\to\infty}
    \left[
    2\pi\int_0^L
    \frac{\mathrm{d}r\,r^{\Delta-1}}{1+4r^2}
    -
    \frac{\pi L^{\Delta-2}}{2(\Delta-2)}
    \right].
\end{equation}
At $d=3$, this gives
\begin{align}
    I(1,1,\Dh_{\hphi^1})
    &\sim 5.1419503\,, \\
    I(1,1,\Dh_{s_-})
    &\sim
    -2.1875087\,,
\end{align}
and hence
\begin{align}
    f_{tt\hphi^1}
    &=
    \frac{1.0598644}{\sqrt N}
    +
    \mathrm{O}(N^{-3/2})\,, \\
    f_{tt s_-}
    &=
    \frac{0.8833166}{\sqrt N}
    +
    \mathrm{O}(N^{-3/2})\,.
    \label{eq:Witten_TiltResults}
\end{align}
The numerical sensitivity of these leading coefficients is negligible compared with the uncertainties in the thermal one-point functions.

\subsection{Three-point functions $f_{\smash{\hphi^1\hphi^1\Oh}}$}
\label{subsec:ThreePointFunctionsfhphi1hphi1Oh}

We now perform a similar calculation for the three-point coefficients $f_{\smash{\hphi^1\hphi^1\Oh}}$.

\paragraph{Cubic effective action.}
Integrating out the fundamental fields and expanding around the conformal saddle gives
\begin{equation}
    S_{\mathrm{eff}}^{(3)}
    =
    \frac{1}{\sqrt{N}}
    \left[
    \frac{1}{6} \operatorname{tr} (G_0\sigma)^3
    +
    \frac{1}{2} C_d \sigma G_0\sigma G_0\sigma C_d
    \right]\,.
    \label{eq:Witten_SingletCubicAction}
\end{equation}
Here $G_0$ is the fundamental propagator on the zero-temperature saddle, and all products include integration over $\mathrm{AdS}_2\times S^{d-2}$.
The first term produces a scalar-loop triangle; the second produces three background-chain contributions.
Both are required at the same order in $1/N$.

\paragraph{Evaluation in three dimensions.}
The result takes the form
\begin{equation}
    \sqrt N\,f_{\smash{\hphi^1\hphi^1\Oh}}
    =
    - N^{3/2}\, \mu_{\smash{s\hphi^1}}^2 \mu_{s\Oh}
    \left[ T(\Dh_{\hphi^1},\Dh_{\hphi^1},\Dh_\Oh)+ R(\Dh_{\hphi^1},\Dh_{\hphi^1},\Dh_\Oh)\right].
\end{equation}
The background contribution reduces to the contact integral:
\begin{equation}
    R(a,a,b)
    =
    2\pi C_3^2\, I(a,a,b)
    \left[
    \frac{1}{a^2(a-1)^2}
    +
    \frac{2}{a(a-1)b(b-1)}
    \right]\,,
    \qquad \text{with }
    C_3^2
    \sim
    0.04441318577\,.
\end{equation}
The triangle $T$ is defined as
\begin{align}
    T(a,a,b)
    &=
    2^b\sum_{\ell=0}^{\infty}d_\ell
    \int\prod_{i=1}^{3}
    \frac{\mathrm{d}\tau_i\,\mathrm{d}r_i}{r_i^2}
    K_a(X_1,0)K_a(X_2,1)K_b(X_3,2)
    \notag \\
    &\phantom{=\ }
    \times
    G_{0,\ell}(X_1,X_2)
    G_{0,\ell}(X_2,X_3)
    G_{0,\ell}(X_3,X_1),
    \qquad
    d_0=1,\quad d_{\ell\geq1}=2.
\end{align}
and evaluated numerically using the angular decomposition of $G_0$, whose $\mathrm{AdS}_2$ propagators are
\begin{equation}
    G_{0,\ell} (X,Y)
    =
    \frac{1}{2\pi}
    Q_{\sqrt{\ell^2+1/4}-1/2}
    \left(
    1
    +
    \frac{(\tau-\tau')^2+(r-r')^2}{2rr'}
    \right)\,.
\end{equation}
Here, $Q_\nu (x)$ designates the Legendre function of the second kind.
The angular sum includes degeneracy one for $\ell=0$ and two for $\ell\geq1$.
For $\Oh=s_-$, the triangle requires a boundary subtraction, with its finite part fixed by analytic continuation.
The separate contributions to the normalized coefficients are
\begin{equation}
    \begin{array}{c|rr}
        & \text{background} & \text{triangle} \\ \hline
        \sqrt N\,f_{\smash{\hphi^1\hphi^1\hphi^1}}
        & 0.663266 & 1.146 \\
        \sqrt N\,f_{\smash{\hphi^1\hphi^1s_-}}
        & -0.806064 & 1.182
    \end{array}.
\end{equation}
Combining them yields
\begin{align}
    f_{\smash{\hphi^1\hphi^1\hphi^1}}
    &=
    \frac{1.81\pm 0.03}{\sqrt{N}}
    +
    \mathrm{O}(N^{-3/2})\,, \\
    f_{\smash{\hphi^1\hphi^1s_-}}
    &=
    \frac{0.38\pm 0.07}{\sqrt{N}}
    +
    \mathrm{O}(N^{-3/2})\,.
    \label{eq:Witten_SingletResults}
\end{align}
The quoted uncertainties reflect variations of the integration, angular cutoff, and subtraction procedure; they are empirical numerical estimates, not rigorous bounds.
The cancellation between the two contributions makes the mixed coefficient less precise than for tilt operators.

\paragraph{Lowest double-trace operator.}
For the unit-normalized primary $s_+=[\hphi^1\hphi^1]_0$, large-$N$ factorization simply gives
\begin{equation}
    f_{\smash{\hphi^1\hphi^1 s_+}}
    =
    \sqrt2 + \mathrm{O}(N^{-1})\,.
\end{equation}

%% file: appendices/Numerics.tex
\section{Details on large $N$ numerics}
\label{app:LargeN_Numerics}

In this appendix, we give details about the numerical computations performed for the large $N$ model in Section~\ref{sec:LargeN}.

\subsection{Mode truncation and asymptotic completion}
\label{subsec:Numerics_ModeSums}

\paragraph{Mode truncation.}
We solve the radial Green equation~\eqref{eq:LargeN_RadialGreenEquation} on a logarithmic radial grid, with boundary conditions determined by the near-defect OPE and the large-distance massive asymptotics.
The homogeneous thermal reference is subtracted mode by mode as in~\eqref{eq:LargeN_ModeSubtraction}.
Low Matsubara and angular modes are evaluated numerically, while the unresolved high modes are completed using WKB asymptotics.
The resolved-mode cutoffs and the radial resolution are then varied to assess the sensitivity of the resulting saddle.

\paragraph{WKB.}
After removing the radial first derivative and using $\log(z)$ as the radial coordinate, the effective frequency is
\begin{equation}
    Q_{n\ell} (z)^2
    =
    \left( \ell+\frac{d-3}{2} \right)^2
    +
    z^2 \left[ (2\pi n)^2 + F_s(z) \right]\,.
\end{equation}
The diagonal Green function $z = z'$ admits the local expansion
\begin{equation}
    \gamma_{n\ell}^{\mathrm{WKB}} (z,z)
    =
    z^{3-d}
    \left[
        \frac{1}{2Q_{n\ell}}
        -
        \frac{(z\partial_z)^2 Q_{n\ell}^2}{16 Q_{n\ell}^5}
        +
        \frac{5\bigl[ z\partial_z Q_{n\ell}^2 \bigr]^2}{64 Q_{n\ell}^7}
        +\ldots
    \right]\,.
\end{equation}
The reference contribution, i.e., the thermal Green function in the absence of the defect, is obtained by replacing $F_s(z)$ with $\mu_d^2$.
We subtract the two expressions and expand their difference through inverse-mode order five.
This should be understood as a joint large-mode expansion at fixed radial position, not a uniform approximation as $z \to 0$.
The remaining mode sums are evaluated on successively larger rectangles and extrapolated using low-degree polynomials in the inverse cutoff scale.

\subsection{Extraction of thermal observables}
\label{subsec:Numerics_Observables}

\paragraph{Bulk profiles and OPE coefficients.}
The profiles $F_\phi$ and $F_s$ are represented by smooth interpolants matched to their UV and IR asymptotic expansions.
Their parameters are determined by solving the mean-field and gap equations self-consistently.
The near-defect fit uses the OPE Ansatz specified in the main text: the dimensions and bulk-defect coefficients are fixed zero-temperature inputs, whereas the thermal one-point coefficients are determined numerically.
The retained multi-particle contributions obey the appropriate large-$N$ factorization relations.

\begin{figure}
    \centering
    \begin{minipage}{.5\textwidth}
        \centering
        \includegraphics[width=0.95\linewidth]{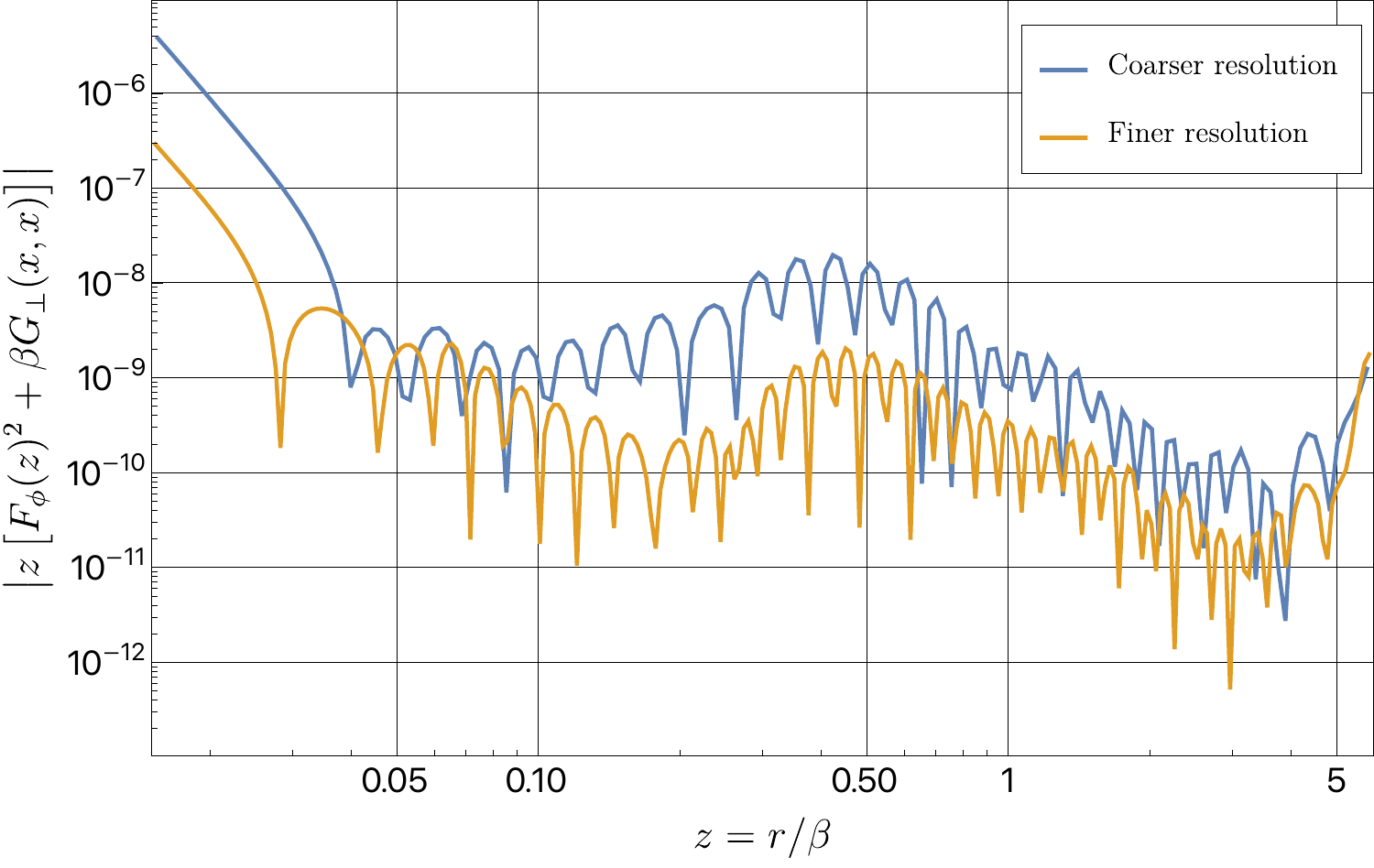}
    \end{minipage}%
    \begin{minipage}{0.5\textwidth}
        \centering
        \includegraphics[width=0.95\linewidth]{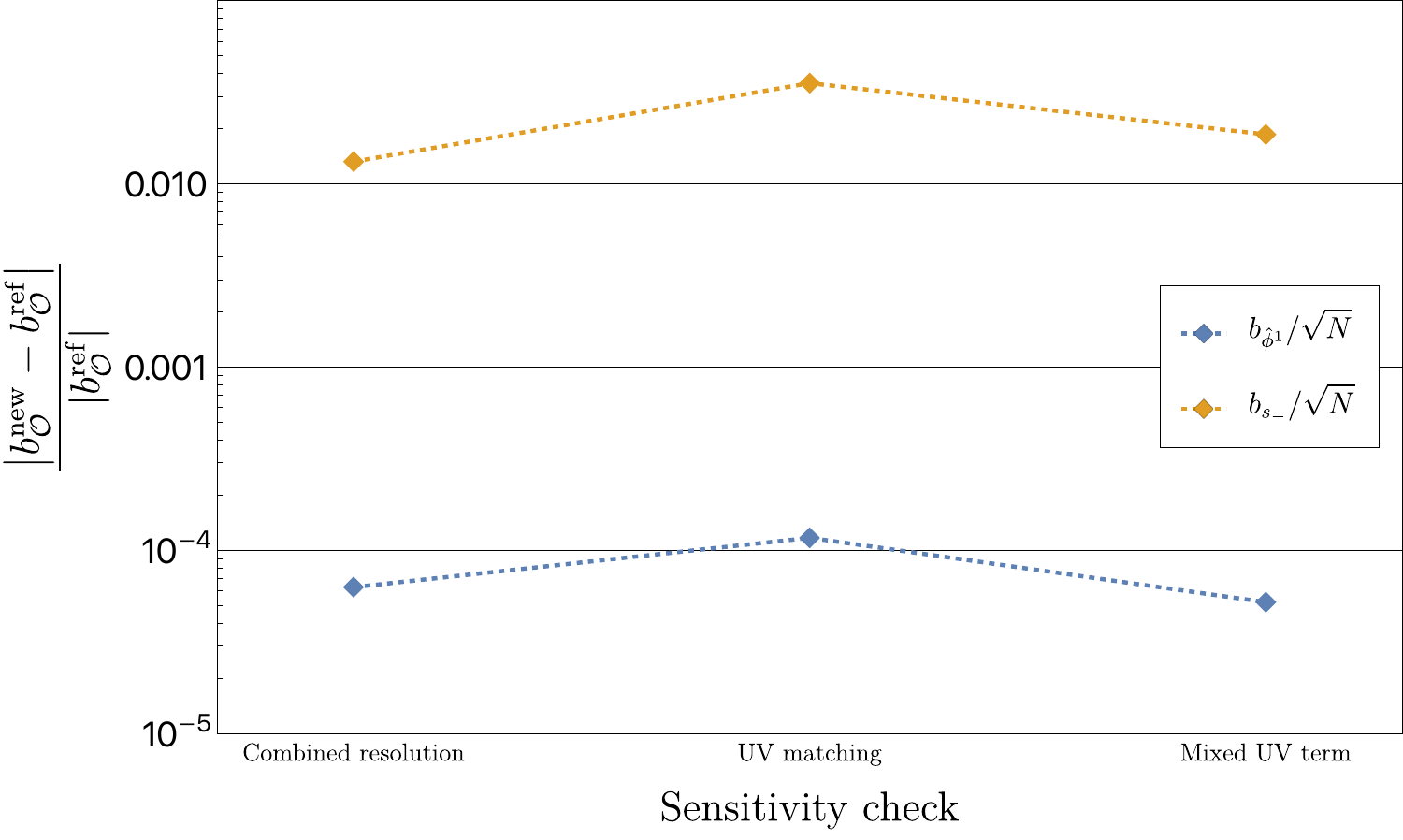}
    \end{minipage}
    \caption{\textbf{Left}: Absolute value of the UV-rescaled gap-equation residual $z[F_\phi(z)^2+\beta G_\perp(x,x)]$ at $d=3$.
    The coarse (blue) and fine (orange) resolutions use $(N_z,n_{\max},\ell_{\max},N_{\mathrm{knots}})=(400,40,30,22)$ and $(640,64,40,28)$, respectively, with fifth-order WKB completion.
    For $z<0.06$, the profiles are represented by the near-defect OPE Ansatz.
    \textbf{Right}: Relative changes in the extracted coefficients $b_{\hphi^1}$ (blue) and $b_{s_-}$ (orange) under three separate variations relative to the same coarse reference: increasing the resolution, moving the UV matching point from $0.06$ to $0.04$, and adding the mixed power $z^{\Dh_{\hphi^1}+\Dh_{s_-}}$.
    The fitted amplitudes are reoptimized in each case.
    Connecting lines guide the eye; these sensitivity checks do not constitute total error bounds.}
    \label{fig:LargeN_Numerics}
\end{figure}

In addition to the operators displayed explicitly in the main text, the numerical Ansatz retains the next scalar, whose dimension at $d=3$ is $\Dh_X \sim 3.7971598$, with a freely fitted amplitude.
The coefficient of the quadratic contribution $z^{2\Dh_{\hphi^1}}$ is constrained by factorization and the bulk equations, using its dimension-dependent relation to the leading amplitude rather than a fixed three-dimensional relation.
We also test the sensitivity to including the mixed power $z^{\Dh_{\hphi^1}+\Dh_{s_-}}$, as illustrated in Figure~\ref{fig:LargeN_Numerics}.
We vary the matching window, interpolation resolution, and retained subleading terms to assess the stability of the extraction.
In particular, note that the coefficient of a subleading OPE power can be substantially less stable than the profile itself.
For this reason, we do not extract OPE coefficients beyond the leading operators $\hphi^1$ and $s_-$.

For the finite-volume points $\beta/R=6,8,10$, we additionally subtract the discretization error evaluated on the known identity-reference profile, while retaining its physical thermal contribution.
The quoted uncertainties include an empirical allowance for sensitivity to the UV Ansatz and this correction procedure, rather than only the observed numerical refinement shifts.

\paragraph{Defect correlators.}
The tilt correlator is obtained by taking the boundary limit of the transverse propagator.
For each Matsubara frequency, we match the IR-decaying radial solution to its near-defect expansion and extract the normalizable coefficient.
We subtract the vacuum contribution before Fourier reconstruction and restore its analytically summed expression $\pi^2\beta^{-2}\csc^2(\pi\tau/\beta)$ afterwards.

The remaining high-frequency tail is determined by the short-distance OPE.
An OPE term $a_{\Oh}(\tau/\beta)^{\Dh_{\Oh}-2}$ in the dimensionless tilt correlator contributes the following large-$n$ Fourier coefficient:
\begin{equation}
    2 a_{\Oh}\,
    \Gamma(\Dh_{\Oh}-1)
    \sin \left( \frac{\pi \Dh_{\Oh}}2 \right)
    (2\pi n)^{\smash{1-\Dh_{\Oh}}}\,.
\end{equation}
Thus the Fourier completion uses OPE powers rather than the local WKB expansion used for the saddle.
For the reported three-dimensional tilt correlator, modes through $n=300$ are retained directly, and the remaining tail uses the $\hphi^1$ and $s_-$ contributions.
The power-law tails are summed analytically using the corresponding Hurwitz-$\zeta$ periodic blocks.
Their coefficients come from the bulk-profile OPE data; agreement with those same inputs is therefore a consistency check rather than an independent determination.

For the longitudinal singlet, the transverse radial problem is replaced by the coupled fluctuation system described in the main text.
The boundary limit gives the connected correlator.
The product of thermal one-point functions is added separately to obtain the full correlator.

\subsection{Precision and convergence checks}
\label{subsec:Numerics_Convergence}

\paragraph{Equation residuals.}
We monitor the mean-field and gap residuals after removing their leading UV powers:
\begin{align}
    &z^{(d+2)/2}
    \left[
        F_\phi''(z)
        +\frac{d-2}{z}F_\phi'(z)
        -F_s(z)F_\phi(z)
    \right],
    \\
    &z^{d-2}
    \left[
        F_\phi(z)^2
        +\beta^{d-2}G_\perp (x,x)
    \right].
\end{align}
Note that a small residual on the fitting grid alone does not establish convergence.
A stronger check consists in evaluating the equations at additional radial points and with increased mode cutoffs.
When the mean-field equation is used to eliminate one profile, its residual primarily tests the radial integration, whereas the gap residual tests self-consistency.

\paragraph{Stability of extracted observables.}
Numerical sensitivity is assessed by varying the radial interval, interpolation resolution, mode cutoffs, asymptotic completion, and OPE matching window.
For defect correlators, we additionally vary the Fourier cutoff and the boundary-extraction window.
Equation residuals and observable stability test different sources of error and must be considered together.

The quoted uncertainties are empirical sensitivity estimates, not rigorous bounds.
In particular, correlator uncertainties obtained on a fixed constrained saddle do not bound errors from omitted terms in the saddle Ansatz.